\RequirePackage{rotating}  

\documentclass[final,5p,times,twocolumn]{elsarticle}

\usepackage{amssymb}
\usepackage{amsmath}

\usepackage{xspace}
\usepackage{color}
\usepackage{soul} 
\usepackage{enumitem} 
\definecolor{brickred}{rgb}{0.8, 0.1, 0.1}

\usepackage{graphicx}

\usepackage{graphicx}
\usepackage[normalem]{ulem}
\usepackage{wrapfig}
\usepackage{siunitx}
\usepackage{textcomp}
\usepackage{amsmath}

\usepackage[dvipsnames]{xcolor}

\usepackage{multirow}

\usepackage{hyperref}

\usepackage{tabularray}

\usepackage{xurl}

\usepackage{natbib}

\begin{document}

\begin{frontmatter}



\title{Virtual Reality in Manufacturing Education: A Scoping Review Indicating State-of-the-Art, Benefits, and Challenges Across Domains, Levels, and Entities}


\author[label1]{Ananya Ipsita\corref{cor1}} 
\ead{aipsita@purdue.edu}
\cortext[cor1]{Corresponding author}

\author[label2,label1]{Ramesh Kaki} 
\author[label1]{Ziyi Liu} 
\author[label1]{Mayank Patel} 
\author[label1]{Runlin Duan} 
\author[label4]{Lakshmi Deshpande} 
\author[label6]{Lin-Ping Yuan} 
\author[label3]{Victoria Lowell} 
\author[label5]{Dr. Ashok Maharaj} 
\author[label9,label10]{Kylie Peppler} 
\author[label8]{Steven Feiner} 
\author[label1,label7]{Karthik Ramani} 

\affiliation[label1]{organization={School of Mechanical Engineering},
            addressline={Purdue University}, 
            city={West Lafayette},
            postcode={47907}, 
            state={IN},
            country={United States}}

\affiliation[label2]{organization={Department of Mechanical Engineering},
            addressline={Birla Institute of Technology and Science, Pilani, Hyderabad Campus},
            city={Hyderabad},
            postcode={500078},
            state={Telangana},
            country={India}}

\affiliation[label7]{organization={School of Electrical and Computer Engineering},
            addressline={Purdue University}, 
            city={West Lafayette},
            postcode={47907}, 
            state={IN},
            country={United States}}

\affiliation[label3]{organization={Learning Design and Technology},
            addressline={Purdue University}, 
            city={West Lafayette},
            postcode={47907}, 
            state={IN},
            country={United States}}

\affiliation[label8]{organization={Department of Computer Science},
            addressline={Columbia University}, 
            city={New York},
            postcode={10027}, 
            state={NY},
            country={United States}}

\affiliation[label9]{organization={School of Information and Computer Sciences},
            addressline={University of California}, 
            city={Irvine},
            postcode={92697}, 
            state={CA},
            country={United States}}

\affiliation[label10]{organization={School of Education},
            addressline={University of California}, 
            city={Irvine},
            postcode={92697}, 
            state={CA},
            country={United States}}

\affiliation[label6]{organization={Hong Kong University of Science and Technology},
            city={Hong Kong SAR},
            country={China}}

\affiliation[label5]{organization={Head},
            addressline={TCS Avapresence- Spatial Intelligence Lab}
            }

\affiliation[label4]{organization={Immersive Design Lead},
            addressline={TCS Avapresence- Spatial Intelligence Lab}
            }

\begin{abstract}
To address the shortage of a skilled workforce in the U.S. manufacturing industry, immersive Virtual Reality (VR)-based training solutions hold promising potential. To effectively utilize VR to meet workforce demands, it is important to understand the role of VR in manufacturing education. Therefore, we conduct a scoping review in the field. As a first step, we used a 5W1H (What, Where, Who, When, Why, How) formula as a problem-solving approach to define a comprehensive taxonomy that can consider the role of VR from all relevant possibilities. Our taxonomy categorizes VR applications across three key aspects: (1) \textcolor{NavyBlue}{\textbf{Domains}}, (2) \textcolor{PineGreen}{\textbf{Levels}}, and (3) \textcolor{Violet}{\textbf{Entities}}. Using a systematic literature search and analysis, we reviewed 108 research articles to find the current state, benefits, challenges, and future opportunities of VR in the field. It was found that VR has been explored in a variety of areas and provides numerous benefits to learners. Despite these benefits, its adoption in manufacturing education is limited. This review discusses the identified barriers and provides actionable insights to address them. These insights can enable the widespread usage of immersive technology to nurture and develop a workforce equipped with the skills required to excel in the evolving landscape of manufacturing.

\end{abstract}




\end{frontmatter}



\section{Introduction}
Manufacturing education plays an important role in preparing a skilled workforce for the U.S. manufacturing sector \cite{todd2001manufacturing, zhao2019developing}. As the industry transitions towards automation, smart manufacturing, and sustainability, there is an urgent need to upskill future generations of workers as well as reskill the current workforce to remain globally competitive \cite{nyt_ai_aging_shift_2024, economist_manufacturing_delusion_2023, nyt_trump_harris_2024}. In addition, the industry faces a demographic shift with an aging workforce and labor shortages \cite{nyt_manufacturing_good_old_days_2024, nyt_ai_aging_shift_2024}, as well as productivity challenges due to slower automation adoption compared to global leaders such as South Korea \cite{economist_manufacturing_inefficiency_2023}. To meet the growing demands of skilled labor, there is a need to introduce widespread skill training solutions. Virtual Reality (VR) technology presents promising potential to enhance such efforts by providing immersive and interactive learning experiences \cite{lowell2024applying, Lowell2024}.

\begin{figure*}[tb]
    \centering
    \includegraphics[width = 0.75\textwidth]{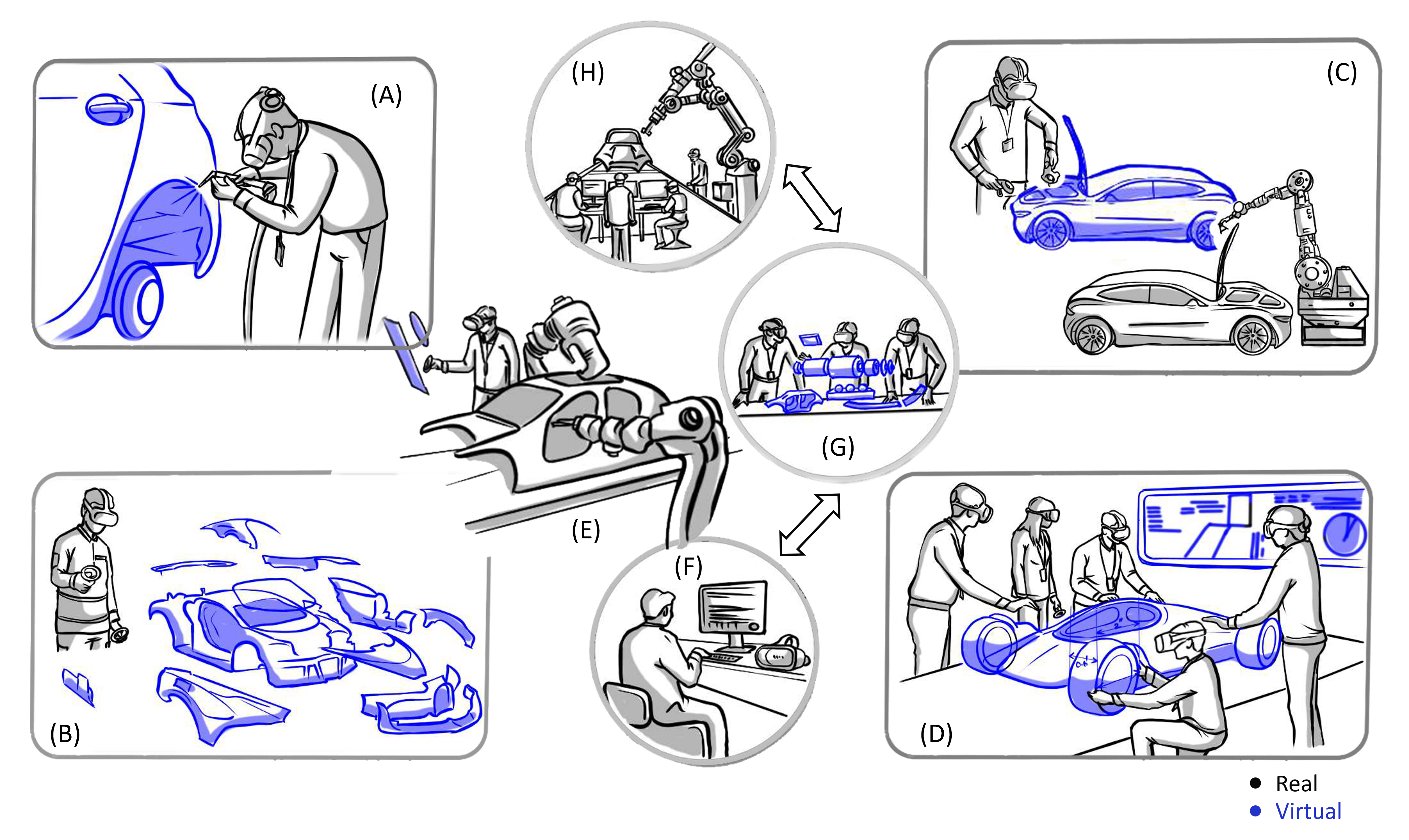}
    \caption{Some examples of VR-based learning applications in the field of manufacturing education include: (A) An industry worker learning welding on a car chassis using VR, (B) A university student learning about the assembly of car parts using VR, (C) A university student programs a robot to perform an automotive operation, and (D) Student team performing collaborative design reviews in their design project. Such immersive applications prepare learners at various levels (university or industry settings) to learn skills (welding or robotics) to prepare them for (E) the requirements in the manufacturing industry. The design of such immersive applications is achieved using collaboration between (F) Developers in charge of creating the content, (G) Implementers in charge of instruction delivery, and (H) Learners utilizing the content to achieve desired goals.}
    \label{fig:ReviewScope}
\end{figure*}

To effectively utilize VR for skill training and learning in manufacturing, it is essential to understand the trends, gaps, and opportunities in the field. Therefore, our research conducts a scoping review on the role of VR in manufacturing education to explore these details. Previous research has reviewed VR applications in industrial training \cite{naranjo2020scoping}, specific manufacturing fields \cite{guo2020applications, leu2013cad, yang2023use, berg2017industry}, and broader educational contexts \cite{radianti2020systematic, rojas2023systematic}. However, these studies lack a comprehensive taxonomy to categorize the role of VR in manufacturing education. As a result, they do not identify the unique opportunities and gaps in this field to meet the demands of the workforce. In contrast, our research fills this gap by developing a comprehensive taxonomy based on a 5W1H (What, Why, Who, Where, When, How) formula that identifies all relevant possibilities to explore the role of VR in manufacturing education. This taxonomy classifies VR applications based on three key aspects: (1) \textcolor{NavyBlue}{Domains} covering conventional manufacturing skills (e.g., welding, machining) and modern manufacturing skills (e.g., automation, robotics); (2) \textcolor{PineGreen}{Levels} distinguishing between formal education (K-12, university, vocational training) and work-based learning (on-the-job-training); and (3) \textcolor{Violet}{Entities} identifying stakeholders involved in developing, implementing, and utilizing VR technologies in education. A brief overview of the relevance of the different aspects is presented in Fig. \ref{fig:ReviewScope}.

Using the taxonomy we developed, a systematic review of 108 articles was conducted to identify the current state, benefits, challenges, and future opportunities of VR in manufacturing education. Our findings show that in recent decades, VR has been explored in several areas of manufacturing education, such as welding \cite{price2019using} and assembly operations \cite{aqlan2019integrating}, automotive manufacturing, design, sustainability \cite{chiou2024virtual}, additive manufacturing (AM) \cite{ostrander2020evaluating}, and more. Its integration has enhanced learning experiences in educational and industrial settings \cite{abele2017learning, carruth2017virtual}. VR offers several benefits \cite{gonzalez2017immersive, carruth2017virtual, price2019using}, such as realistic experiences without physical equipment or exposure to potential hazards \cite{poyade2021transferable}, increased student motivation and interest in learning content \cite{azzam2024virtual}, and spatial interaction with 3D models that facilitate understanding of complex manufacturing processes \cite{el2016assessment}. Despite these benefits, several barriers hinder its widespread adoption to meet the growing demand for skilled professionals in the labor market today \cite{badamasi2022drivers, scott2020investigation, jalo2021state, fernandez2017augmented, liagkou2019realizing}. Some examples include high costs of implementation on a large scale, limited infrastructure, and resistance to shifting from traditional instructional methods \cite{badamasi2022drivers, scott2020investigation, jalo2021state, fernandez2017augmented, liagkou2019realizing, nyt_trump_harris_2024}. These findings finally contribute to understanding the main barriers to VR adoption and developing actionable insights to address them. We believe such insights can be useful in enhancing the training potential of VR by making immersive and interactive experiences more accessible to learners.

The structure of the paper is as follows. Section \hyperref[sec:2]{2} provides a related work section that compares the current review process with existing studies. Section \hyperref[sec:3]{3} provides the methodology followed to conduct the scoping review. Sections \hyperref[sec:4]{4}, \hyperref[sec:5]{5}, and \hyperref[sec:6]{6} present the results of the review process, highlighting the current state, benefits, challenges, and future opportunities across Domains, Levels, and Entities, respectively. Section \hyperref[sec:7]{7} provides a discussion on the review results. Finally, Section \hyperref[sec:8]{8} provides a conclusion of the work.

\label{sec:2}
\section{Comparison with Existing Review Literature}

Researchers have conducted some reviews on VR-based skill training and learning in the field of manufacturing. For example, VR-based industrial training reviews such as that of Naranjo et al. \cite{naranjo2020scoping} assess the applications of VR in multiple industries. Broader Extended Reality (XR) training reviews such as Doolani et al. \cite{doolani2020review} examine VR and Augmented Reality (AR) applications for manufacturing training. Although these reviews are important to provide an overview of VR-based training applications in the industry, they face challenges in identifying the unique opportunities and barriers of VR adoption to address the shortage of skilled labor. This is in part due to the limited scope of the review which lacks a comprehensive taxonomy to define the nature of virtual reality in manufacturing education.

Some reviews have also considered field-specific manufacturing applications. Guo et al. \cite{guo2020applications} systematically review the role of VR in industrial maintenance during the product lifecycle. Their primary focus is on equipment maintenance, predictive maintenance, and workforce efficiency. Leu et al. \cite{leu2013cad} focus on simulation and planning of virtual assemblies based on CAD. Industry-oriented reviews conducted by Yang et al. \cite{yang2023use} and Berg et al. \cite{berg2017industry} explore the adoption of VR in product design and industrial manufacturing. Dreyfus et al. \cite{dreyfus2022virtual} emphasize virtual metrology and predictive quality estimation. Cali et al. \cite{cali2022opportunities} explore metaverse applications and digital transformation strategies. Basu et al. \cite{basu20226g} focus on 5G and smart CPS applications. Although these reviews highlight important technological advances, they lack an explicit focus on manufacturing education. General reviews on VR in education conducted by Radianti et al. \cite{radianti2020systematic} and Rojas-S\'{a}nchez et al. \cite{rojas2023systematic} also provide insights into instructional design and learning theories. However, it is worth noting that they do not address the unique challenges and opportunities within manufacturing training.

In contrast to previous work, our research fills this gap by developing a comprehensive taxonomy and then conducting a scoping review using the taxonomy to identify trends, gaps, and opportunities in the field. Using the review's findings, we try to answer the following research question.

\emph{What are the different barriers that hinder the adoption of VR in manufacturing education? What efforts can address these barriers to facilitate greater adoption of VR in manufacturing education?}

\label{sec:3}
\section{Research Methodology}
The methodology followed to achieve the research objectives was carried out in two steps: (1) Taxonomy Development and (2) Systematic Review.

\subsection{Taxonomy Development}

\begin{figure*}[tb]
    \centering
    \includegraphics[width = \textwidth]{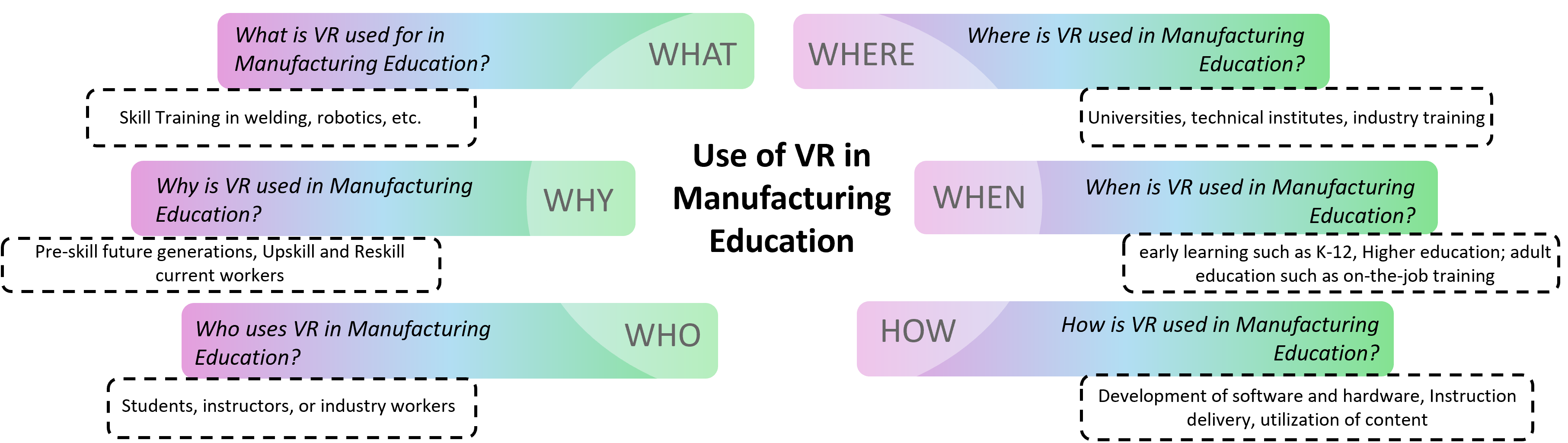}
    \caption{5W1H formula used to develop the primary taxonomy}
    \label{fig:Research}
\end{figure*}

To holistically situate the role of VR in manufacturing education, we used a 5W1H formula to answer specific questions as shown in Fig. \ref{fig:Research}. The 5W1H method, an innovative and attractive problem-solving approach, is widely used to provide a comprehensive analysis of a specific problem or knowledge \cite{awal2018ontology}. Although the answers to these questions were sometimes repetitive, we found that the answers could be grouped primarily under the three major aspects that formed the basis of the primary taxonomy used in our review analysis. To make it easier for the reader to follow, the primary classification of aspects is described in this section. The secondary classification for each aspect is explained in the corresponding section.

\textcolor{NavyBlue}{Domains:} Manufacturing education involves instructing people in both conventional and modern manufacturing skills. Conventional manufacturing education programs include understanding production processes, materials management, and quality assurance, in addition to mastering a wide variety of technical skills, which include expertise in mechanics, fabrication (welding and machining), and industrial troubleshooting \cite{elshennawy2015manufacturing}. Modern manufacturing education programs have evolved beyond the traditional emphasis on the Industrial Age of metalworking and welding \cite{ManufacturingEducation, raman2014manufacturing, groover2020fundamentals}. They encompass a broader scope that includes areas such as system engineering, emerging technologies such as the Internet of Things (IoT), digital toolsets, and more \cite{mourtzis2018cyber}. To encompass the broad spectrum of manufacturing skills, we categorize the relevant papers into two domains: (A) \emph{\textcolor{NavyBlue}{Conventional}}, which encompass traditional skills such as machining and welding, and (B) \emph{\textcolor{NavyBlue}{Modern}}, which encompasses skills such as digital manufacturing and automation. This aspect aligns with the dimensions ``What'' and ``Why'' of the 5W1H framework, as shown in Fig. \ref{fig:Research}. It captures the specific skills and knowledge areas that VR-based manufacturing education aims to develop. Both sets of skills are vital, and modern manufacturing builds on the foundational knowledge of conventional methods to innovate and improve manufacturing processes.

\textcolor{PineGreen}{Levels:} Furthermore, the nature of the manufacturing education programs is found to be quite diverse. While some may prepare employees for manufacturing roles with on-the-job training, there are opportunities for students in fields related to science, technology, engineering, and mathematics (STEM) \cite{spak2013us}. This shows the diverse pathways available within the manufacturing sector catering to individuals with varying educational backgrounds and skill sets \cite{chryssolouris2013manufacturing}. The second aspect, Levels, corresponds to the ``When'' and ``Why'' dimensions of the 5W1H framework, as shown in Fig. \ref{fig:Research}. It considers time factors, learning progression, and the integration of VR at different stages of education and professional development. By focusing on the type of education and training delivery, the classification of manufacturing education can be broadly consolidated into two main categories: (A) \emph{\textcolor{PineGreen}{Formal Education}}, and (B) \emph{\textcolor{PineGreen}{Work-based Learning}}.

\textcolor{Violet}{Entities:} Finally, VR-based learning content must be tailored to the specific needs of different user groups and learning environments. Several entities are involved in making immersive content effective for the students \cite{gaspar2020research, cassola2021novel}. These may range from the development of the content to the ones implementing and delivering the content, and finally to the ones that utilize the content. Design-based iteration loops through these entities make the content effective in achieving the target learning outcomes. The third aspect, Entities, addresses the ``Who'', ``Where'' and ``How'', focusing on the stakeholders involved in the development, implementation, and utilization of VR applications. We review and categorize the works into three entities: (A) \emph{\textcolor{Violet}{Development}}, (B) \emph{\textcolor{Violet}{Implementation}}, and (C) \emph{\textcolor{Violet}{Utilization}}, as shown in Fig. \ref{fig:Research}. The categorization of entities helps to understand the collaborative ecosystem of VR in manufacturing education. This highlights how different stakeholders contribute to its success from development to implementation and adoption.

\begin{figure*}[tb]
    \centering
    \includegraphics[width = 0.75\textwidth]{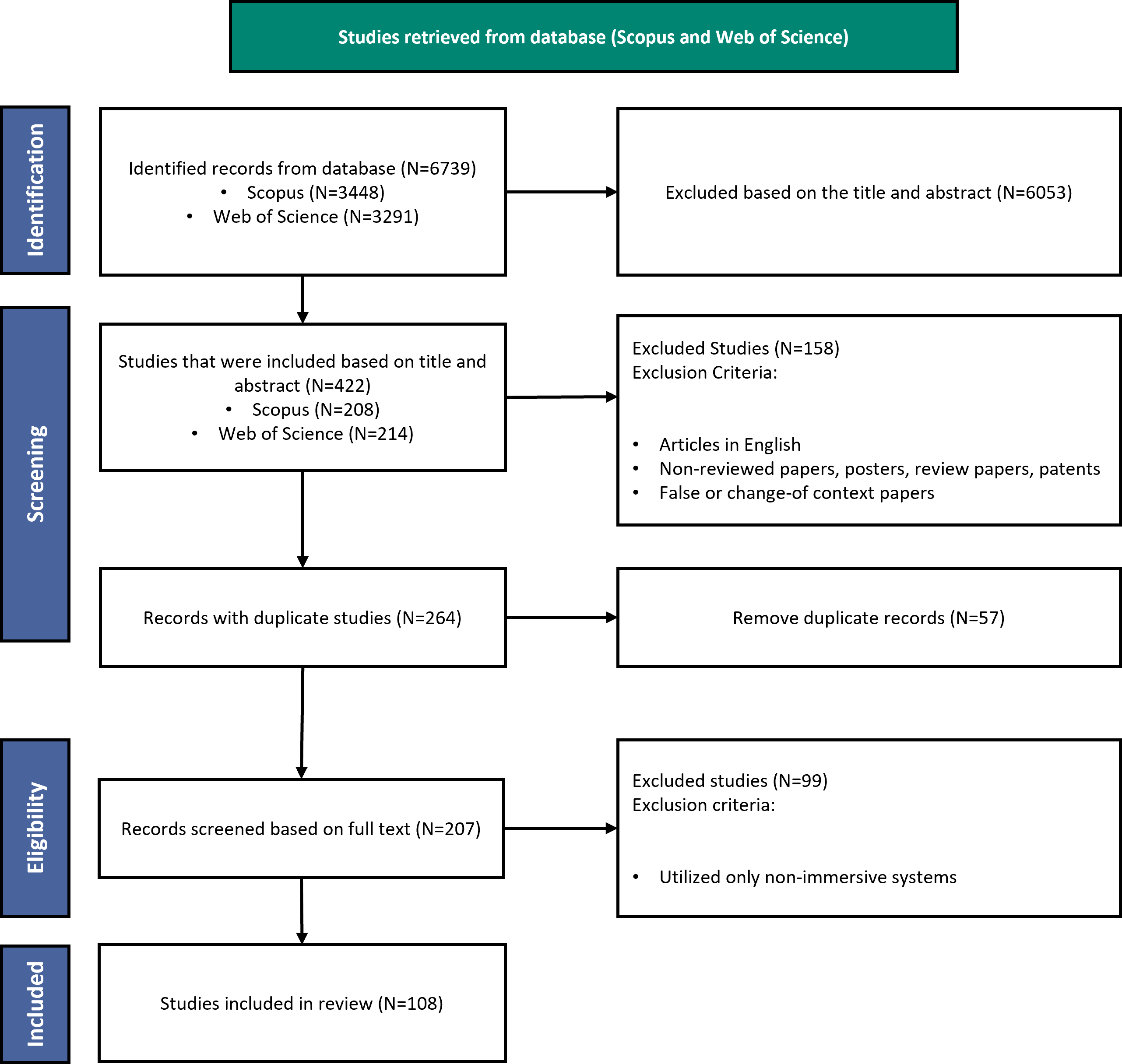}
    \caption{Process followed to find the relevant articles for final review}
    \label{fig:ArticleProcessing}
\end{figure*}

\subsection{Systematic Review} After developing the taxonomy, we proceeded to perform a systematic review and analysis using the PRISMA guideline \cite{page2021prisma} : (1) Search Strategy, (2) Identification, (3) Screening, (4) Coding, and (5) Analysis. Each of these methods is described in detail as follows:

\subsubsection{Search Strategy:} The first step to finding relevant articles included: (1) identifying keywords for data retrieval, (2) defining the affiliated country for search, (3) the publication timeline, and (4) the type of document of the publication. The identification of keywords was based on their relevance to the topic, which primarily focused on three independent areas: VR, education, and manufacturing. Although manufacturing holds global significance, our review primarily focuses on the US manufacturing sector. Therefore, the search restricted the affiliation to enable a more targeted analysis of technological trends specific to the US. It was also observed that this accounted for the largest number of publications, which highlights the research focus in this area. Furthermore, 2010 was chosen as the starting point due to the rise of immersive VR technology following advances in VR hardware after that year. Finally, only peer-reviewed and published articles from journals, conference proceedings, and workshop papers were planned to be considered for the review.

\subsubsection{Identification:} We conducted a search on Scopus and Web of Science platforms to collect articles relevant to the research topic. We adopted a slightly different technique for the data retrieval process using the two platforms due to restrictions on the query length and to improve the efficiency of data collection.

For Scopus, we used a keyword search in \emph{Article Title}, \emph{Abstract} and \emph{Keywords}, and applied filter criteria for \emph{Subject Area}, affiliated \emph{Country} as United States, \emph{Language} as English, \emph{Document Type} as Conference Proceedings and Journals, \emph{Publication Type} as Final and \emph{Publication Year} ranging from 2010 to 2025. The keywords used in the search can be found in Table \ref{tab:review_keywords}. In addition to using various expressions and abbreviations of VR, education, and manufacturing, additional keywords were defined focusing on any specific manufacturing skill. This skill-specific information was obtained using information collected from standard educational and industry frameworks that will be discussed later in Section 4. This decision was made for two reasons: (1) the search for keywords in \emph{Article Title}, \emph{Abstract}, and \emph{Keywords} would be more efficient than a search conducted in all fields. (2) Restricting the search to these fields may omit relevant papers that focused on the use of VR in any specific manufacturing skill but did not use the standard expressions of manufacturing in the fields. Furthermore, the refinement criteria also excluded topics from the \emph{Subject Area}, that were considered outside the scope of the review. The Scopus search process yielded 3448 articles.

A similar process was followed to obtain relevant articles from the Web of Science. Due to limitations on search query length, the initial keyword search was conducted in all fields. Applying refinement criteria for \emph{Country}, \emph{Year of Publication}, \emph{Language}, and \emph{Document Type} yielded 3,291 articles.

\subsubsection{Screening:} The initial corpus of 6739 articles was then divided into 3 equal parts, each of which was individually screened by three researchers. The authors read the title and abstract of the paper and screened the paper based on the following inclusion and exclusion criteria: (1) We focused on including articles that were only written in English. (2) We included peer-reviewed conference papers such as journals, conferences, and workshop papers. (3) We excluded non-reviewed papers (e.g. arXiv papers), posters, literature review papers, and patents. (4) We removed false and change-of-context papers that were not focused on VR, education, and manufacturing, but were present because of keywords. If the authors were unsure about the paper, a discussion was held on Zoom to decide its consideration for the review. In case confusion remained, the paper was included to move onto the next step in the review process. This process resulted in 264 papers, which were then checked for duplicates across database platforms. After removing 57 duplicate articles, a final set of 207 papers was downloaded for detailed review.

\begin{table*}[tb]
\centering
\caption{Search Query utilized to retrieve data from Scopus and Web of Science}
\label{tab:review_keywords}
\resizebox{\textwidth}{!}{%
\begin{tabular}{|cllll|}
\hline
\multicolumn{5}{|c|}{\textbf{Scopus}} \\ \hline
\multicolumn{5}{|c|}{\begin{tabular}[c]{@{}c@{}}( TITLE-ABS-KEY ( "virtual reality" OR "VR" ) AND TITLE-ABS-KEY ( "educat*" OR "learn*" OR "train*" OR "skill*" OR "instruct*" OR "teach*" ) AND TITLE-ABS-KEY \\ ( ( "manufactur*" OR "industr*" OR "production" ) OR ( "machin*" OR "turn*" OR "mill*" OR "drill*" OR "grind*" OR "form*" OR "forg*" OR "stamp*" OR "bend" * OR "cast*" \\ OR "weld*" OR "join*" OR "braz*" OR "solder*" OR "tool*" OR ( "die-making" OR "die making" ) OR ( "quality control*" OR "quality-control*" ) OR "inspect*" OR ( "process \\ control*" OR "process-control*" ) OR "SPC" ) OR ( "CAD" OR "CAM" OR ( "computer-aided" OR "computer aided" ) OR ( "additive manufactur*" OR "additive-manufactur*" ) \\ OR ( "3D print*" OR "3D-print*" ) OR "robot*" OR ( "Automated Guided Vehicle*" OR "Automated-Guided-Vehicle*" OR "Automated-Guided Vehicle*" ) OR "AGV*" OR "CNC*"\\  OR "automat*" OR ( "smart manufactur*" ) OR ( "Internet of thing*" OR "Internet-of-thing*" ) OR "IoT" OR ( "cyber physical system*" OR "cyber-physical system*" OR\\  "cyber-physical-system*" ) OR ( "big data" OR "bigdata" OR "big-data" ) OR "analytic*" OR "composite*" OR ( "nano material*" OR "nano-material*" OR "nanomaterial*" ) \\ OR "sustainab*" OR ( "green manufactur*" OR "green-manufactur*" ) OR "energy" OR ( "waste manage*" OR "waste-manage*" ) ) ) ) AND PUBYEAR \textgreater 2009 AND \\ PUBYEAR \textless 2026 AND ( LIMIT-TO ( PUBSTAGE,"final" ) ) AND ( LIMIT-TO ( AFFILCOUNTRY,"United States" ) ) AND ( EXCLUDE ( SUBJAREA,"MEDI" ) OR EXCLUDE \\ ( SUBJAREA,"BIOC" ) OR EXCLUDE ( SUBJAREA,"HEAL" ) OR EXCLUDE ( SUBJAREA,"NURS" ) OR EXCLUDE ( SUBJAREA,"AGRI" ) OR EXCLUDE ( SUBJAREA,"PHAR" )\\  OR EXCLUDE ( SUBJAREA,"IMMU" ) OR EXCLUDE ( SUBJAREA,"ECON" ) OR EXCLUDE ( SUBJAREA,"DENT" ) OR EXCLUDE ( SUBJAREA,"VETE" ) OR EXCLUDE \\ ( SUBJAREA,"Undefined" ) OR EXCLUDE ( SUBJAREA,"NEUR" ) OR EXCLUDE ( SUBJAREA,"PHYS" ) OR EXCLUDE ( SUBJAREA,"MATH" ) OR EXCLUDE ( SUBJAREA,"EART" ) \\ OR EXCLUDE ( SUBJAREA,"CENG" ) OR EXCLUDE ( SUBJAREA,"CHEM" ) ) AND ( LIMIT-TO ( DOCTYPE,"cp" ) OR LIMIT-TO ( DOCTYPE,"ar" ) ) AND ( LIMIT-TO\\  ( LANGUAGE,"English" ) )\end{tabular}} \\ \hline
\multicolumn{5}{|c|}{\textbf{Web of Science}} \\ \hline
\multicolumn{5}{|c|}{\begin{tabular}[c]{@{}c@{}}((ALL=("virtual reality" OR "VR")) AND ALL=("educat*" OR "learn*" OR "train*" OR "skill*" OR "instruct*" OR "teach*")) AND ALL=((“manufactur*” or “industr*” or “production”)) and \\ 2025 or 2012 or 2024 or 2023 or 2022 or 2021 or 2020 or 2019 or 2018 or 2017 or 2016 or 2015 or 2014 or 2013 or 2011 or 2010 (Publication Years) and USA (Countries/Regions) and Article \\ or Proceeding Paper or Review Article or Early Access (Document Types) and English (Languages)\end{tabular}} \\ \hline
\end{tabular}%
}
\end{table*}

\subsubsection{Coding:} In the context of the current research, VR-based environments refer only to immersive systems such as head-mounted devices (HMD) and heads-up display (HUD)-based systems, CAVE environments, and stereoscopic systems. These environments are designed to fully immerse users in a simulated setting, thereby inducing a compelling sense of presence through embodied interactions. Therefore, in our coding process, we followed this additional exclusion criterion: (1) We excluded the papers that discussed only non-immersive systems (desktop-based, tablet-based, and mobile-based).

The coding process was achieved in two stages. First, two authors independently read the articles to categorize them based on the type of VR system: HUD/HMD, stereoscopic systems, CAVE, or desktop-based systems. If immersiveness exclusion criteria were not met, the reviewer subsequently categorized the article based on the review taxonomy that focused on the three main aspects.

\subsubsection{Analysis:} Following the coding process, an inter-rater reliability analysis was conducted to evaluate the agreement between the authors' classifications. The analysis revealed high values of agreement for all categories. For papers with differing categorizations, the authors held discussions via Zoom to resolve inconsistencies, ultimately achieving full agreement. This process resulted in a total of 108 papers being included in the final review.

A decline in number of papers was found for the period between 2021-2023 in almost all categories. This suggests that external factors such as the COVID-19 pandemic influence research efforts in this field. The trends in data for devices suggested that new immersive technologies (such as HMDs) are gaining more traction over traditional CAVE-based environments, which are relatively more expensive and complicated to set up. The findings of the review and analysis for each aspect are presented in detail in Sections \hyperref[sec:4]{4}, \hyperref[sec:5]{5} and \hyperref[sec:6]{6}.

\label{sec:4}
\textcolor{NavyBlue}{\section{Domains of Manufacturing Education}}
First, the secondary taxonomy of domains is explained. Then the findings from the review and analysis are presented in terms of current state, benefits, challenges, and future opportunities of VR in the corresponding category.

\subsection{\textcolor{NavyBlue}{Domain Taxonomy: }}
\begin{figure*}[tb]
    \centering
    \includegraphics[width = 0.75\textwidth]{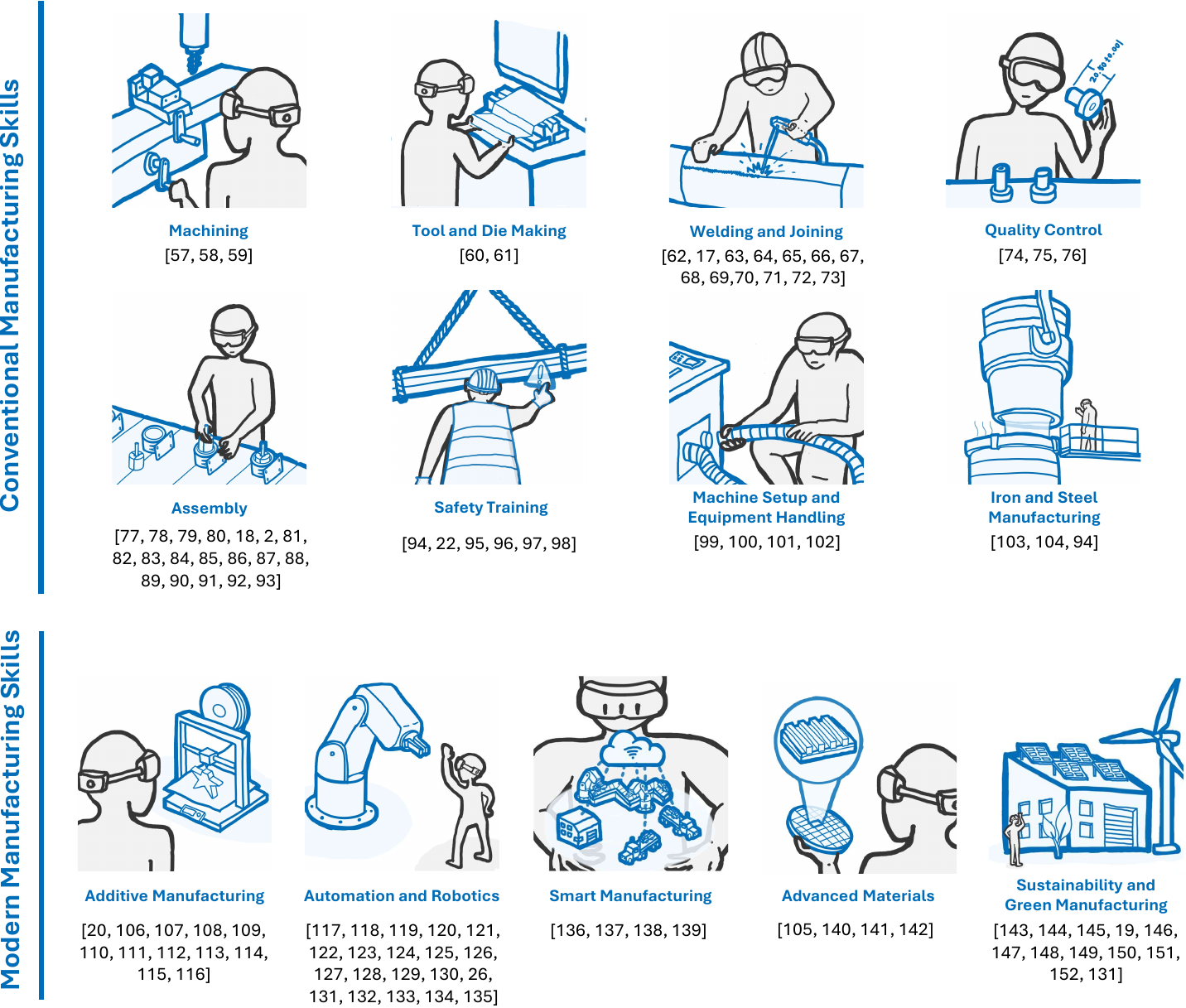}
    \caption{Virtual Reality enables education in conventional skills like welding and assembly, as well as emerging technologies such as smart manufacturing and robotics, preparing learners for the modern manufacturing landscape.}
    \label{fig:DomainClassification}
\end{figure*}

The manufacturing education domains are grouped into Conventional and Modern Manufacturing Skills as shown in Fig. \ref{fig:DomainClassification}. This association is based on educational standards and frameworks such as the Classification of Instructional Programs (CIP) \cite{CIP2025} and International Standard Classification of Education (ISCED) \cite{ISCED2025} and Industry Classification frameworks such as North American Industry Classification System (NAICS) \cite{NAICS2025} through its comprehensive coverage of skills. Conventional skills such as machining, welding, and inspection remain fundamental for manufacturing operations \cite{elmaraghy2021evolution, mourtzis2018development}. Meanwhile, modern manufacturing skills emphasize automation, AM, cyberphysical systems (CPS), and sustainability to optimize smart factories and improve environmental sustainability \cite{oztemel2020literature, hozdic2023evolution, adel2024convergence}. This classification captures the shift in workforce demands where traditional expertise provides stability and advanced technologies drive adaptability in an increasingly digitalized and data-driven manufacturing landscape \cite{li2024reskilling, mavrikios2013industrial}. 

\subsubsection{\textcolor{NavyBlue}{Conventional Manufacturing Education Skills:}} Conventional manufacturing education focuses on hands-on skills and familiarity with physical processes and materials. It includes (A) \emph{\textcolor{NavyBlue}{Machining}}, such as proficiency in operating lathes, mills, drills, and grinders; (B) \emph{\textcolor{NavyBlue}{Forming}}, involving knowledge of forging, stamping, bending, and casting; (C) \emph{\textcolor{NavyBlue}{Welding and Joining}}, such as competence in arc welding, gas welding, brazing, and soldering; (D) \emph{\textcolor{NavyBlue}{Tool and Die Making}}, involving the ability to design and fabricate tools, dies, and molds; (E) \emph{\textcolor{NavyBlue}{Inspection and Quality Control}} to ensure products meet quality standards; (F) \emph{\textcolor{NavyBlue}{Assembly}}, involving knowledge of assembling products and components; (G) \emph{\textcolor{NavyBlue}{Safety Training}} in safety protocols and accident prevention; (H) \emph{\textcolor{NavyBlue}{Machine Setup and Equipment Handling}}, involving familiarity with setting up machinery and handling industrial equipment; and (I) \emph{\textcolor{NavyBlue}{Iron and Steel Manufacturing}}, involving knowledge of manufacturing processes specific to iron and steel.

\subsubsection{\textcolor{NavyBlue}{Modern Manufacturing Education Skills:}} Modern manufacturing education emphasizes digital tools, automation, systems thinking, and sustainability. It typically includes (A) \emph{\textcolor{NavyBlue}{Additive Manufacturing}} that includes proficiency in CAD (Computer-Aided Design) and CAM (Computer-Aided Manufacturing) software for designing and producing parts and understanding of AM (3D printing) principles and applications; (B) \emph{\textcolor{NavyBlue}{Automation and Robotics}} that includes knowledge of programming and operating automated machinery and robots, including CNC (Computer Numerical Control) machines and AGVs (Automated Guided Vehicles); (C) \emph{\textcolor{NavyBlue}{Smart Manufacturing}} that includes understanding of Industry 4.0 concepts, including IoT (Internet of Things) integration, cyber-physical systems, and the ability to analyze big data for decision-making in manufacturing; (D) \emph{\textcolor{NavyBlue}{Advanced Materials}} that includes familiarity with the properties and manufacturing techniques for composites, and nanomaterials; and (E) \emph{\textcolor{NavyBlue}{Sustainability and Green Manufacturing}} that includes knowledge of energy-efficient manufacturing processes, waste management practices, and the selection of sustainable materials to minimize environmental impact.

\subsection{\textcolor{NavyBlue}{Findings from Review Analysis for Conventional Manufacturing Education Skills: }}

\subsubsection{\textcolor{NavyBlue}{State of the art:}} From the analysis, it was found that 49 articles were associated with the Conventional manufacturing domain: \emph{\textcolor{NavyBlue}{Machining}} (4), \emph{\textcolor{NavyBlue}{Tool and Die Making}} (2), \emph{\textcolor{NavyBlue}{Welding and Joining}} (13), \emph{\textcolor{NavyBlue}{Quality Control}} (3), \emph{\textcolor{NavyBlue}{Assembly}} (19), \emph{\textcolor{NavyBlue}{Safety Training}} (6), \emph{\textcolor{NavyBlue}{Machine Setup and Equipment Handling}} (5), \emph{\textcolor{NavyBlue}{Iron and Steel Manufacturing}} (3). Table \ref{tab:conventional_domain_classification} shows the distribution of papers reviewed across the different categories of conventional manufacturing skills. Overall, the trends in data showed a growing trend in conventional manufacturing over the observed period. Assembly skills appear frequently. This highlights a consistent demand for these skills within the context. Overall, there is a fluctuating but apparent interest in welding skills. This may be possibly due to the significance of the skill in manufacturing and the demand for experiential training in welding. 

{
\footnotesize
\begin{longtblr}[
  caption = {Distribution of papers reviewed across Conventional Manufacturing Education Skills},
  label={tab:conventional_domain_classification}
]{
  width = \linewidth,
  colspec = {Q[188]Q[752]},
  row{1} = {c},
  row{2} = {c},
  row{6} = {c},
  row{8} = {c},
  row{21} = {c},
  row{25} = {c},
  row{45} = {c},
  row{52} = {c},
  row{56} = {c},
  cell{2}{1} = {c=2}{0.94\linewidth},
  cell{6}{1} = {c=2}{0.94\linewidth},
  cell{8}{1} = {c=2}{0.94\linewidth},
  cell{21}{1} = {c=2}{0.94\linewidth},
  cell{25}{1} = {c=2}{0.94\linewidth},
  cell{45}{1} = {c=2}{0.94\linewidth},
  cell{52}{1} = {c=2}{0.94\linewidth},
  cell{56}{1} = {c=2}{0.94\linewidth},
  hlines,
  vlines,
}
\textbf{Reference}                                                                                             & \textbf{Short description indicating the use of VR for skill training}                                                                                                                                                                                                                              \\
\textbf{\textit{\textcolor{NavyBlue}{Machining}}}                                                                                    &                                                                                                                                                                                                                                                                                                     \\
Studer et al. \cite{studer2024open}                                                           & VR training for drilling with a 3-axis milling machine, emphasizes anopen-ended approach for psychomotor skill learning with multiple pathways to achieve the goal.                                                                                                                                 \\
Lie et al. \cite{lie2023training}                                                             & Comparative study showing VR and video-based training improves drilling task performance (higher rate, fewer mistakes, less time) over paper and video-based methods.                                                                                                                               \\
Park et al. \cite{park2023work}                                                               & VR-based mechanical engineering lab for machining; repeated practice of creating hammer parts using milling machines, drill presses, lathes.                                                                                                                                                        \\
\textbf{\textit{\textcolor{NavyBlue}{Forming}}}                                                                                      &                                                                                                                                                                                                                                                                                                     \\
{Moreland et al. \\ \cite{moreland2020integrating, moreland2022development}}                  & Developed an interactive virtual die casting simulator; integrates fluid flow simulations with machine parameters to improve operator understanding.                                                                                                                                                \\
\textbf{\textit{\textcolor{NavyBlue}{Welding and Joining}}}                                                                          &                                                                                                                                                                                                                                                                                                     \\
White et al. \cite{white2011low}                                                              & Simulated realistic welding using standard hardware (helmet and gun) and audio and visuals based on numerical heat transfer; accuracy is verified against real welds.                                                                                                                               \\
Price et al. \cite{price2019using}                                                            & Used VR to introduce entry-level industrial distribution undergraduate students to welding for hands-on real welding in labs.                                                                                                                                                                       \\
Ipsita et al. \cite{ipsita2022towards}                                                        & Developed a low-cost VR welding training system using backward design; higher learning gains and skill transfer compared to video training; role of haptics during virtual training is highlighted for training that requires physical tools.                                                       \\
Mclaurin et al.   \cite{mclaurin2012comparison}                                               & Compared fully virtual and integrated VR training for welding; integrated VR is necessary for higher difficulty tasks.                                                                                                                                                                              \\
Stone et al. \cite{stone2013full}                                                             & Compared fully virtual and integrated VR training for welding; integrated VR is necessary for higher difficulty tasks.                                                                                                                                                                              \\
Stone et al. \cite{stone2011physical}                                                         & Evaluated the various impacts of VR integrated training against traditional welding training methods; VR integrated training leads to improved performance and higher certification rates.                                                                                                          \\
Stone et al. \cite{stone2011virtual}                                                          & Evaluated the various impacts of VR integrated training against traditional welding training methods; integrated training achieved superior training outcomes, higher levels of team interaction, and reduced material costs compared to those trained solely through traditional methods.          \\
Byrd et al. \cite{byrd2015use}                                                                & Used VRTEX® 360 welding simulator to assess novice and experienced   welders' skills; welding experience had a large effect on the quality score for each weld type.                                                                                                                                \\
Lassiter et al.   \cite{lassiter2023welding}                                                  & Interviewed instructors on integration of AR, VR, and AI in welding training; emphasizes the need for hands-on instruction and a balanced, inclusive approach to preparing future welders.                                                                                                          \\
Ipsita et al. \cite{ipsita2024design,   ipsita2021vrfromx}                                    & Developed a VR prototyping system for welding to enable subject matter experts to author virtual environments from real-world scans; expertise with VR programming was not required to author content using the system.                                                                             \\
Ye et al. \cite{ye2023robot}                                                                  & Developed a robot-assisted motor skill training system to digitalize expert motor skills and guide novice trainees through proprioceptive and haptic feedback.                                                                                                                                      \\
Ye et al. \cite{ye2024user}                                                                   & Discussed robot-assisted VR welding training and compared effects of static, visual and haptic guidance on cognitive load during training.                                                                                                                                                          \\
\textbf{\textit{\textcolor{NavyBlue}{Quality Control}}}                                                                              &                                                                                                                                                                                                                                                                                                     \\
Belga et al. \cite{belga2022carousel}                                                         & Introduced the Carousel method for VR-based inspection training by allowing users to control relevant inspection possibilities.                                                                                                                                                                     \\
Srinivasa et al.   \cite{srinivasa2021virtual}                                                & Used VR for training in materials testing for manufacturing.                                                                                                                                                                                                                                        \\
Bowling et al.   \cite{bowling2010evaluating}                                                 & Providing feedforward information in a VR simulation of an aircraft cargo bay enhances visual inspection performance in terms of process measures (fixation points, fixation durations, and area covered).                                                                                          \\
\textbf{\textit{\textcolor{NavyBlue}{Assembly}}}                                                                                     &                                                                                                                                                                                                                                                                                                     \\
Dodoo et al. \cite{dodoo2018evaluating}                                                       & Created a virtual assembly training application for a mock aircraft wing.                                                                                                                                                                                                                           \\
Adas et al. \cite{adas2013virtual}                                                            & Used VR/AR to enable personalized learning in machine assembly design   with step-by-step instructions.                                                                                                                                                                                             \\
Cecil et al. \cite{cecil2013virtual}                                                          & Used semi-immersive VR environments to teach micro-assembly concepts with improved student learning outcomes in work cell design and genetic algorithm operators.                                                                                                                                   \\
Fitton et al. \cite{fitton2024watch}                                                          & Demonstrated the effectiveness of observational learning in VR training for psychomotor tasks; When combined with hands-on practice can lead to better skill transfer to real-world; avatar similarity does not enhance fine motor learning, and skills decay without continued training.           \\
Aqlan et al. \cite{aqlan2019integrating}                                                      & Developed a VR simulation for craft production to design and build a toy car assembly based on given parts and customer requirements.                                                                                                                                                               \\
Zhao et al. \cite{zhao2019developing}                                                         & Presented a player modeling technique for delivering adaptive experience for each student where an AI assistant can display hints based on the player's past actions.                                                                                                                               \\
Aqlan et al. \cite{aqlan2020multiplayer}                                                      & Explored multiplayer VR simulations during collaborative design and assembly tasks to teach teamwork and Lean manufacturing principles.                                                                                                                                                             \\
Zhu et al. \cite{zhu2021eye}                                                                  & Integrated eye tracking in VR simulation for craft production to enhance and evaluate problem-solving skills in manufacturing settings.                                                                                                                                                             \\
Zhu et al. \cite{zhu2022sensor}                                                               & Developed an analytical model that integrates signal detection theory with Conflict \textbackslash{} Error to quantify problem-solving skills; the proposed joint signal detection theory with the Conflict \textbackslash{} Error model is more effective than using only signal detection theory. \\
Hartleb et al. \cite{hartleb2023exploring}                                                    & Introduced “magic interactions,” such as copying, pasting, and holographic visualization of objects. Usability evaluation indicates increase in user performance and collaboration.                                                                                                                 \\
Kim et al. \cite{kim2024behavioral}                                                           & Used a behavioral modeling method to assess collaborative problem-solving skills using sensor-based behavioral data.                                                                                                                                                                                \\
Dwivedi et al. \cite{dwivedi2018manual}                                                       & Developed a tool to streamline the assembly planning of complex systems e.g. zero-G treadmill on spacecraft; immersive VR training can be much faster and more accurate than training on a 2D screen for certain tasks.                                                                             \\
Etemadpour et al. \cite{etemadpour2019visualization}                                          & Developed a web-based visualization tool to support interactive analytic process of data collected from the assembly planning of complex systems.                                                                                                                                                   \\
De et al. \cite{de2019effects}                                                                & Tested the influence of stereopsis and immersion on execution of a bimanual assembly task; immersive stereoscopic platforms are the most promising contenders followed by non-immersive non-stereoscopic solutions that use larger screens (like the TV set).                                       \\
Ma et al. \cite{ma2020approach}                                                               & Discussed the effects of matching VR training protocols to students' motor skill levels to improve performance in assembly operations; haptic features provided effective skill training to desired levels, especially for moderate and low performers.                                             \\
Chang et al. \cite{chang2024efficient}                                                        & Virtual replicas enhance task efficiency and user preference for VR–AR remote collaboration in complex assembly tasks as compared to hand gestures and 3D drawing.                                                                                                                                  \\
Al-Ahmari et al. \cite{al2016development}                                                     & Used a virtual environment to create an interactive workbench that can be used for evaluating assembly decisions and training assembly; system provides visual, auditory, tactile and kinesthetic feedback while providing collaborative or individual interaction modes.                           \\
Sharma et al. \cite{sharma2019collaborative}                                                  & Presented a collaborative virtual assembly environment that aims to allow scientists and engineers to discuss a concept design in a real-time VR environment so that they can interact with objects and review their work before it is deployed.                                                    \\
Kim et al. \cite{kim2010itrain}                                                               & Combined computer-based and immersive training methods for assembly simulations, using ontologies and Sharable Content Object Reference Model (SCORM) to enhance effectiveness and integration.                                                                                                     \\
\textbf{\textit{\textcolor{NavyBlue}{Safety Training Skills and Ergonomics}}}                                                        &                                                                                                                                                                                                                                                                                                     \\
Moreland et al.   \cite{moreland2021development}                                              & Provided VR-based safety training for hazard awareness and prevention in the steel industry.                                                                                                                                                                                                        \\
Carruth et al. \cite{carruth2017virtual}                                                      & Developed VR training tools for workplace safety and tool use training for novice workers based on the learning objectives and the details of the tasks of interest.                                                                                                                                \\
Hannah et al. \cite{hannah2024results}                                                        & Integrated short VR exercises into computer-based training on knowledge retention for adult learners in contractor safety training for the energy industry in understanding confined space awareness concepts.                                                                                      \\
Bushra et al. \cite{bushra2018comparative}                                                    & Tablet-based interfaces are more effective than floating menus for risk assessment training in a VR machine shop environment.                                                                                                                                                                       \\
Islam et al. \cite{ISLAM2024103648}                                                           & Demonstrated the effectiveness of VR forklift simulators for training novice drivers; monitoring kinematic patterns to track skill acquisition is highlighted.                                                                                                                                      \\
Tang et al. \cite{tang2024evaluation}                                                         & VR training boosts urgency and perception of realism in emergency preparedness scenarios but requires careful design to align with specific educational goals and manage technological and cognitive demands.                                                                                       \\
\textbf{\textit{\textcolor{NavyBlue}{Machine Setup and Equipment Handling}}}                                                         &                                                                                                                                                                                                                                                                                                     \\
{Gupta et al. \cite{gupta2019viis} \\ Dpplani et al. \cite{doolani2020vis} } & Used storytelling in VR-based vocational training to train users for mechanical micrometer use; VR is more engaging and effective for long-term retention compared to text-based methods, but equally effective as video-based training.                                                            \\
Azzam et al. \cite{azzam2023virtual}                                                          & Used VR to teach hydraulic gripper design in engineering courses; over 90 percent of students reported increased engagement, excitement, and practicality in learning.                                                                                                                              \\
Mccusker et al. \cite{mccusker2018virtual}                                                    & Develop a virtual electronics laboratory where students can interact with electronic bench equipment and simple electronic components resembling a real laboratory environment.                                                                                                                     \\
\textbf{\textit{\textcolor{NavyBlue}{Iron and Steel Manufacturing}}}                                                                 &                                                                                                                                                                                                                                                                                                     \\
Zhou et al. \cite{zhou2016comprehensive}                                                      & Integrated VR with computational fluid dynamics models to simulate blast furnace operations for training.                                                                                                                                                                                           \\
Chen et al. \cite{chen2010virtual}                                                            & Developed a solution to convert and visualize numerical models and simulation results for blast furnace training.                                                                                                                                                                                   \\
Moreland et al. \cite{moreland2021development}                                                & Utilized AR and VR to provide safety and maintenance information in the steel industry.                                                                                                                                                                                                             
\end{longtblr}
}

\subsubsection{\textcolor{NavyBlue}{Benefits for Conventional Manufacturing Education:}} \hfill\\
\emph{\textcolor{NavyBlue}{Hands-on experiences in a spatial and situated environment promote skill development:}} Conventional manufacturing often emphasizes manual and psychomotor skill training, which is well-supported by the ability of VR to simulate physical tasks and provide repetitive practice in safe, controlled environments \cite{ma2020approach, fitton2024watch, studer2024open, park2023work}. Self-guided learning along with the provision of various kinds of sensory guidance and feedback facilitates the development of foundational skills applicable to traditional manufacturing processes (e.g., machining, assembly, welding) \cite{al2016development, stone2013full, park2023work, lie2023training, ye2024user, white2011low}. Studies have shown that guided training in VR develops perceptual, cognitive, and motor skills critical for hands-on tasks. It helps learners build spatial awareness, improve task comprehension, and retain information effectively, which boosts confidence in applying these skills in real-world scenarios \cite{kim2024behavioral}. The integration of visual-haptic feedback using realistic tools enables learners to refine psychomotor skills such as hand-eye coordination, precise hand and arm movements, and correct body positioning \cite{ipsita2022towards, ye2024user}.

\emph{\textcolor{NavyBlue}{A safe environment minimizes risk to learners and workplace environment:}} Conventional manufacturing deals with scenarios such as welding, machining, and heavy machinery operation where physical risks are higher due to hazardous equipment and processes. Studies have shown that VR is useful to simulate dangerous situations without risk. VR provides a safe and controlled environment for students to practice hazardous tasks and learn essential safety protocols such as identifying Personal Protective Equipments (PPEs) \cite{al2016development, moreland2021development, ipsita2022towards}. The system teaches proper equipment usage, safe navigation practices, and evacuation techniques. This can ensure that learners are well-prepared to handle workplace hazards \cite{bushra2018comparative, carruth2017virtual}. Virtual avatars are used to simulate real-time safety drills to demonstrate the consequences of unsafe practices without actual risks. Such an approach helps learners internalize safety precautions, develop hazard management skills, and build confidence in applying safety measures in real-world manufacturing settings \cite{hannah2024results, tang2024evaluation}.

\emph{\textcolor{NavyBlue}{Resource consumption is reduced in terms of instructor time and effort, consumables, machine availability, and travel requirements for in-person training:}} Conventional manufacturing processes such as welding and machining demand significant resources, including raw materials, consumables, machinery, and dedicated workspace, which contribute to high training costs \cite{ipsita2022towards}. In-person training further requires substantial instructor time and faces geographical constraints when students and instructors need to travel. Machine availability is often limited, making it costly to provide hands-on experience for every student. VR addresses these challenges by offering a clean, resource-efficient training solution that reduces material wastage, eliminates the need for physical consumables \cite{el2016assessment, white2011low, price2019using, stone2011virtual, park2023work}, and minimizes dependency on expensive machinery. This approach not only lowers costs but also provides scalable, accessible, and environmentally friendly training opportunities \cite{ipsita2022towards}. Physics-based simulations help illustrate irreversible changes, such as welding or chemical reactions, allowing repeated practice without material loss. Additionally, defect simulations allow learners to refine their work practices, improving performance, minimizing errors, and all while reducing resource waste \cite{stone2013full, stone2011virtual, price2019using}.

\emph{\textcolor{NavyBlue}{Realistic and immersive visualizations improve comprehension and understanding of detailed processes:}} By using VR-based simulations, learners can gain a comprehensive understanding of intricate, multi-step processes such as metal forming and casting without needing to observe these processes firsthand \cite{jalilvand2024vr, moreland2020integrating}. Animations that demonstrate how things work are particularly effective in teaching spatial relationships and functions in assemblies \cite{etemadpour2019visualization, adas2013virtual}. Visualizations of procedural instructions, such as machine setup and equipment operation, assembly, and maintenance can enable students to easily grasp complex tasks. In such cases, avatar demonstrations have been shown to enhance procedural learning and psychomotor skill acquisition \cite{fitton2024watch, dwivedi2018manual, chang2024efficient}.

\subsubsection{\textcolor{NavyBlue}{Challenges and Future Opportunities for Conventional Manufacturing Education:}} \hfill\\ 
\emph{\textcolor{NavyBlue}{Creating physics-based simulations for detailed processes can be challenging:}} Creating highly detailed and realistic simulations of conventional manufacturing processes can be challenging and resource-intensive. Accurate representation of the processes is essential for learning. Advances in AI can play an important role in creating and rendering realistic physical environments and simulations \cite{ipsita2024design}. This can result in enhancing the fidelity and immersion of VR-based training.

\emph{\textcolor{NavyBlue}{Sensory perception in the form of auditory, visual, haptic, olfactory feedback is essential for building muscle memory:}} For conventional manufacturing tasks requiring hands-on experience, there is concern whether VR can replicate the complexities of working with physical materials and tools. A major challenge lies in replicating feedback associated with physical processes such as welding, machining, or assembly. Research emphasizes the importance of physical tools and associated auditory and haptic feedback in manufacturing training \cite{ipsita2022towards, white2011low}, for instance, the role of a physical welding gun in VR training. This feedback is critical for developing muscle memory and psychomotor skills \cite{ipsita2022towards}. One way of addressing this challenge involves integrating actual tools and objects that mimic the behavior of real-world equipment in virtual environments. This can ensure that learners gain the necessary and close-to-real hands-on experience in a simulated setting. In this context, efforts have integrated physical tools into VR-based training \cite{ipsita2021vrfromx, stone2013full, ipsita2022towards}.

\emph{\textcolor{NavyBlue}{Limitations in user evaluations raise concerns about the effectiveness of VR training for long-term skill retention:}} Another challenge is ensuring that skills learned in a virtual environment are transferrable to real-world applications. Most studies focus only on evaluating immediate and short-term results. As many skills associated with conventional manufacturing practices require sustained practice, evaluations are essential to assess the long-term effectiveness of VR training \cite{bowling2010evaluating, de2019effects, price2019using, park2023work}. Additionally larger-scale studies should compare various training methods \cite{lie2023training}. For instance, this can include fully VR-based programs and those combining VR with physical practice to determine their relative effectiveness \cite{stone2011physical}. Such comparisons will provide insights into when and how these tools can be most effectively integrated into training curricula. Studies should also focus on skill application in real-world scenarios to evaluate how well VR-based training translates to real-world skill transfer.

\subsection{\textcolor{NavyBlue}{Findings from Review Analysis for Modern Manufacturing Education Skills}}

\subsubsection{\textcolor{NavyBlue}{Current State:}} 58 articles were included in the Modern Manufacturing Education Skills: \emph{\textcolor{NavyBlue}{Additive Manufacturing}} (14), \emph{\textcolor{NavyBlue}{Automation and Robotics}} (20), \emph{\textcolor{NavyBlue}{Smart Manufacturing}} (6), \emph{\textcolor{NavyBlue}{Advanced Materials}} (4), \emph{\textcolor{NavyBlue}{Sustainability and Green Manufacturing}} (12). Table \ref{tab:modern_domain_classification} shows the distribution of papers reviewed across the different categories of modern manufacturing skills. Overall, the trends in data for Domains show a growing trend in \textcolor{NavyBlue}{Modern Manufacturing Education Skills} over the observed period, especially from 2017 to the present indicating the growing focus on enhancing these skills. The data suggests that \textcolor{NavyBlue}{Sustainability and Green Manufacturing} Skills are gaining relevance within the period analyzed. This coincides with global trends towards environmental consciousness. \textcolor{NavyBlue}{Automation and robotics} skills appear before \textcolor{NavyBlue}{smart manufacturing}. This indicates the early adoption of basic automation before the integration of more complex, data-driven smart manufacturing systems. However, a rising trend in data across both fields suggests a gradual transformation within industries towards the integration of automation and smart technologies. A clear increase from 2020 to 2024 is observed for \textcolor{NavyBlue}{additive manufacturing} which suggests future potential in the field. Organizations and educational institutions may consider investing in training and development in AM technologies, developing curricula focused on CAD, and related technologies to match industry trends and demand.

{
\footnotesize
\begin{longtblr}[
  caption = {Distribution of papers reviewed across the various categories of Modern Manufacturing Education Skills},
  label={tab:modern_domain_classification}
]{
  width = \linewidth,
  colspec = {Q[150]Q[792]},
  row{1} = {c},
  row{2} = {c},
  row{15} = {c},
  row{35} = {c},
  row{39} = {c},
  row{44} = {c},
  cell{2}{1} = {c=2}{0.942\linewidth},
  cell{15}{1} = {c=2}{0.942\linewidth},
  cell{35}{1} = {c=2}{0.942\linewidth},
  cell{39}{1} = {c=2}{0.942\linewidth},
  cell{44}{1} = {c=2}{0.942\linewidth},
  hlines,
  vlines,
}
\textbf{Reference}                                                                                                          & \textbf{Short description indicating the use of VR for skill training}                                                                                                                                                                                                                                                                                                                                                                                                              \\
\textbf{\textit{\textcolor{NavyBlue}{Additive Manufacturing}}}                                                                                    &                                                                                                                                                                                                                                                                                                                                                                                                                                                                                     \\
Ostrander et al. \cite{ostrander2020evaluating}                                                            & use VR to teach introductory concepts in AM. Results from a comparative study show that both interactive and passive VR can effectively teach these concepts with no significant difference in learning outcomes compared to physical environments.                                                                                                                                                                                                                                 \\
Mathur et al. \cite{mathur2022identifying}                                                                 & explore the effects of immersion on design for AM (DfAM) evaluation. They measured DfAM scores, evaluation time, confidence, and cognitive load across three mediums: Computer-Aided Engineering, VR, and real environments. Results showed no significant differences in evaluation outcomes or cognitive load between immersive and non-immersive mediums, although the work points out the limitations of using a relatively simple material extrusion process for the analysis. \\
Mathur et al. \cite{mathur2023designing}                                                                   & provide immersive training experiences that can empower design students with skills to create and evaluate digital artifacts for AM.                                                                                                                                                                                                                                                                                                                                                \\
Mathur et al. \cite{mathur2024effects}                                                                     & evaluate VR and computer-aided instruction focusing on material extrusion and powder bed fusion processes.                                                                                                                                                                                                                                                                                                                                                                          \\
Mathur et al. \cite{mathur2024using}                                                                       & The team also investigates how VR immersion impacts additive manufacturability outcomes and experiential factors in solving build-with-AM problems. This work shows that VR helps designers achieve more favorable manufacturability outcomes for complex designs by reducing build time, support material usage, and evaluation effort without increasing cognitive load.                                                                                                          \\
Aryal et al. \cite{aryal2024imvr}                                                                          & explore AM and DfAM features.                                                                                                                                                                                                                                                                                                                                                                                                                                                       \\
Mogessie et al. \cite{mogessie2020work}                                                                    & develop a VR training system for metal AM machines to guide students in learning the necessary safety procedures, equipment components, and operational steps for setting up the EOS M290 for 3D printing. The work plans to create a generalized framework for broader applicability to other machines.                                                                                                                                                                            \\
Rafa et al. \cite{rafa2024enhancing}                                                                       & developed and evaluated a VR-based training platform for AM to allow students to gain hands-on experience with process parameters, safety measures, and 3D printing operations.                                                                                                                                                                                                                                                                                                     \\
Novoa et al. \cite{novoa2022new}                                                                           & integrate immersive technologies, 3D CAD, and 3D printing for virtual collaboration from design sketching to production.                                                                                                                                                                                                                                                                                                                                                            \\
Conesa et al. \cite{conesa2023influence}                                                                   & introduce a collaborative VR application designed to improve spatial skills through 3D model-building activitis. The application allows users to interact simultaneously in a shared virtual scene with audiovisual communication. Results show significant improvements in spatial test scores and high usability ratings.                                                                                                                                                         \\
Kobir et al. \cite{kobir2023human}                                                                         & develop a virtual 3D printing laboratory for workforce training in AM to optimize workplace layout design using human factors approaches such as link analysis. A 53.46\% reduction in user movement was observed while operating the 3D printing system. Moreover learning experience was enhanced by using hands-on training in safety, operational processes, printing parameters, and decision-making for 3D printing.                                                          \\
Ismail et al. \cite{ismail2024immersive}                                                                   & address cybersecurity in AM.                                                                                                                                                                                                                                                                                                                                                                                                                                                        \\
\textbf{\textit{\textcolor{NavyBlue}{Automation, Robotics and Human-Robot Collaboration}}}                                                        &                                                                                                                                                                                                                                                                                                                                                                                                                                                                                     \\
Ertekin et al. \cite{ertekin2023board}                                                                     & discuss robotics and CNC machining processes, and integrate VR simulations and offline programming for industrial robots in manufacturing environments.                                                                                                                                                                                                                                                                                                                             \\
Tram et al. \cite{tram2023intuitive}                                                                       & present a virtual simulation framework for robotic systems that allows operators to train robots in VR to execute tasks on real robots. The system enhances 3D spatial understanding, simplifies robot programming, and facilitates flexible automation assignments without needing physical robots.                                                                                                                                                                                \\
Kuts et al. \cite{kuts2022digital}                                                                         & conduct a study evaluation to compare traditional teach pendants and VR-based digital twin interfaces for controlling industrial robots. Results indicate that VR offers comparable task efficiency while enhancing overall performance, but causes higher stress in users as compared to traditional methods.                                                                                                                                                                      \\
Shi et al. \cite{shi2020affordable}                                                                        & developed a low-cost, open-source VR system to allow room-scale interaction with controllable robot models and off-the-shelf VR hardware for robotics education.                                                                                                                                                                                                                                                                                                                    \\
{Chang and Devine \\ \cite{robotlearning2018}}                                                             & use VR to investigate how the use of VR can affect the learner’s behavior and performance during the design verification stage of industrial robot programming.                                                                                                                                                                                                                                                                                                                     \\
Chang et al. \cite{chang2020exploring}                                                                     & use programming by demonstration to enable students to program industrial robots by virtually moving a robot end effector to generate tool paths.                                                                                                                                                                                                                                                                                                                                   \\
Chang et al. \cite{chang2021using}                                                                         & also used a quasi-experimental mixed-method approach to explore the potential of the VR add-on to address challenges faced by novice users such as high mental load and lack of holistic task understanding. Preliminary findings highlight the promise of VR for improving task performance and learning.                                                                                                                                                                          \\
Theofanidis et al. \cite{theofanidis2017varm}                                                              & describe a novel teleoperation interface system to enable users with no prior knowledge of robotics to safely program mechanical manipulators.                                                                                                                                                                                                                                                                                                                                      \\
Darmoul et al. \cite{darmoul2015virtual}                                                                   & develop a virtual environment for a robotic cell that can allow users to validate layout designs and conduct implementation planning for real robotic cells. A virtual teach pendant is included to control the robotic movements using which a teacher can train students without risking the equipment, as well as without injuries.                                                                                                                                              \\
Mitchell et al. \cite{mitchell2020safety}                                                                  & investigates safety perceptions and behaviors during human-robot interaction in virtual environments. Results from a comparative study with physical environments showed that VR simulations can replicate key safety-related outcomes observed in real environments. This suggests the use of VR as a low-cost, low-risk alternative for worker training in human-robot interaction tasks.                                                                                         \\
Srinivasan et al. \cite{srinivasan2021biomechanical}                                                       & developed a VR simulation testbed and evaluated its efficacy as an occupational trainer for human-robot collaborative tasks. Results showed minimal differences in human movement kinematics and kinetics between real and virtual tasks. This suggests the capabilities of VR to transfer skills to real world settings.                                                                                                                                                           \\
Lor et al. \cite{lor2024enabling}                                                                          & create personalized and adaptive learning environments by utilizing data such as eye gaze, hand movements, and self-reported confidence levels. By fine-tuning a compact LSTM model with minimal data, they show significant improvements in estimating learner confidence.                                                                                                                                                                                                         \\
Ryan et al. \cite{ryan2022immersive}                                                                       & focus on immersive VR training for CNC milling setup processes with an error management approach. The focus is to teach setup procedures, identify and manage errors, and provide interactive practice.                                                                                                                                                                                                                                                                             \\
{Rogers et al. \cite{rogers2018assessment} \\ El-Mounayri et al. \cite{el2016assessment}} & utilize VR to teach CNC milling machine operation as cost-effective, accessible alternatives to physical labs. Results from the survey highlight the immersive and beneficial aspects of immersive technology.                                                                                                                                                                                                                                                                      \\
Chiou et al. \cite{chiou2019virtual}                                                                       & utilize VR simulation of industrial robots for undergraduate curricula to teach students about the application of robotics in ultrasonic welding.                                                                                                                                                                                                                                                                                                                                   \\
Wang et al. \cite{wang2019virtual}                                                                         & use a VR system for robot-assisted welding and hidden Markov model-based human intention recognition to improve welding performance of operators. The teleoperation facilitated by VR and a robot results in smoother intended movements while achieving operator safety.                                                                                                                                                                                                           \\
Wang et al. \cite{wang2020digital}                                                                         & introduce a bi-directional VR-based system that connects human users and robots for human–robot interactive welding and welder behavior analysis. The system's effectiveness was validated with skilled and unskilled welders in which 94.44\% accuracy was achieved in classifying welder behavior using a data-driven approach.                                                                                                                                                   \\
Abujelala et al. \cite{abujelala2018collaborative}                                                         & train and assess human–robot collaboration skills in assembly tasks within manufacturing environments.                                                                                                                                                                                                                                                                                                                                                                              \\
Cecil et al. \cite{cecil2024study}                                                                         & conduct a study in robotic assembly using digital twins. The findings show that prior experience shapes perception, audio cues enhance understanding, and improved perception significantly boosts cognition and knowledge acquisition.                                                                                                                                                                                                                                             \\
\textbf{\textit{\textcolor{NavyBlue}{Smart Manufacturing}}}                                                                                       &                                                                                                                                                                                                                                                                                                                                                                                                                                                                                     \\
Zhu et al. \cite{zhu2023learniotvr}                                                                        & focus on IoT education within a VR environment through immersive design, programming, and exploration of real-world environments empowered by IoT.                                                                                                                                                                                                                                                                                                                                  \\
{Jun et al. \cite{jun2021human} \\ Yun et al. \cite{yun2022immersive}}                    & develop an immersive and interactive cyber-physical system using VR for intuitive and real-time collaboration between human, machines and autonomy. The work aims to augment human skills, reduce errors, and translate human actions into autonomous robot programs.                                                                                                                                                                                                               \\
Osti et al. \cite{osti_10293626}                                                                          & presented a small-scale implementation for Industry 4.0 in an educational setting to teach students the concepts of Industry 4.0 by integrating robotics, automation, 3D printing, and VR.                                                                                                                                                                                                                                                                                          \\
\textbf{\textit{\textcolor{NavyBlue}{Advanced Materials}}}                                                                                        &                                                                                                                                                                                                                                                                                                                                                                                                                                                                                     \\
Jalilv et al. \cite{jalilvand2024vr}                                                                       & developed an adaptive and interactive digital twin interface in both VR and AR environments to simulate a complex thermoforming process used in thermoplastic manufacturing in real-time. The focus is to facilitate immersive training for novice operators.                                                                                                                                                                                                                       \\
Rahman et al. \cite{rahman2023workforce}                                                                   & utilize VR to train personnel in composite manufacturing processes, such as compression molding.                                                                                                                                                                                                                                                                                                                                                                                    \\
Kamali et al. \cite{kamali2020virtual}                                                                     & develop a VR curriculum with five interactive laboratory modules including photolithography, sputter deposition, etching, and characterization which was tested in summer camps. The work aims to overcome challenges while making nanotechnology education more accessible to universities and even high school programs.                                                                                                                                                          \\
Wang et al. \cite{wang2020towards}                                                                         & presents a VR training system including a hint system for operating machines in a microfabrication laboratory covering wafer preparation procedures (wafer cleaning, photoresist coating, and baking). Pilot testing showed improved learning speed, independent learning ability, and error reduction.                                                                                                                                                                             \\
\textbf{\textit{\textcolor{NavyBlue}{Sustainability and Green Manufacturing}}}                                                                    &                                                                                                                                                                                                                                                                                                                                                                                                                                                                                     \\
Frank et al. \cite{frank2021green}                                                                         & use VR to teach renewable energy concepts such as water electrolysis fundamentals, fuel cell characterization, solar power generation parameters and effects, and wind turbine parameters and operation. The work aims to help students visualize and conceptualize various abstract topics, such as energy and mass conservation principles as renewable energy systems to engineering design.                                                                                     \\
Chiou et al. \cite{chiou2021developing}                                                                    & develop a VR-based laboratory module to teach solar energy efficiency and its environmental impact. They aim to give students hands-on experience about heat, tilt angle, shading, and maximum power point remotely, without the need for physical equipment.                                                                                                                                                                                                                       \\
Chiou et al. \cite{chiou2020project}                                                                       & incorporate project-based learning with VR in green energy manufacturing to allow students to experience the whole wind turbine design process, from 3D design all the way to VR analysis.                                                                                                                                                                                                                                                                                          \\
Chiou et al. \cite{chiou2024virtual}                                                                       & educate students on green manufacturing processes. They specifically focus on the intricacies of wind turbine efficiency.                                                                                                                                                                                                                                                                                                                                                           \\
{Borst et al. \cite{borst2016virtual} \\ Ritter et al. \cite{ritter2016work}}             & introduce a VR environment that models a real energy facility to enable virtual field trips and self-guided exploration to teach concentrating solar power technology.                                                                                                                                                                                                                                                                                                              \\
Ritter et al. \cite{ritter2018virtual}                                                                     & test the effectiveness of the application by performing case studies using high school and university students.                                                                                                                                                                                                                                                                                                                                                                     \\
Borst et al. \cite{borst2018teacher}                                                                       & offer two VR field trip approaches: one with a live teacher using depth camera imagery and another standalone version with prerecorded teacher content. Results indicate the live networked approach leads to greater learning gains and better student engagement, while the standalone approach offers scalability and simplicity.                                                                                                                                                \\
Woodworth et al. \cite{woodworth2023study}                                                                 & compare three eye-tracking visual cues to a baseline without cues in the same environment to evaluate the effectiveness in maintaining attention. It was observed that eye-tracking visual cues successfully guided gaze and maintained attention although there was a disconnect in the student feedback among the types of visual cues.                                                                                                                                           \\
Kula et al. \cite{kula2024development}                                                                     & developed a VR-based energy assessment process for industrial and commercial buildings. Participant feedback indicates that VR can be a substitute for real-world assessments for training purposes.                                                                                                                                                                                                                                                                                \\
Earle et al. \cite{earle2021wicked}                                                                        & develop a conceptual model to integrate design thinking and traditional instruction practices with AR and VR to address sustainability challenges. They acknowledge the immersive technology as a complementary, not standalone, solution.                                                                                                                                                                                                                                          \\
Chiou et al. \cite{chiou2019virtual}                                                                       & utilize VR to teach ultrasonic welding with robotic systems. They highlight importance of the fast, clean process that does not require adhesives, binding agents, solders, fluxes, or solvents.                                                                                                                                                                                                                                                                                    
\end{longtblr}
}

\subsubsection{\textcolor{NavyBlue}{Benefits for Modern Manufacturing Education:}} \hfill\\
\emph{\textcolor{NavyBlue}{VR enhances accessibility to advanced interdisciplinary topics:}} VR enables students to engage with advanced manufacturing technologies such as robotics, automation, and digital manufacturing which are often inaccessible in traditional classrooms. It facilitates learning these concepts at various levels and supports immersive visual programming interfaces and digital twins to simplify complex and interdisciplinary topics \cite{robotlearning2018, kuts2022digital, chang2020exploring, jun2021human}. For example, VR equips students with hands-on experience in Industry 4.0 technologies, such as IoT-enabled systems, cyber-physical systems, and big data analytics \cite{zhu2023learniotvr, osti_10293626, jun2021human, yun2022immersive}. Such experiences can provide students with essential skills for smart factories. Studies have also shown that users can perform design reviews with enhanced visualization of manufacturability features and process integration for AM \cite{mathur2023designing, mathur2022identifying, aryal2024imvr}. Additionally, VR enhances accuracy in teleoperation tasks thus improving human-robot collaboration \cite{wang2019virtual, wang2020digital}.

\emph{\textcolor{NavyBlue}{VR facilitates sustainability through sustainable learning practices and teaching sustainability concepts:}} VR simulations promote sustainability by modeling green manufacturing processes to teach students about renewable energy sources and waste management practices \cite{earle2021wicked, frank2021green, chiou2021developing}. This can enable students to (1) understand the environmental impacts of their work and explore eco-friendly alternatives \cite{chiou2024virtual, han2023virtual} and (2) adopt strategies to reduce resource consumption and wastage. Additionally, VR supports remote collaboration and training. This minimizes unnecessary travel and further contributes to sustainable practices.

\emph{\textcolor{NavyBlue}{VR supports co-located and distributed collaboration and teamwork:}} VR fosters collaboration among interdisciplinary teams and students across different locations by enabling team-based activities within shared virtual manufacturing environments \cite{shetty2018strategies, aqlan2020multiplayer}. This mirrors the collaborative nature of modern manufacturing operations where distributed teams work together to solve real-world problems \cite{conesa2023influence, mcgrath2014cloud}. These activities enhance critical thinking, team-building, and communication skills, while also fostering design and entrepreneurial skills. Studies have shown that VR supports product design by improving communication in teams through immersive visualizations \cite{aqlan2020multiplayer, evans2021prototyping, ashour2021connected}. By using factory simulations and digital twins, VR reduces unnecessary travel and enables on-demand teamwork. Additionally, VR allows teams to test minimum viable product prototypes before actual implementation. This results in confirming the feasibility of their ideas and reducing development costs \cite{kobir2023human, shi2020affordable, novoa2022new}.

\subsubsection{\textcolor{NavyBlue}{Challenges and Future Opportunities for Modern Manufacturing Education:}} \hfill\\
\emph{\textcolor{NavyBlue}{It is challenging to create effective content for interdisciplinary digital skills:}} Modern manufacturing education often focuses on cutting-edge technologies such as automation, robotics, and smart manufacturing \cite{mogessie2020work, rafa2024enhancing, osti_10293626}. These complex topics require an understanding of multiple disciplines, including computer science, electronics, and mechanical engineering concepts \cite{zhu2023learniotvr, jalilvand2024vr}. Creating VR experiences that effectively integrate these varied fields to provide a cohesive learning experience can be challenging. VR content and simulations also need to be regularly updated to reflect the latest developments which would require ongoing investment and development. As these topics emphasize the development of cognitive and logic-building skills, there can be possibilities where learners find it difficult to comprehend abstract concepts using only immersive visualizations and hands-on experiences. To provide insights in this area, future studies can explore the cognitive benefits of using immersive platforms in learning such topics as compared to traditional methods. Furthermore, studies should also explore the benefits of integrating both methods by carefully weighing the benefits of each.

\label{sec:5}
\textcolor{PineGreen}{\section{Levels in Manufacturing Education}}
First, the secondary taxonomy for levels is explained. Then the findings from the review and analysis is presented in terms of the current state, benefits, challenges, and future opportunities of VR in the corresponding category.

\subsection{\textcolor{PineGreen}{Level Taxonomy: }}
\begin{figure*}[tb]
    \centering
    \includegraphics[width = 0.75\textwidth]{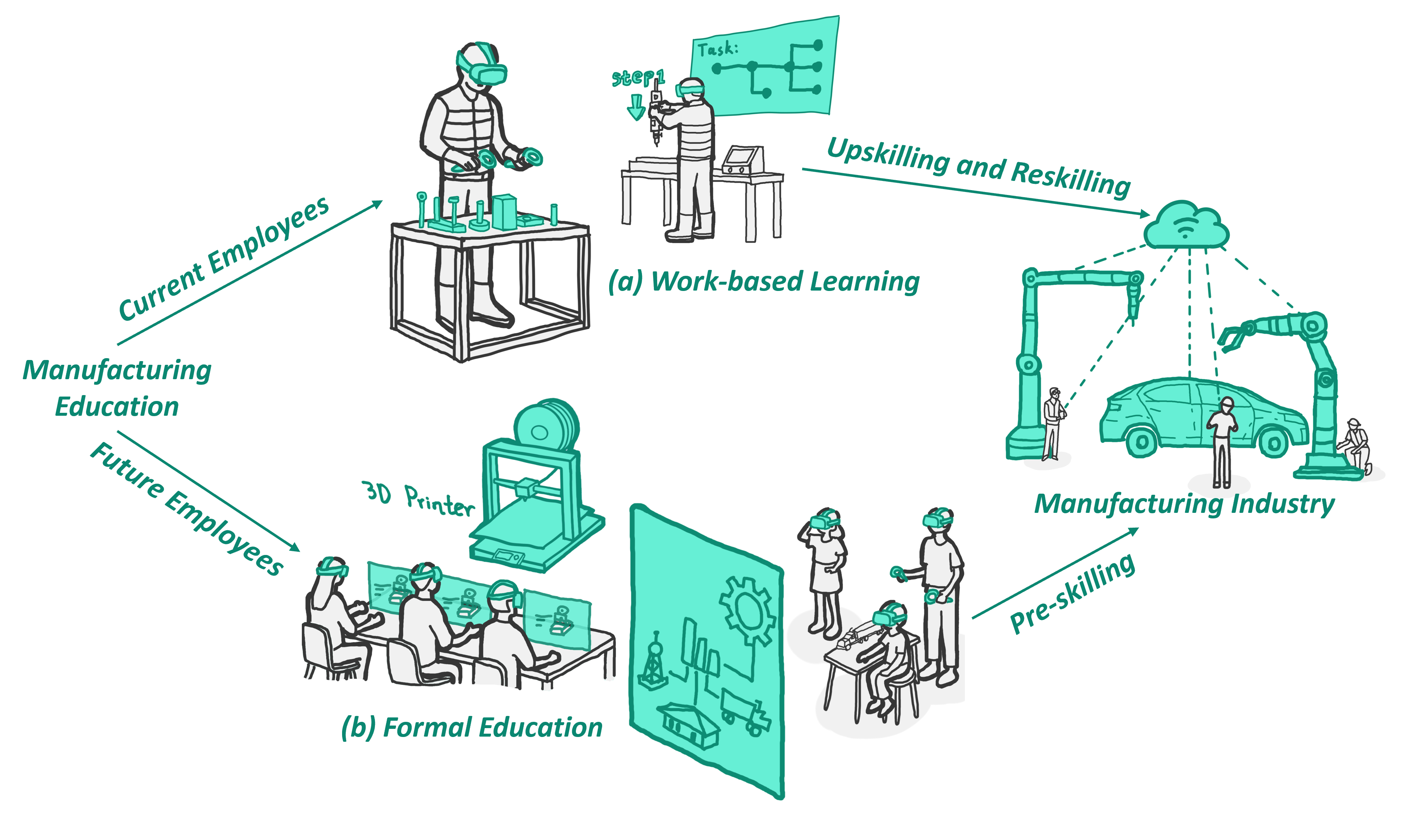}
    \caption{Different levels of manufacturing education are shown, highlighting efforts to upskill and reskill the current workforce through (a) Work-Based Learning. Simultaneously, the education system emphasizes preskilling future employees via (b) Formal Education, preparing them for the evolving demands of the manufacturing industry.}
    \label{fig:LevelClassification}
\end{figure*}

The levels are classified into two main categories: (1) \emph{\textcolor{PineGreen}{Formal Education}}, and (2) \emph{\textcolor{PineGreen}{Work-based Learning}}. As shown in Fig. \ref{fig:LevelClassification}, the category \emph{\textcolor{PineGreen}{Formal Education}} focuses on preskilling individuals for the future of manufacturing, while \emph{\textcolor{PineGreen}{Work-based Learning}} focuses on upskilling and reskilling of the current workforce.

\subsubsection{\textcolor{PineGreen}{Formal Education: }} The categories are: (1) secondary and high schools (K--12), (2) university and college programs (higher education), and (3) community colleges and trade schools. This level includes traditional educational institutions that offer structured programs ranging from secondary to post-secondary and higher education levels. Programs are designed to provide foundational knowledge, technical skills, and degrees or certificates in manufacturing and related fields. To determine the appropriate level for a paper, we examined whether it addressed any specific tier of formal education, such as higher education or K--12 learning.

\subsubsection{\textcolor{PineGreen}{Work-Based Learning:}}
The categories are: (1) worker training which includes on-the-job training, apprenticeships, and corporate training programs, and (2) Nature of Skills. Worker training category involves learning that occurs in a work setting, offering practical experience and skills development directly relevant to the manufacturing industry. For example, a VR application is developed for operators training and reskilling \cite{jalilvand2024vr} to upskill employees in industrial assembly and maintenance procedures such as disassembling and reassembling a digital dosing pump. To classify a paper at this level, we assessed whether it focused on (1) training, skill transfer, or the evaluation of operators or workers in workplace environments, and/or (2) emphasized skills essential for workplace applications. For instance, Wang et al. \cite{wang2019virtual} primarily explored how workers' skills could be analyzed through the teleoperation of a robot in a virtual environment. Similarly, Byrd et al. \cite{byrd2015use} highlighted the significance of welding in the industry and demonstrated how VR can provide accessible training solutions for welders.

\subsection{\textcolor{PineGreen}{Findings from the Review Analysis for Formal Education: }} From the analysis, 55 papers were found to be associated with the Formal Education category. Overall, formal education show growth trends during the time period. Table \ref{tab:level_classification} shows the distribution of papers reviewed across the different categories of formal education.

\subsubsection{\textcolor{PineGreen}{Current State:}}
In K--12, most efforts have been found to teach renewable energy concepts \cite{woodworth2023study, borst2018teacher, borst2016virtual, ritter2018virtual}. Self-guided exploration, virtual teacher-guided exploration, and eye-tracking-based attention techniques have been explored to guide students during virtual field trips. Accessibility to advanced concepts such as IoT in Industry 4.0 \cite{osti_10293626}, AM \cite{renner2015virtual} and nanotechnology education \cite{kamali2020virtual} has also been explored in K--12, as well as higher education and community college settings.

Higher education has seen a diverse set of systems and evaluation spread across multiple skills. For instance, in machining to understand operations using equipment such as lathes and milling machines \cite{lie2023training, park2023work}; in welding for entry-level introduction \cite{price2019using} and robotic ultrasonic welding \cite{chiou2019virtual}; in material testing laboratories \cite{srinivasa2021virtual}; in assembly for manufacturing systems \cite{kim2024behavioral, zhu2022sensor} and micro-assembly focusing on topics such as work cell design and genetic algorithm operators \cite{cecil2013virtual}; and for machine operations such as hydraulic grippers in fluid power education \cite{azzam2024virtual}. The most efforts are seen in AM in a variety of topics. Some of these include layout design and equipment arrangement for reducing operator movement \cite{kobir2023human}, cybersecurity in AM \cite{ismail2024immersive}, complex and less accessible AM processes like material extrusion and powder bed fusion \cite{mathur2024effects}, process parameters and safety measures in 3D printing operations \cite{rafa2024enhancing} and metals AM \cite{mogessie2020work}, introductory AM concepts \cite{ostrander2020evaluating}, DfAM \cite{mathur2022identifying}, 3D model building activities \cite{conesa2023influence} and solving build-with AM problems \cite{mathur2024using}. VR has also been used to teach industrial robotics concepts using RoboDK simulations \cite{ertekin2023board} and industrial robot programming \cite{robotlearning2018, chang2021using}. Additionally, VR has been found useful in teaching sustainable practices such as energy assessments in industrial and commercial buildings \cite{kula2024development}, continuous improvement in manufacturing \cite{parsley2024enhancing} and renewable and green energy manufacturing \cite{frank2021green, chiou2020project, chiou2024virtual}. In \textcolor{PineGreen}{undergraduate} category, the rising trend suggests increased funding and initiatives to adopt VR in education.

VR has been explored in community colleges and trade schools in the field of welding. Studies have collected perspectives of welding instructors on integrating technology on integrating VR to complement traditional welding training \cite{lassiter2023welding}. Low-cost immersive training systems have been designed for welding training \cite{white2011low}. VR-based integrated training has been shown to enhance training outcomes, reduce cost, and decrease material consumption \cite{stone2011virtual}.

\subsubsection{\textcolor{PineGreen}{Benefits of VR for Formal Education:}}
\hfill\\
\emph{\textcolor{PineGreen}{Development of skills and learning outcomes in Formal Education:}} Many studies have shown improvement in learning outcomes and skill development such as eye-hand coordination and spatial rotation abilities for novice learners \cite{robotlearning2018, chang2021using}, problem solving \cite{ertekin2023board}, critical thinking \cite{parsley2024enhancing}, and computation \cite{ismail2024immersive}. Improvements have been noted in collaborative problem-solving skills using sensor-based behavioral data and personalized interventions \cite{kim2024behavioral} and spatial skills in collaborative VR environments \cite{conesa2023influence}. VR has been shown to increase confidence, proactive participation, and better performance as compared to a control group \cite{park2023work, ritter2018virtual, cecil2013virtual}. Studies have reported increased engagement, excitement, and practicality in learning through immersive experiences \cite{azzam2024virtual}. Open-ended learning has shown to increase learning rates and reduce mistakes \cite{lie2023training}. Studies have reported improved understanding of machine operations and process parameters \cite{kobir2023human}. VR training reduces spatial load and show comparable knowledge gains to conventional instruction and significant learning gains in AM processes and is thus considered as a viable alternative to high-barrier-to-entry systems \cite{mathur2024effects}. 

\emph{\textcolor{PineGreen}{Promotes low resource usage:}}
VR reduces build time, supports material usage and evaluation effort without increasing cognitive load, and enhances AM design iterations by enabling faster, lower-cost solutions \cite{mathur2024using}. VR can be used for introductory AM concepts avoiding barriers such as limited course offerings and high costs associated with hands-on AM education \cite{ostrander2020evaluating}. It provides a safe environment for allowing hands-on practice \cite{rafa2024enhancing}. VR has been used in layout design for increasing efficiency \cite{kobir2023human} and for training purposes in energy assessments for industrial and commercial buildings \cite{kula2024development}. 

\emph{\textcolor{PineGreen}{Interactive method, a great way to learn for young students:}}
VR has also been shown to be an interactive method of learning to come up with creative and innovative solutions, appealing for younger generations and motivating for slow learners \cite{chiou2020project}.

\subsubsection{\textcolor{PineGreen}{Challenges and future opportunities of VR for Formal Education:}}
\hfill\\
\emph{\textcolor{PineGreen}{Cost and Accessibility Issues:}}
One of the primary concerns for the adoption of VR in formal education is due to high costs and limited accessibility, primarily in in under-resourced settings \cite{ostrander2020evaluating}. Studies call for low-cost solutions \cite{price2019using, ostrander2020evaluating, mogessie2020work} while highlighting the importance of passive VR learning using more accessible and affordable headsets such as Google Cardboard (discontinued since 2021 and software is open-sourced; some companies still make compatible headsets). 

\emph{\textcolor{PineGreen}{Scalability and inclusion Barriers:}}
Scalable solutions for VR content creation can increase the accessibility of VR in educational settings. Such efforts may include reinforcement learning for virtual environment generation \cite{ashour2021connected}, modularized systems for broader applicability across machines \cite{mogessie2020work}, and authoring efforts using programming by demonstration \cite{chang2020exploring}. Utilizing established theories from learning sciences can be useful in designing an effective curriculum. Studies highlight the integration of cognitive and constructivist theories \cite{ismail2024immersive} to align VR with educational goals. Inclusivity studies emphasize improving outcomes for underrepresented groups \cite{ismail2024immersive, srinivasa2021virtual, rafa2024enhancing} while advocating for pedagogical approaches accommodating diverse learning styles. Collaborative efforts among academia, industry, and policymakers are essential for addressing inclusion and scalability barriers \cite{ma2019efficacy}. 

\emph{\textcolor{PineGreen}{Limitations in User Evaluations:}}
The adoption of VR in formal education can also benefit from an evaluation using larger sample sizes, curriculum integration, and comparative analysis with conventional programming methods \cite{chang2021using}.

\begin{table*}[]
\centering
\caption{Distribution of papers used in the review found across the different categories of Levels}
\label{tab:level_classification}
\resizebox{0.75\textwidth}{!}{%
\begin{tabular}{|cl|cl|}
\hline
\multicolumn{2}{|c|}{\textbf{Formal Education}} & \multicolumn{2}{c|}{\textbf{Work-based learning}} \\ \hline
\multicolumn{1}{|c|}{\textbf{\begin{tabular}[c]{@{}c@{}}Higher \\ Education\end{tabular}}} & \begin{tabular}[c]{@{}l@{}}\cite{kobir2023human, osti_10293626, robotlearning2018, kim2024behavioral, ertekin2023board, ismail2024immersive, ashour2021connected, zhao2019developing, kula2024development, mathur2024effects}\\ \cite{ma2019efficacy, parsley2024enhancing, rafa2024enhancing, ostrander2020evaluating, zhu2021eye, frank2021green, mathur2022identifying, aqlan2019integrating, mccusker2018virtual}\\ \cite{aqlan2020multiplayer, chiou2020project, zhu2022sensor, lopez2020click, conesa2023influence, lie2023training, chang2021using, mathur2024using, price2019using}\\ \cite{cecil2013virtual, srinivasa2021virtual, azzam2024virtual, chiou2019virtual, kamali2020virtual, chiou2024virtual, park2023work, mogessie2020work}\\ \cite{yamada2017educational, chiou2021developing, hartleb2023exploring, chang2020exploring, evans2021prototyping, earle2021wicked, wang2020towards, adas2013virtual, ritter2018virtual}\end{tabular} & \multicolumn{1}{c|}{\textbf{\begin{tabular}[c]{@{}c@{}}Worker \\ Training\end{tabular}}} & \begin{tabular}[c]{@{}l@{}}\cite{renner2015virtual, etemadpour2019visualization, ma2020approach, studer2024open, belga2022carousel, ISLAM2024103648, bushra2018comparative, zhou2016comprehensive}\\  \cite{hoover2021designing, al2016development, moreland2021development, moreland2022development, kuts2022digital, wang2020digital, lor2024enabling}\\ \cite{dodoo2018evaluating, bowling2010evaluating, tang2024evaluation, stone2013full, ryan2022immersive, moreland2020integrating, tram2023intuitive, kim2010itrain}\\ \cite{dwivedi2018manual, rumsey2020manufacturing, stone2011physical, hannah2024results, mitchell2020safety, cecil2024study, srinivasan2021biomechanical, ipsita2024design, de2019effects}\\ \cite{byrd2015use, ipsita2022towards, gupta2019viis, chen2010virtual, carruth2017virtual, darmoul2015virtual, stone2011virtual, wang2019virtual, doolani2020vis, ipsita2021vrfromx, fitton2024watch, mogessie2020work}\end{tabular} \\ \hline
\multicolumn{1}{|c|}{\textbf{K-12}} & \cite{osti_10293626, renner2015virtual, woodworth2023study, borst2018teacher, borst2016virtual, kamali2020virtual, ritter2018virtual} & \multicolumn{1}{c|}{\multirow{3}{*}{\textbf{\begin{tabular}[c]{@{}c@{}}Nature \\ of \\ Skills\end{tabular}}}} & \multirow{3}{*}{\cite{sharma2019collaborative, kula2024development, chang2024efficient, jun2021human, yun2022immersive, ye2023robot, lie2023training, ye2024user, theofanidis2017varm, abujelala2018collaborative}} \\ \cline{1-2}
\multicolumn{1}{|c|}{\textbf{\begin{tabular}[c]{@{}c@{}}Community colleges \\ and trade schools\end{tabular}}} & \cite{white2011low, stone2011virtual, osti_10293626, lassiter2023welding} & \multicolumn{1}{c|}{} &  \\
\multicolumn{1}{|c|}{} &  & \multicolumn{1}{c|}{} &  \\ \hline
\end{tabular}%
}
\end{table*}

\subsection{\textcolor{PineGreen}{Findings from the Review Analysis for Work-Based Learning:}} From the analysis, 50 papers were identified as focusing on work-based learning. Overall, work-based learning shows growth trends during the time period. The substantial rise in work-based learning by 2024 indicates a growing emphasis on practical, on-the-job learning. This might be possibly due to labor market demands. Table \ref{tab:level_classification} shows the distribution of papers reviewed across the different categories of Work-based Learning.

\subsubsection{\textcolor{PineGreen}{Current State:}}
The majority of the efforts have been in the field of welding, closely followed by robotics and automation, and assembly tasks. In the welding industry, VR is used for skill assessment in welders \cite{byrd2015use}. Studies have focused on the evaluation of cognitive and physical impact \cite{stone2011physical}, training potential, team learning, material consumption, and cost implications \cite{stone2011virtual} on welding training. It is noted that performance is dependent on task difficulty \cite{stone2013full}. Robot-assisted welding systems are used to improve welding performance using remote control for smoother intended movements \cite{wang2019virtual} and to digitize expert motor skills and guide trainees through feedback \cite{ye2023robot, ye2024user}. Some systems have focused on the content creation such as authoring virtual welding scenes from real world scans \cite{ipsita2021vrfromx, ipsita2024design} and curriculum design using backwards design \cite{ipsita2022towards}.

In the field of robotics and automation, systems have been developed to assess worker skills in human-robot collaboration tasks \cite{srinivasan2021biomechanical} such as assembly \cite{abujelala2018collaborative, cecil2024study}, safety perception and behavior \cite{mitchell2020safety, darmoul2015virtual}, CNC machining training \cite{ryan2022immersive} and welding \cite{wang2020digital}. Teleoperation interfaces have also been developed to program industrial robots \cite{theofanidis2017varm, kuts2022digital} and train robots in VR to execute tasks on real robots \cite{tram2023intuitive}.

In assembly operations, efforts have been made to develop systems for assembly training \cite{dodoo2018evaluating, dwivedi2018manual, etemadpour2019visualization}. A few use cases involve an assembly environment for product design \cite{sharma2019collaborative}, bimanual assembly tasks \cite{de2019effects}, and VR--AR remote collaboration in complex assembly tasks \cite{chang2024efficient}. Some studies have evaluated feedback \cite{al2016development} such as haptic-VR training modules for motor skill training \cite{ma2020approach}. There are also efforts to develop ontology \cite{kim2010itrain} systems for training.

There are some efforts dispersed around a few other skills such as safety, quality control, machine operations, iron and steel manufacturing, sustainability, AM, machining, forming and smart manufacturing. VR systems have been developed for safety \cite{carruth2017virtual}, risk assessment \cite{bushra2018comparative}, emergency preparedness scenarios to balance stress levels and cognitive load \cite{tang2024evaluation}, and contractor safety for the energy industry \cite{hannah2024results}. In quality control, studies have evaluated the effects of feedforward information during inspection tasks \cite{bowling2010evaluating} and assessed the inspection knowledge using Carousel techniques \cite{belga2022carousel}. In machine setup and equipment handling, VR is used to learn micrometers \cite{gupta2019viis, doolani2020vis} and forklift simulators for training novice drivers \cite{ISLAM2024103648}. In iron and steel manufacturing, VR is used for visualizing a blast furnace \cite{chen2010virtual}, with CFD simulations \cite{zhou2016comprehensive}. It is also used for safety training in the steel industry \cite{moreland2021development}. In sustainability, VR is used for energy assessment for industrial and commercial buildings \cite{kula2024development}. VR is used in AM training \cite{renner2015virtual}, with a few studies focusing on customizable training solutions for metals AM machines \cite{mogessie2020work}. In machining, open-ended training has been developed for machining skills such as drilling \cite{lie2023training, studer2024open}. In forming, VR is used with CFD simulations for high-pressure die casting \cite{moreland2020integrating, moreland2022development}. In smart manufacturing, VR is also used to prepare operators for Industry 5.0 \cite{jun2021human, yun2022immersive}.

\subsubsection{\textcolor{PineGreen}{Benefits of VR for Work-based Learning:}}
\hfill\\
\emph{\textcolor{PineGreen}{Increased gains in productivity and learning:}}
Some research emphasized teaching safety practices to employees, while others concentrated on developing hands-on skills to build muscle memory in learners. These skills could then be applied in workplace environments to enhance productivity. Studies highlight the use of VR-based learning environments to allow active exploration and alignment with workplace learning \cite{studer2024open}. Haptic guidance in immersive VR improves performance and reduces errors in motor skill tasks such as assembly \cite{ma2020approach}.

\emph{\textcolor{PineGreen}{Reduced cost of training:}}
VR reduces material consumption and minimizes training time. Remote training capabilities can provide hands-on experience at low-cost. While allowing learners to engage with training activities from a distance, this is also environment-friendly \cite{ye2023robot}.

\emph{\textcolor{PineGreen}{Facilitates collaboration between humans and autonomy:}}
HMD systems simplify robot programming for collaborative tasks \cite{tram2023intuitive}. They facilitate human-robot collaboration and prepare operators for Industry 5.0 \cite{yun2022immersive, kuts2022digital} and collaborative assembly training \cite{sharma2019collaborative}. VR is also useful for training robots to automate routine tasks with operators leveraging these technologies to monitor production processes remotely. VR expands capabilities, enabling robot swarm control and virtual laboratories \cite{guerrero2023integration, darmoul2015virtual}.

\emph{\textcolor{PineGreen}{Safety and Hazard Prevention:}}
Hazard prevention is crucial in workplace environments such as in steel manufacturing, welding, etc. \cite{moreland2021development, bushra2018comparative}. VR can provide preliminary experiences in such cases for hazard identification and emergency preparedness.

\subsubsection{\textcolor{PineGreen}{Challenges and Future Opportunities of VR for Work-based Learning:}}
\hfill\\
\emph{\textcolor{PineGreen}{Infrastructure and organizational resistance:}}
Broader investigations examine VR adoption in manufacturing and organizational contexts. While HMDs improve flexibility and cost compared to CAVE systems, implementation barriers such as infrastructure and organizational resistance persist \cite{rumsey2020manufacturing}. In welding, VR supports skill assessments \cite{byrd2015use, wang2020digital} and complements in-person training, though it cannot fully replace traditional methods \cite{lassiter2023welding}. 

\emph{\textcolor{PineGreen}{Limited evidence on long-term retention of skills:}}
Most VR studies that have shown effect on learning outcomes have focused immediate transfer of skills \cite{ye2023robot}, although long-term transferability remains a question \cite{ipsita2022towards, lassiter2023welding}. 

\emph{\textcolor{PineGreen}{Limited efforts in personalized and adaptive training systems:}}
Studies also call for adaptive XR training systems by analysing learner behavior \cite{hoover2021designing} and data such as eye gaze, hand movements, and self-reported confidence levels \cite{lor2024enabling}. 

\emph{\textcolor{PineGreen}{Concerns with Hygiene and user preferences:}}
Comparative studies show that non-immersive setups are viable where hygiene or user preferences are concerns \cite{de2019effects}. 

\emph{\textcolor{PineGreen}{Issues with content development:}}
Efforts emphasize support for knowledge sharing through ontology systems offering new pathways for collaborative development, support, and learning \cite{kim2010itrain}.

\label{sec:6}
\textcolor{Violet}{\section{Entities in Manufacturing Education}}
First, the secondary taxonomy for entities is explained. Then the findings from the review and analysis are presented in terms of the current state, benefits, challenges, and future opportunities of VR in the corresponding category.

\subsection{\textcolor{Violet}{Entity Taxonomy:}}
\begin{figure*}[tb]
    \centering
    \includegraphics[width = \textwidth]{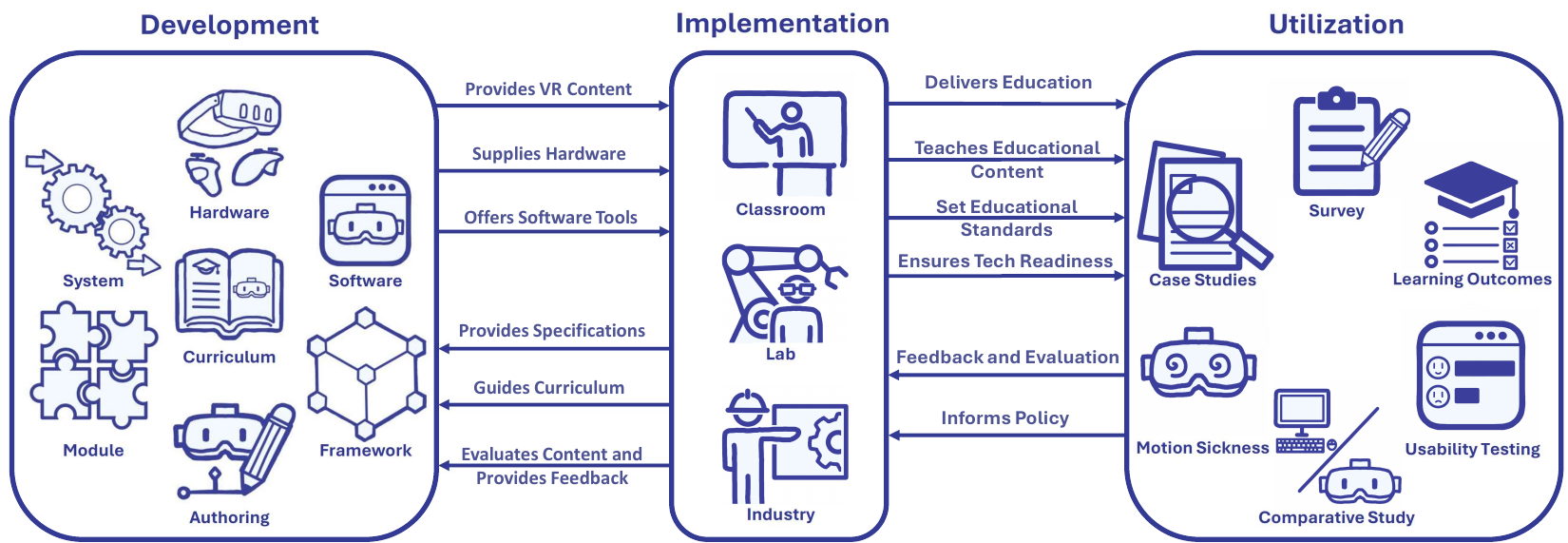}
    \caption{The entities in manufacturing education are categorized into three main groups: (A) Development, (B) Implementation, and (C) Adoption. The image highlights the focus of the various entity groups inside the rectangular boxes and illustrates the functional relationships between these entity groups using the arrows.}
    \label{fig:EntityClassification}
\end{figure*}

The entities involved in the use of VR in manufacturing education can be broadly grouped into three main groups based on their primary functions and roles: (A) \emph{\textcolor{Violet}{Development}}, (B) \emph{\textcolor{Violet}{Implementation}}, and (C) \emph{\textcolor{Violet}{Utilization}}. These entities collaborate in an iterative manner to prepare goal-oriented learning experiences. As shown in Fig. \ref{fig:EntityClassification}, the focus of each entity group is highlighted inside the rectangular boxes and the collaboration between the entity groups is highlighted using the arrows. Across the categories in entities, there is a general upward trend over the years. Development has a steep increase. This suggests more resources have been allocated towards it. There is a correlation between development and utilization trends which indicates that as more content is developed, they are also being increasingly utilized. Implementation lags behind development and utilization trends. This suggests potential barriers to translating developments into implemented solutions.

\subsubsection{\textcolor{Violet}{Development:}} This entity is responsible for creating the technology and content to be used for educational purposes. The primary contributors include designers and developers of educational content. The output is achieved in form of hardware devices, software applications, systems (including user interfaces, interactions, and virtual environments), learning modules, curricula, and frameworks. Their efforts aim to ensure the technology is accessible, user-friendly, and tailored to meet educational needs.

\subsubsection{\textcolor{Violet}{Implementation: }} This entity focuses on implementing the VR technology in education. They are responsible for integrating VR technologies into teaching and learning processes, and ensuring the content is relevant and aligned with educational and industry standards. The primary contributors include subject matter experts guiding the design of the learning activities, and instructors integrating the learning experiences into educational settings and setting the learning outcomes for learners. The focus of this entity is to implement VR content in classrooms, labs, and industry settings.

\subsubsection{\textcolor{Violet}{Utilization: }} This entity is involved in evaluating VR technology in manufacturing education, assessing the impact on learning outcomes, and establishing guidelines for its effective use. The primary contributors include learners/students who are the end-users of the learning experiences; research institutions studying the impact and effectiveness of VR; and policymakers and educational authorities setting standards and policies for VR use in education. The focus of this entity is to participate in surveys and user studies and provide feedback.

\subsection{\textcolor{Violet}{Findings from the review analysis for Development:}} From the review analysis, 77 papers were found to be associated with this group in the field of manufacturing education. Table \ref{tab:entity_classification} shows the distribution of papers reviewed across the different categories of the Development entity.

\subsubsection{\textcolor{Violet}{Current State:}}
Among the development efforts, 48 papers developed system workflows in the form of virtual environments, interfaces, and interactions. 8 papers provided novel designs of hardware interfaces. 13 papers provided details on the software development aspects of the VR applications. 21 papers highlighted learning modules. 4 papers provided curricula on manufacturing use cases. 13 papers developed frameworks for integrating VR in manufacturing education. 7 papers showed authoring efforts.

The development of VR-based systems span across a variety of domains, including welding \cite{white2011low, wang2019virtual, wang2020digital, ipsita2021vrfromx, ipsita2022towards, ye2023robot, ipsita2024design, ye2024user}, machining \cite{ryan2022immersive, lie2023training}, forming \cite{moreland2020integrating, moreland2022development}, assembly \cite{adas2013virtual, al2016development, dodoo2018evaluating, dwivedi2018manual, etemadpour2019visualization, sharma2019collaborative, aqlan2020multiplayer, zhu2021eye, zhu2022sensor, hartleb2023exploring, chang2024efficient}, additive manufacturing \cite{renner2015virtual}, safety \cite{bushra2018comparative, moreland2021development, kobir2023human, mathur2024using}, quality control \cite{belga2022carousel}, robotics \cite{darmoul2015virtual, theofanidis2017varm, robotlearning2018, shi2020affordable, chang2020exploring, srinivasan2021biomechanical, ertekin2023board, tram2023intuitive}, sustainability \cite{borst2016virtual}, iron and steel \cite{moreland2021development}, smart manufacturing \cite{jun2021human, yun2022immersive, zhu2023learniotvr}. These efforts provide interfaces \cite{bushra2018comparative}, interactions \cite{ye2024user, hartleb2023exploring, belga2022carousel, fitton2024watch}, environments \cite{lor2024enabling, ye2024user, sharma2019collaborative, borst2016virtual, darmoul2015virtual, ryan2022immersive}, training simulators \cite{dodoo2018evaluating, al2016development, white2011low, ipsita2022towards, ryan2022immersive, moreland2020integrating, moreland2021development, moreland2022development, ye2023robot, lie2023training, ye2024user, rafa2024enhancing} and assessment techniques \cite{bushra2018comparative, wang2020digital, zhu2021eye,belga2022carousel, zhu2022sensor, ye2023robot, ye2024user, kim2024behavioral, lor2024enabling}. Innovative approaches include teleoperation interfaces \cite{theofanidis2017varm, chang2020exploring, wang2019virtual, tram2023intuitive}, visualization tool \cite{etemadpour2019visualization}, collaborative and multiplayer environments \cite{sharma2019collaborative, kim2024behavioral, chang2024efficient, aqlan2020multiplayer, hartleb2023exploring, kim2024behavioral}, gaming systems \cite{aqlan2020multiplayer}, CPS system using VR \cite{jun2021human, yun2022immersive} and adaptive and personalized training systems \cite{lor2024enabling} and layout optimization \cite{kobir2023human}. Developers leverage powerful engines like Unity \cite{ shi2020affordable, sharma2019collaborative, zhao2019developing, chiou2021developing, ma2019efficacy, dodoo2018evaluating, zhu2021eye, frank2021green, mccusker2018virtual, dwivedi2018manual, aqlan2020multiplayer, chiou2020project, wang2020towards, theofanidis2017varm, borst2016virtual, chiou2019virtual, chiou2024virtual} and Unreal Engine, which have advanced to support the physics-intensive nature of manufacturing processes. Widely used VR devices, such as Oculus and HTC Vive headsets, are supported by libraries and SDKs that streamline virtual environment development. Real-time data collection through sensor tracking, including eye-tracking, hand movement, and body posture monitoring, enables precise user assessment and interaction. Collaborative systems built with Photon networking in Unity enhance multi-user experiences in virtual environments.

Hardware-specific solutions have also been introduced to address specialized manufacturing training needs. Some of these include hardware development for immersive system \cite{chen2010virtual, white2011low, wang2020digital, ye2023robot, ipsita2022towards}, teleoperation interfaces \cite{wang2019virtual}, interaction hardware \cite{renner2015virtual}, physical prototyping \cite{adas2013virtual} and haptic integration \cite{white2011low, al2016development, wang2020digital, ye2023robot, ipsita2022towards}. Software development aspects have focused on details regarding code workflows \cite{adas2013virtual, cecil2013virtual, renner2015virtual, al2016development, chiou2024virtual}, ontology systems \cite{kim2010itrain}, visualizations of numerical models using CFD \cite{chen2010virtual, zhou2016comprehensive}, algorithms for virtual simulation \cite{white2011low}, intent recognition \cite{wang2019virtual}, skill assessment \cite{wang2020digital, ye2023robot} and guidance mechanisms \cite{woodworth2023study}.

Learning modules that target specific learning objectives are focused in specific topics such as CNC machining \cite{el2016assessment}, electronics lab \cite{mccusker2018virtual}, craft production \cite{zhao2019developing}, manufacturing system design \cite{ma2019efficacy}, robotic ultrasonic welding \cite{chiou2019virtual}, collaborative design and assembly \cite{aqlan2020multiplayer}, microfabrication \cite{wang2020towards}, nanotechnology \cite{kamali2020virtual}, green manufacturing \cite{chiou2021developing, frank2021green}, welding \cite{ipsita2022towards}, 3D printing \cite{kobir2023human}, robotics and machining \cite{ertekin2023board}, drilling \cite{lie2023training}, equipment training \cite{park2023work},  machining \cite{studer2024open}, continuous improvement techniques \cite{parsley2024enhancing}, energy assessment \cite{kula2024development}, safety and emergency training \cite{tang2024evaluation}, hydraulic grippers in fluid power course \cite{azzam2024virtual} while extensive curriculum are focused in certain works \cite{ashour2021connected, kamali2020virtual, ertekin2023board, ipsita2022towards} in the field of welding, machining, robotics and naontechnology.

Frameworks have been developed to define foundations and provide design guidelines for integrating VR into manufacturing education. Efforts include educational frameworks design using learning theories \cite{studer2024open, ipsita2022towards}, CPS framework using VR for collaboration between humans, machines, and autonomy \cite{yun2022immersive, jun2021human}, design recommendations for adaptive training systems \cite{hoover2021designing}, interactive storytelling for vocational training systems \cite{doolani2020vis, gupta2019viis}, integrating VR into learning pipeline for sustainability education \cite{chiou2020project, frank2021green}, teaching framework for industry 4.0 \cite{osti_10293626} and craft production \cite{aqlan2019integrating}, assessment frameworks for human-robot collaboration in assembly tasks \cite{abujelala2018collaborative}, and scalable and customizable training for additive manufacturing \cite{mogessie2020work}.

Authoring platforms like programming by demonstration and WYSIWYG (What You See Is What You Get) tools allow users to develop and test content without programming expertise \cite{ chang2020exploring, yun2022immersive, tram2023intuitive, ipsita2024design, theofanidis2017varm, ipsita2021vrfromx}. AI-based methods, such as reinforcement learning \cite{ ashour2021connected} and other deep learning approaches support innovative and adaptive content creation.

In Development, \textcolor{Violet}{Systems} has seen an increase throughout the time period. The role of \textcolor{Violet}{hardware} and \textcolor{Violet}{authoring} has remained relatively stagnant. These trends show a stable demand for such capabilities. The data also suggests a growing use of \textcolor{Violet}{modules}.

\begin{table*}[]
\centering
\caption{Distribution of papers used in the review found across the different categories of Entities}
\label{tab:entity_classification}
\resizebox{\textwidth}{!}{%
\begin{tabular}{|cl|cl|cl|}
\hline
\multicolumn{2}{|c|}{\textbf{Development}} & \multicolumn{2}{c|}{\textbf{Implementation}} & \multicolumn{2}{c|}{\textbf{Utilization}} \\ \hline
\multicolumn{1}{|c|}{\textbf{System}} & \begin{tabular}[c]{@{}l@{}}\cite{kobir2023human, robotlearning2018, renner2015virtual, etemadpour2019visualization, shi2020affordable, kim2024behavioral}\\ \cite{ertekin2023board, belga2022carousel, sharma2019collaborative, ashour2021connected, al2016development, moreland2021development}\\ \cite{moreland2022development, wang2020digital, chang2024efficient, lor2024enabling, rafa2024enhancing, dodoo2018evaluating, hartleb2023exploring}\\ \cite{chang2020exploring, zhu2021eye, jun2021human, yun2022immersive, ryan2022immersive, moreland2020integrating, tram2023intuitive, kim2010itrain}\\ \cite{zhu2023learniotvr, white2011low, dwivedi2018manual, aqlan2020multiplayer, ye2023robot, zhu2022sensor, srinivasan2021biomechanical, lopez2020click}\\ \cite{ipsita2024design, ipsita2022towards, lie2023training, ye2024user, mathur2024using, theofanidis2017varm, adas2013virtual, borst2016virtual}\\ \cite{darmoul2015virtual, wang2019virtual, ipsita2021vrfromx, fitton2024watch, bushra2018comparative}\end{tabular} & \multicolumn{1}{c|}{\multirow{2}{*}{\textbf{Classroom}}} & \begin{tabular}[c]{@{}l@{}}\cite{ismail2024immersive, renner2015virtual, ashour2021connected}\\ \cite{parsley2024enhancing, robotlearning2018, rafa2024enhancing}\\ \cite{mccusker2018virtual, white2011low, aqlan2020multiplayer}\\ \cite{stone2011physical, chiou2020project}\\ \cite{borst2018teacher, lopez2020click, lie2023training}\\ \cite{cecil2013virtual, stone2011virtual}\end{tabular} & \multicolumn{1}{c|}{\textbf{Survey}} & \begin{tabular}[c]{@{}l@{}}\cite{zhou2016comprehensive, hoover2021designing, mathur2022identifying, chang2021using, price2019using, cecil2013virtual, chiou2019virtual}\\ \cite{kamali2020virtual, lassiter2023welding}\end{tabular} \\ \cline{1-2} \cline{5-6} 
\multicolumn{1}{|c|}{\textbf{Hardware}} & \cite{renner2015virtual, al2016development, wang2020digital, white2011low, ye2023robot, adas2013virtual, chen2010virtual, wang2019virtual} & \multicolumn{1}{c|}{} &  & \multicolumn{1}{c|}{\textbf{Case Studies}} & \cite{dwivedi2018manual, rumsey2020manufacturing, theofanidis2017varm, ritter2018virtual} \\ \hline
\multicolumn{1}{|c|}{\textbf{Software}} & \begin{tabular}[c]{@{}l@{}}\cite{renner2015virtual, zhou2016comprehensive, al2016development, wang2020digital, kim2010itrain, white2011low, ye2023robot, woodworth2023study}\\ \cite{adas2013virtual, cecil2013virtual, chen2010virtual, wang2019virtual, chiou2024virtual}\end{tabular} & \multicolumn{1}{c|}{\multirow{2}{*}{\textbf{Lab}}} & \multirow{2}{*}{\begin{tabular}[c]{@{}l@{}}\cite{kobir2023human, el2016assessment, ertekin2023board}\\ \cite{zhu2021eye, price2019using, srinivasa2021virtual}\\ \cite{azzam2024virtual, chiou2019virtual}\\ \cite{chiou2024virtual, park2023work, mogessie2020work}\end{tabular}} & \multicolumn{1}{c|}{\textbf{\begin{tabular}[c]{@{}c@{}}Learning Outcomes \\ (Testing)\end{tabular}}} & \begin{tabular}[c]{@{}l@{}}\cite{ma2020approach, yamada2017educational, el2016assessment, ismail2024immersive, ISLAM2024103648, ashour2021connected, kula2024development}\\ \cite{wang2020digital, mathur2024effects, ma2019efficacy, chang2024efficient, parsley2024enhancing, rafa2024enhancing, bowling2010evaluating}\\ \cite{ostrander2020evaluating, tang2024evaluation, stone2013full, mccusker2018virtual, zhu2023learniotvr, aqlan2020multiplayer, stone2011physical}\\ \cite{chiou2020project, evans2021prototyping, hannah2024results, cecil2024study, borst2018teacher, lopez2020click, de2019effects, conesa2023influence, byrd2015use}\\ \cite{wang2020towards, ipsita2022towards, lie2023training, chang2021using, mathur2024using, price2019using, adas2013virtual, cecil2013virtual}\\ \cite{srinivasa2021virtual, stone2011virtual, chiou2019virtual, kamali2020virtual, ritter2018virtual, fitton2024watch, park2023work}\end{tabular} \\ \cline{1-2} \cline{5-6} 
\multicolumn{1}{|c|}{\textbf{Module}} & \begin{tabular}[c]{@{}l@{}}\cite{kobir2023human, studer2024open, el2016assessment, ertekin2023board, zhao2019developing, chiou2021developing, kula2024development, ma2019efficacy}\\ \cite{parsley2024enhancing, tang2024evaluation, zhu2021eye, frank2021green, mccusker2018virtual, aqlan2020multiplayer, wang2020towards, ipsita2022towards}\\ \cite{lie2023training, azzam2024virtual, chiou2019virtual, kamali2020virtual, park2023work}\end{tabular} & \multicolumn{1}{c|}{} &  & \multicolumn{1}{c|}{\textbf{Motion Sickness}} & \cite{ma2019efficacy, mitchell2020safety, de2019effects, ipsita2022towards} \\ \hline
\multicolumn{1}{|c|}{\textbf{Curriculum}} & \cite{ertekin2023board, ashour2021connected, ipsita2022towards, kamali2020virtual} & \multicolumn{1}{c|}{\multirow{6}{*}{\textbf{Industry}}} & \multirow{6}{*}{\cite{hannah2024results}} & \multicolumn{1}{c|}{\textbf{Usability Testing}} & \begin{tabular}[c]{@{}l@{}}\cite{kobir2023human, etemadpour2019visualization, belga2022carousel, zhao2019developing, kula2024development, kuts2022digital, ma2019efficacy}\\ \cite{chang2024efficient, hartleb2023exploring, zhu2021eye, ryan2022immersive, tram2023intuitive, zhu2023learniotvr, dwivedi2018manual, ye2023robot}\\ \cite{zhu2022sensor, woodworth2023study, srinivasan2021biomechanical, ipsita2024design, conesa2023influence, ipsita2022towards, ye2024user}\\ \cite{borst2016virtual, srinivasa2021virtual, wang2019virtual, doolani2020vis, ipsita2021vrfromx, bushra2018comparative}\end{tabular} \\ \cline{1-2} \cline{5-6} 
\multicolumn{1}{|c|}{\textbf{Framework}} & \begin{tabular}[c]{@{}l@{}}\cite{osti_10293626, studer2024open, hoover2021designing, mathur2023designing, frank2021green, jun2021human, yun2022immersive, aqlan2019integrating}\\ \cite{chiou2020project, gupta2019viis, doolani2020vis, mogessie2020work, abujelala2018collaborative}\end{tabular} & \multicolumn{1}{c|}{} &  & \multicolumn{1}{c|}{\multirow{5}{*}{\textbf{Comparative Study}}} & \multirow{5}{*}{\begin{tabular}[c]{@{}l@{}}\cite{ma2020approach, ismail2024immersive, belga2022carousel, kuts2022digital, wang2020digital, mathur2024effects, ma2019efficacy, chang2024efficient}\\ \cite{ostrander2020evaluating, tang2024evaluation, hartleb2023exploring, mathur2022identifying, ryan2022immersive, mccusker2018virtual, zhu2023learniotvr}\\ \cite{dwivedi2018manual, aqlan2020multiplayer, evans2021prototyping, hannah2024results, mitchell2020safety, zhu2022sensor, woodworth2023study}\\ \cite{borst2018teacher, srinivasan2021biomechanical, lopez2020click, ipsita2024design, de2019effects, byrd2015use, ipsita2022towards, lie2023training}\\ \cite{ye2024user, mathur2024using, theofanidis2017varm, cecil2013virtual, srinivasa2021virtual, stone2011virtual, ritter2018virtual, park2023work, bushra2018comparative}\end{tabular}} \\ \cline{1-2}
\multicolumn{1}{|c|}{\multirow{4}{*}{\textbf{Authoring}}} & \multirow{4}{*}{\cite{sharma2019collaborative, chang2020exploring, yun2022immersive, tram2023intuitive, ipsita2024design, theofanidis2017varm, ipsita2021vrfromx}} & \multicolumn{1}{c|}{} &  & \multicolumn{1}{c|}{} &  \\
\multicolumn{1}{|c|}{} &  & \multicolumn{1}{c|}{} &  & \multicolumn{1}{c|}{} &  \\
\multicolumn{1}{|c|}{} &  & \multicolumn{1}{c|}{} &  & \multicolumn{1}{c|}{} &  \\
\multicolumn{1}{|c|}{} &  & \multicolumn{1}{c|}{} &  & \multicolumn{1}{c|}{} &  \\ \hline
\end{tabular}%
}
\end{table*}

\subsubsection{\textcolor{Violet}{Benefits of VR for Development entity:}}
\hfill\\
\emph{\textcolor{Violet}{Availability of resources for VR development:}}
The development is based on computer graphics and game development, with resources in using gamified learning for education. In that context, resources are available. Firstly, the availability of powerful game engines like Unity and Unreal Engine enables developers to create versatile VR environments tailored to diverse training scenarios. The integration of robust sensor tracking and real-time data analytics not only enhances the quality of the content but also provides developers with insights for iterative improvements \cite{lor2024enabling, zhu2021eye, zhu2022sensor, woodworth2023study}. Collaborative frameworks, such as Photon networking, streamline the development of multi-user systems, reducing development time and complexity for team-based VR solutions. Tech giants such as Meta and Google also keep developing SDKs and device upgrades which are normally open source, so resources are available for development.

\emph{\textcolor{Violet}{Advancement in technology (5G, AI, wearables) facilitates authoring:}}
The modularity of VR systems, including adaptive frameworks and AI-based methods, ensures that content can be updated and expanded to align with evolving educational and industry needs. Finally, the increasing diversity in VR environments, including realistic representations of users, helps developers cater to broader audiences and enhance relatability, which can further elevate the adoption and success of their solutions \cite{ ipsita2021vrfromx}.

\subsubsection{\textcolor{Violet}{Challenges and Future Opportunities of VR for Development entity:}}
\hfill\\
\emph{\textcolor{Violet}{Collaboration barriers:}}
Collaborative efforts among programmers, subject matter experts, and industry stakeholders are often time-intensive and costly. Additionally, aligning VR content with learning theories, such as backward design, requires developers to collaborate closely with subject matter experts, which can slow down production \cite{ipsita2022towards}. The need to update and reuse content to adapt to changing learning objectives adds complexity to the development process. The prototyping phase also faces issues in terms of conveying requirements between stakeholders. Finally, AI-driven methods, while promising, require further refinement to effectively capture user intent and provide seamless interactive experiences for content creators \cite{ipsita2021vrfromx}.

\emph{\textcolor{Violet}{Disconnected development and testing platforms:}}
Immersive authoring platforms empower developers and content creators, especially subject matter experts, by enabling iterative testing and real-time adjustments during the creation process \cite{chang2020exploring, yun2022immersive, ipsita2024design, theofanidis2017varm, ipsita2021vrfromx}. While immersive authoring tools are designed to reduce dependency on programming expertise, their adoption is still limited, and their usability for large-scale development remains a concern. 

\emph{\textcolor{Violet}{Standardization issues in software and hardware:}} Due to frequent upgrades, the code base needs to get updated. When previous versions get outdated, it requires considerable changes in the code base and rework.
Frameworks for adaptive training and diverse content creation are available but lack standardization, making it difficult to ensure consistent outcomes \cite{rumsey2020manufacturing}.

\emph{\textcolor{Violet}{Difficult to replicate environment and feedback:}} It can be difficult to replicate the virtual environment close to physical settings due to limited resources to mimic physics and environmental conditions such as temperature, smell, taste, etc. \cite{ipsita2024design}. In such cases, a combination of virtual and physical training can take place, where virtual training can complement the relatively easy-to-replicate parts of physical training \cite{stone2011virtual}.

\subsection{\textcolor{Violet}{Findings from the Review Analysis for Implementation: }} From the analysis, 30 papers were found to be associated with this entity in the field of manufacturing education. Table \ref{tab:entity_classification} shows the distribution of papers reviewed across the different categories of the Implementation entity.

\subsubsection{\textcolor{Violet}{Current State:}} Among the implementation efforts, 16 studies focused on classroom settings, 11 targeted laboratory settings, and only 1 was conducted in an industry setting. The limited implementation in industry likely stems from copyright and proprietary concerns which may restrict the publication of findings. In contrast, classroom and laboratory settings benefit from the proximity of research institutions to student participants. This makes implementation in such settings relatively easier as compared to industry.

In classroom settings, the implementation vary across a number of courses such as welding training in community colleges and trade schools \cite{white2011low, stone2011physical, stone2011virtual}, in higher education in micro-assembly \cite{cecil2013virtual}, additive manufacturing \cite{renner2015virtual, rafa2024enhancing}, robot programming \cite{robotlearning2018}, electronic workbench \cite{mccusker2018virtual}, sustainability education \cite{borst2018teacher, chiou2020project}, collaborative design and assembly \cite{aqlan2020multiplayer}, manufacturing system \cite{lopez2020click}, drilling \cite{lie2023training}, cybersecurity in additive manufacturing \cite{ismail2024immersive}, and Lean Systems Engineering courses \cite{parsley2024enhancing}. The implementation efforts in laboratory settings range across topics such as the operation of CNC milling machine \cite{el2016assessment}, welding \cite{price2019using}, ultrasonic welding \cite{chiou2019virtual}, additive manufacturing \cite{mogessie2020work, kobir2023human}, manufacturing tasks \cite{zhu2021eye}, material testing \cite{srinivasa2021virtual}, robotics and automation \cite{ertekin2023board}, equipment training \cite{park2023work}, green manufacturing \cite{chiou2024virtual}, and hydraulic gripper in a fluid power course \cite{azzam2024virtual}. In industry settings, VR short exercises are used to teach contractor safety training for the energy industry \cite{hannah2024results}.

\subsubsection{\textcolor{Violet}{Benefits of VR for Implementation entity:}} The implementation of VR in manufacturing education presents several benefits. 
\hfill\\
\emph{\textcolor{Violet}{Student enthusiasm and proximity facilitates implementation and testing:}}
First, the immersiveness of VR captures student enthusiasm \cite{mathur2024effects}. This enthusiasm enhances participation and student motivation to learn manufacturing concepts that can make the integration of such tools more effective. Direct testing with students makes it easier for educators to assess the transferability of skills. 

\emph{\textcolor{Violet}{Resource optimization in terms of cost and time:}}
VR also minimizes resource wastage and consumable costs, making it a cost-effective option for institutions \cite{ipsita2022towards, yamada2017educational}. Furthermore, VR reduces instructor time requirements and allows educators to focus on higher-level teaching goals. 

\emph{\textcolor{Violet}{Remote teaching can be interactive, collaborative and engaging:}}
The technology also supports distance and collaborative learning, providing hands-on experiences without geographical constraints \cite{aqlan2020multiplayer, stone2011virtual, lowell2023virtual}. This becomes valuable in remote education scenarios or during disruptions such as the COVID-19 pandemic.

\subsubsection{\textcolor{Violet}{Challenges and Future Opportunities of VR for Implementation entity:}}
\hfill\\
\emph{\textcolor{Violet}{Behavioral resistance and lack of awareness:}} Unlike desktop-based platforms that are well-integrated into learning systems, VR demands a shift in user behavior. Employees in traditional industries may feel reluctant to wear headsets or adapt to new digital interactions. Furthermore, educators remain unaware of the benefits associated with VR. A solution can be the development of technical workshops to spread awareness of immersive training and its impact on learning outcomes \cite{ismail2024immersive}.

\emph{\textcolor{Violet}{Cost and Infrastructure Concerns:}} While HMD costs have been reduced, the cost to implement VR on a large scale is high. Setting up the necessary hardware and environments for specific tasks adds to the high cost. Keeping up with advancements in changes in software and hardware is costly, many companies may prefer to hold onto older hardware rather than upgrading frequently. High costs of advanced hardware, such as CAVE systems \cite{renner2015virtual, yamada2017educational, ismail2024immersive, zhou2016comprehensive, rumsey2020manufacturing, chen2010virtual}, and integration of sophisticated tracking technologies also pose financial and logistical challenges.  

\emph{\textcolor{Violet}{Standardization concerns and Interoperability issues:}} The interactions offered by VR applications such as gesture-based controls vary across different headsets. Interoperability is another issue. Companies that train employees on one VR platform hesitate to transition to another due to inconsistencies in interfaces and workflows. The diverse needs of users may also necessitate the use of multiple platforms which can complicate standardization and content delivery \cite{rumsey2020manufacturing}. 

\emph{\textcolor{Violet}{Accessibility issues with content creation:}} If the VR experience is poorly designed, users may experience nausea or discomfort. Authoring and collaboration for content creation require time, effort, and financial investment. A solution in this space can be the use of authoring tools and AI-assisted digital content creation. These tools enable the development of prototypes by eliminating complex software and platforms \cite{ipsita2024design, ipsita2024authoring, ipsita2021vrfromx}.

\emph{\textcolor{Violet}{Inclusivity and scalability barriers:}} Incorporating established learning theories and frameworks can further ease adoption by aligning VR content with established pedagogical practices. Policies are needed to guide content evaluation, ensure privacy, and promote widespread adoption. Furthermore, efforts should be made to enable inclusivity across different groups \cite{ismail2024immersive, chen2010virtual}. 

\emph{\textcolor{Violet}{Policy and IT Support:}} Maintaining and updating content to suit targeted learning outcomes is essential \cite{ismail2024immersive}. Learning management systems thus, should adapt to deliver VR content effectively. Unlike universities, which have dedicated research programs such as in HCI and computing education, industry lacks structured divisions such as dedicated IT support and policy support focused on exploring the potential of VR to reach a “must-have” status \cite{rumsey2020manufacturing}.

\subsection{\textcolor{Violet}{Findings from the Review Analysis for Utilization: }} From the analysis, 75 papers were found to be associated with this entity group. Table \ref{tab:entity_classification} shows the distribution of papers reviewed across the different categories of the Utilization entity.

\subsubsection{\textcolor{Violet}{Current State:}} Among the utilization efforts, 9 papers conducted survey in classrooms to get student perception of the technology and learning outcomes. 4 papers focused on conducting case studies on any manufacturing use cases. 45 papers focused on how VR can help in the development of learning outcomes. 4 papers studies motion sickness using VR. 28 papers tested the usability of the system on an environment developed in VR. 39 papers performed comparative studies which compared user interfaces (menus, interaction interfaces), training systems (HMD, CAVE, in-person, online, integrated, video, desktop), and skill development (expertise, haptics).

Surveys are conducted to collect data regarding student perception and performance \cite{cecil2013virtual, price2019using, chiou2019virtual, kamali2020virtual, chang2021using, mathur2022identifying}, interviews conducted with experts for adaptive training \cite{hoover2021designing} and using VR in welding training \cite{lassiter2023welding}. Case studies have explored virtual technologies in manual assembly \cite{dwivedi2018manual}, solar energy center \cite{ritter2018virtual}, and industrial robotics \cite{theofanidis2017varm}. Another case study also looked into the organizational impact of VR on human performance \cite{rumsey2020manufacturing}.

Studies indicate improvement in learning outcomes using VR learning in the form of skill development \cite{ma2020approach, wang2020digital, ipsita2022towards, zhu2023learniotvr, tang2024evaluation, fitton2024watch}, better engagement \cite{borst2018teacher, kamali2020virtual, rafa2024enhancing, tang2024evaluation}, reduced error rates \cite{wang2020towards, ipsita2022towards, lie2023training}, higher quiz scores \cite{cecil2013virtual}, enhanced knowledge \cite{ma2019efficacy, lie2023training, mathur2024effects}, improved learning speed \cite{wang2020towards}, independent learning ability \cite{wang2020towards}, self-efficacy \cite{srinivasa2021virtual}, positive attitude \cite{srinivasa2021virtual}, realistic experience \cite{ipsita2022towards}, improved task performance \cite{bowling2010evaluating, stone2011physical, stone2013full, byrd2015use, ipsita2022towards, park2023work, rafa2024enhancing}, creative and innovative solutions \cite{chiou2020project, yamada2017educational, evans2021prototyping}, better user experience \cite{ritter2018virtual}, increased confidence \cite{park2023work}, proactive participation \cite{park2023work}, increased motivation \cite{ma2019efficacy, chiou2020project, lopez2020click, ashour2021connected, chang2021using}, stimulated interest \cite{park2023work}, reduced struggles \cite{park2023work}, enhanced problem-solving willingness \cite{parsley2024enhancing, park2023work}, improved spatial and computational skills \cite{conesa2023influence, ismail2024immersive, parsley2024enhancing}, reduced spatial cognitive load \cite{chang2021using, mathur2024effects}, better practical problem-solving and critical thinking \cite{parsley2024enhancing}, decreased task completion time \cite{ipsita2022towards, lie2023training, ISLAM2024103648}, improved movement precision over repeated trials \cite{ISLAM2024103648}, higher task efficiency \cite{chang2024efficient}, greater user preference \cite{chang2024efficient}, knowledge retention \cite{hannah2024results, fitton2024watch}, increased urgency \cite{tang2024evaluation}, enhanced perception of realism \cite{el2016assessment, tang2024evaluation}, better comprehension and understanding \cite{mccusker2018virtual, price2019using, ostrander2020evaluating, srinivasa2021virtual, cecil2024study}, improved learning transfer \cite{ipsita2022towards, fitton2024watch}, and reduced build time \cite{mathur2024using}, support material usage \cite{mathur2024using}, and evaluation effort without increasing cognitive load \cite{mathur2024using}, increased team-based learning \cite{stone2011virtual, aqlan2020multiplayer}. The outcomes were dependent on guidance shown to the user \cite{bowling2010evaluating}, expertise levels of learner and task difficulty \cite{stone2013full, byrd2015use, ma2020approach, wang2020digital}.

Many studies indicates high usability ratings of the systems and methods developed for training purposes using System Usability Scores (SUS) evaluation \cite{borst2016virtual, dwivedi2018manual, bushra2018comparative, etemadpour2019visualization, zhao2019developing, ma2019efficacy, wang2019virtual, doolani2020vis, zhu2021eye, srinivasan2021biomechanical, srinivasa2021virtual, ipsita2021vrfromx, belga2022carousel, kuts2022digital, ryan2022immersive, zhu2022sensor, ipsita2022towards, kobir2023human, hartleb2023exploring, tram2023intuitive, zhu2023learniotvr, ye2023robot, woodworth2023study, conesa2023influence, ipsita2024design, ye2024user, kula2024development, chang2024efficient}. Some studies also show low scores of simulator sickness in immersive environments \cite{ma2019efficacy, de2019effects, mitchell2020safety, ipsita2022towards}.

Comparative studies are conducted to compare effectiveness of VR with traditional training such as physical training using in-person instruction and real equipment, desktop-based methods, computer-aided instruction, other types of immersive platforms such as CAVE and integrated VR training \cite{stone2011virtual, cecil2013virtual, theofanidis2017varm, mccusker2018virtual, dwivedi2018manual, ritter2018virtual, ma2019efficacy, ostrander2020evaluating, mitchell2020safety, lopez2020click, srinivasan2021biomechanical, srinivasa2021virtual, kuts2022digital, mathur2022identifying, ipsita2022towards, lie2023training, mathur2024effects, tang2024evaluation, hannah2024results, mathur2024using, de2019effects, evans2021prototyping, ryan2022immersive, zhu2023learniotvr, ismail2024immersive, mccusker2018virtual}. A few studies compare the effectiveness of VR training based on skill level of learners \cite{byrd2015use, ma2020approach, wang2020digital, ipsita2024design} while others compare user experience and usability of techniques and system features including user interfaces such as menus and selections \cite{bushra2018comparative, belga2022carousel, chang2024efficient}, interactions \cite{hartleb2023exploring}, types of content such as networked vs. standalone teaching \cite{borst2018teacher}, guidance techniques \cite{woodworth2023study, ye2024user} and assessment techniques \cite{zhu2022sensor}.

In utilization, the data indicates a growing trend in \textcolor{Violet}{Learning Outcomes} and \textcolor{Violet}{Usability Testing}. This suggests these areas have become increasingly important over the years.

\subsubsection{\textcolor{Violet}{Benefits of VR for Utilization entity:}} 
\hfill\\
\emph{\textcolor{Violet}{Improved retention due to higher engagement and motivation:}}
The \emph{immersive and realistic} nature of VR can significantly increase student \emph{engagement and motivation}, making learning more interactive and enjoyable, and can lead to improved retention rates and a deeper understanding of subject matter. Students reported higher engagement, motivation, and the ability to experiment safely without resource-intensive activities \cite{ismail2024immersive, kim2010itrain}. 

\emph{\textcolor{Violet}{Situated and experiential learning facilitates skill development:}}
The situated and experiential learning environment fosters hands-on learning, enhances spatial reasoning, and bridges the gap between theory and practical application. By allowing learners to simulate complex systems and troubleshoot in real time, it allows for safe experimentation and immediate feedback, fostering critical thinking \cite{parsley2024enhancing, ismail2024immersive} and problem-solving skills \cite{kim2024behavioral}. 

\emph{\textcolor{Violet}{Multiple pathways for varied objectives and personalized learning:}}
VR provides the flexibility to tailor learning experiences to individual learners' needs and skill levels, enhancing personalized education \cite{ipsita2024authoring, renner2015virtual}. Design frameworks such open-ended design enables learners to explore multiple pathways, fostering deeper engagement and better error management during skill acquisition \cite{studer2024open, ryan2022immersive}.

\subsubsection{\textcolor{Violet}{Challenges and Future Opportunities of VR for Utilization entity:}} 
\hfill\\
\emph{\textcolor{Violet}{Privacy and Security concerns:}}
VR, as a spatial platform, inherently involves the collection of extensive user data through devices that track eye movement, facial expressions, body movements, and other interactions \cite{zhu2021eye, kim2010itrain, ismail2024immersive, rafa2024enhancing}. This data can reveal significant information about users, raising privacy concerns. Private companies are continuously developing measures to address these issues by ensuring sensitive information is not misused and by obtaining explicit user permission before collecting such data. Institutions utilizing VR must adopt similar practices, securing deliberate approval from students before engaging in any form of tracking to safeguard their privacy.

\emph{\textcolor{Violet}{User experience and preferences:}}
On the other hand, there are health concerns such as motion sickness, nausea, and obstacle collisions, associated with interactions in VR \cite{azzam2024virtual}. Developers and hardware manufacturers are working towards improving resolution, reducing latency, and increasing the field of view, to enhance the user experience. Additionally, advancements in environment and scene understanding features can help avoid collisions with obstacles, making immersive environments safer and more practical for users.

\label{sec:7}
\section{Discussion}
We discuss two topics in this section: (1) the distribution of different aspects, and (2) the barriers and opportunities of VR adoption. In the first topic, the connections between the categories of the aspects are analyzed to find any interesting trends. In the second topic, the initial findings from the review analysis are analyzed to identify the major barriers in VR adoption in manufacturing education. Potential ways to address these barriers including technological impact are thought through to provide actionable insights.

\begin{figure*}[tb]
    \centering
    \includegraphics[width = 0.75\textwidth]{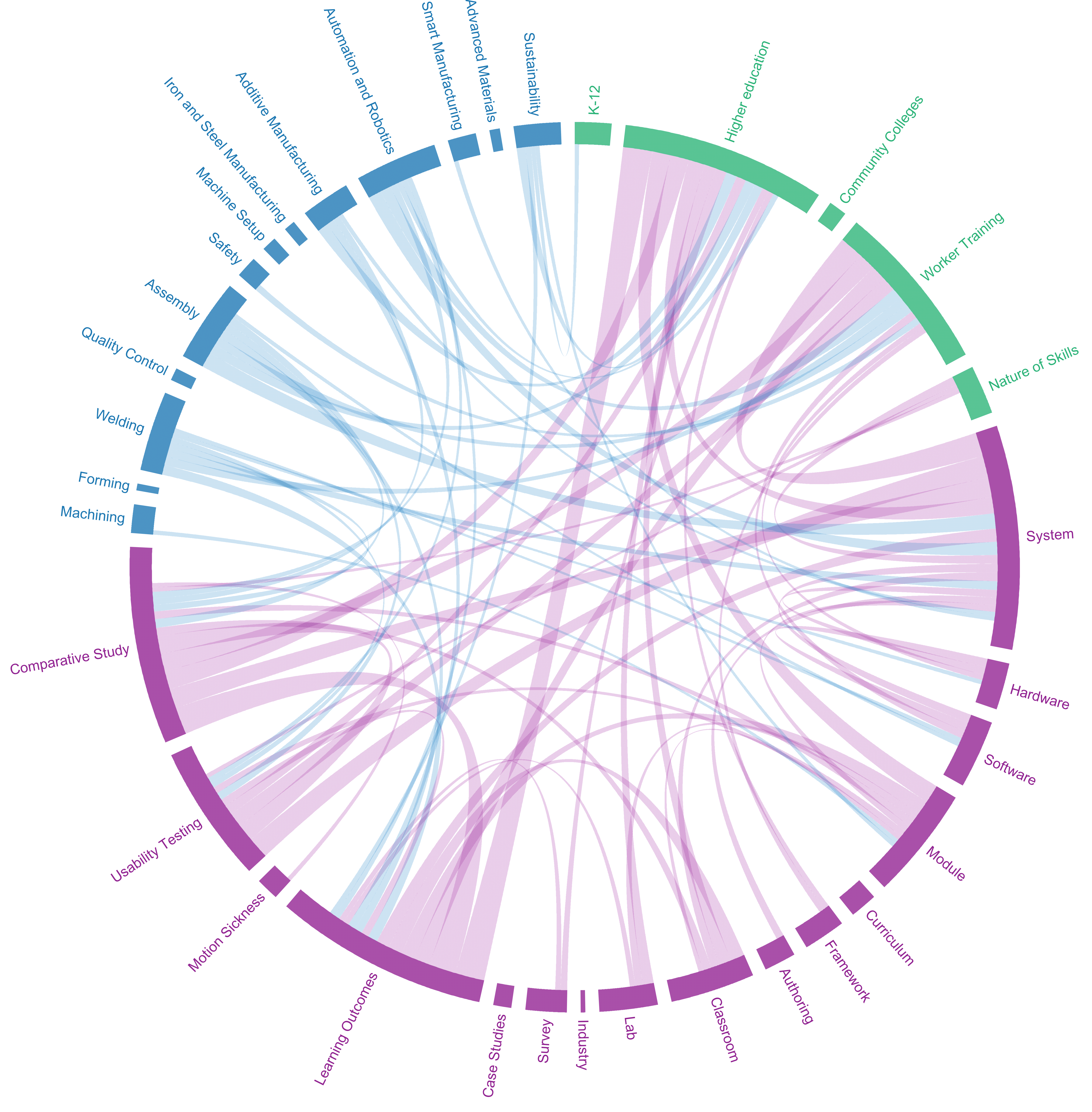}
    \caption{Chord Diagram showing the distribution of manufacturing education domains across various levels and entities involved in education (only the connections above the 75 percentile are shown for clarity purposes).}
    \label{fig:EntityDistribution}
\end{figure*}

\subsection{Distribution across Different Aspects: }

Figure \ref{fig:EntityDistribution} presents a Chord Diagram showing the interconnected distribution between the different aspects of manufacturing education. Below, we highlight key patterns observed from the analysis, with the number of interconnections in parentheses.

The number of connections between any two categories (as visualized by the thickness of links in Figure \ref{fig:EntityDistribution}) can indicate their mutual significance. Some of these values will be discussed below to provide interesting insights.

\textcolor{PineGreen}{Higher education} tops the chart with its strong connections across the different categories of the Utilization entity, such as \textcolor{Violet}{Learning Outcomes} (27), \textcolor{Violet}{Modules} (17), \textcolor{Violet}{Comparative Testing} (17), and \textcolor{Violet}{Systems} (15). This may be obvious because of the nature of the assessment at this level. \textcolor{PineGreen}{Higher Education} easily finds scope in terms of users, use cases, space, and proximity to research institutions to test the system and developed modules. Similarly, the distribution in the case of implementation settings also shows a similar data distribution. \textcolor{Violet}{Classroom} has strong connections with \textcolor{Violet}{Learning Outcomes} (13) and \textcolor{PineGreen}{Higher Education} (11). \textcolor{Violet}{Lab} shows strong connections with \textcolor{PineGreen}{Higher Education} (10). It shows that training applications can also be implemented easily in classroom and lab settings, as indicated by their connections with \textcolor{PineGreen}{Higher Education}. Educators who want to introduce immersive technology in their classrooms can also find advantages in terms of student enthusiasm about the new technology. Furthermore, the data collected can be used to compare against conventional methods (such as desktop-based methods, in-person training with real equipment, etc.) or integrated training (combined VR and physical) in comparative testing methods.

\textcolor{PineGreen}{Worker Training} follows a close second with strong connections across \textcolor{Violet}{System} (22), \textcolor{Violet}{Comparative Testing} (16), \textcolor{Violet}{Learning Outcomes} (14), and \textcolor{Violet}{Usability Testing} (13). Short-term evaluations often lead to usability testing and comparative evaluation with traditional methods, which can be observed from the chart. However, industry implementations may focus on the results in productivity, efficiency, and gains instead of reporting results from long-term usage of VR. Furthermore, controlled studies may affect productivity in industries. There is a need to find the outcomes of this to more effectively determine the benefits of VR.

\textcolor{PineGreen}{K-12} and \textcolor{PineGreen}{Community College} have relatively lower values across most categories, indicating less focus on these education levels. This may be due to the challenges of developing and implementing solutions at these levels. These may include limited funding, less awareness about the technology, and health concerns associated with headsets, particularly for young students at the K-12 level.
Another level that has shown limited influence but can benefit significantly from the advantages of VR is \textcolor{PineGreen}{Informal and Continuing Education} \cite{roussou2000immersive, ghafar2024practical}. The state of informal education can be improved with more educators in the field creating immersive content for manufacturing. The content can be imparted through workshops, enrichment programs, and vocational courses to expose learners to new emerging trends and create an interest in the field of manufacturing. Continuing education can benefit from Massive Open Online Courses (MOOCs) to help professionals in the field gain new hands-on experiences and skills from the benefits of VR \cite{serna2022leveraging, chandramouli2019mooc}. These courses can be developed through the collective efforts of universities, research institutions, and industry partners. Online repositories could contain content ready for anyone to download and learn manufacturing skills at their own pace.

Across domains, \textcolor{NavyBlue}{Assembly} shows the highest connections across \textcolor{Violet}{System} (14), \textcolor{PineGreen}{Higher Education} (9), \textcolor{Violet}{Comparative Study} (8), and \textcolor{PineGreen}{Worker Training} (8). The connections show the importance of the skills at both levels of higher education and worker training. This signifies the variety of manufacturing processes that require assembly skills.

\textcolor{NavyBlue}{Automation and Robotics} show strong connections across \textcolor{Violet}{System} (12) and \textcolor{PineGreen}{Worker Training} (10). \textcolor{NavyBlue}{Welding} shows strong connections with \textcolor{Violet}{Learning Outcomes} (9), \textcolor{Violet}{System} (8), and \textcolor{PineGreen}{Worker Training} (8). The impact of robotics and automation in the industry has been noticed over the past decades. This may have caused the growing requirement to upskill workers to learn these new skills, stay updated on industry standards, and contribute effectively to the output. These fields have seen more efforts in preparing workers compared to formal education. This can be related to the importance of welding in the manufacturing industry and the nature of training required to train workers in this field.

\textcolor{NavyBlue}{Additive Manufacturing} shows strong connections with \textcolor{PineGreen}{Higher Education} (10). \textcolor{NavyBlue}{Sustainability} shows strong connections with \textcolor{PineGreen}{Higher Education} (8). This indicates the growing relevance of these fields. Though these fields have seen a growing emphasis, their industry adoption is still at an early stage. Therefore, these fields are slowly being introduced in formal education settings to develop interest in the minds of young students. Also, the research carried out in these fields is ongoing, which may also contribute to the higher impact at formal education levels.

\subsection{Barriers and Opportunities for VR Adoption in Manufacturing Education: }

Based on our findings from the review, we try to answer the research question that was posed before: \emph{What are the different barriers that hinder the adoption of VR in manufacturing education? What efforts can address these barriers to facilitate greater adoption of VR in manufacturing education?} As shown in Table \ref{tab:barriers_to_VR}, we now talk about the major barriers identified from our findings. For each barrier, we also discuss action guidelines and technological affordances that can provide insights in addressing the barrier.

\textbf{Cost:} The cost to develop VR content, build infrastructural support, and implement VR at scale is high \cite{rumsey2020manufacturing}. Beyond the expenses of setting up hardware, software, and environments, additional investment is required to support collaboration between stakeholders as well as for ongoing maintenance and content updates. These large upfront costs, with uncertain returns on investment, often fall outside of standard budgets. Unless there is a guarantee of positive outcomes for learning and productivity, organizations tend to be skeptical about these high investments \cite{lowell2024applying}. In such cases, stakeholders from industry and academia should develop cost-benefit analysis models to demonstrate long-term returns on investment.

In academic settings, research grants in this area typically prioritize VR-based research over infrastructure and accessibility. In addition, a lack of clear policies that align academic programs with industry demands in manufacturing has affected the role of academia in workforce development. As a result, industries, facing an immediate need for specialized skills, often bypass academia and develop their own training programs that are faster and more cost-effective. While this approach reduces short-term training costs for industry, it weakens future workforce development by limiting the exposure of academia to current industry requirements. As a result, preskilling initiatives suffer, making it harder to transfer expertise from retiring and aging workers to future employees.

Therefore, organizational policies in both industry and academia should prioritize structured collaborations that integrate VR-based manufacturing training into existing curricula. Additionally, policies should promote technology-supported learning and clearer adoption regulations to secure budget allocations and dedicated IT support \cite{rumsey2020manufacturing}. At the government level, funding should be allocated to support VR in education and promote industry-academic collaboration. Such initiatives can have positive outcomes, helping the U.S. remain globally competitive, boost national productivity, and reduce reliance on external supply chains by promoting in-house production.

\textbf{Content:} Developing VR-based training applications for manufacturing requires stacks of multiple software platforms, such as Unity, Unreal Engine, AutoCAD, and Blender.. The need for programming expertise prevents subject matter experts in manufacturing from creating their own instructional content. As a result, content is often developed by programmers who lack manufacturing expertise. The resulting skill gap not only affects the relevance of content but also acts as a direct barrier to effectively transferring content from manufacturing experts to learners. To enhance accessibility to VR instruction, authoring systems should allow educators to create their own instruction. Approaches such as programming by demonstration and immersive authoring platforms should be encouraged to enable manufacturing experts to build applications using no-code solutions \cite{ chang2020exploring, tram2023intuitive, ipsita2024design}. From an instructional standpoint, workflows, and learning management systems should also allow creating diverse training pathways to accommodate different learning objectives and learner needs. In such cases, modular development of learning resources could promote long-term sustainability by enabling content reuse and reducing rework. For instance, standardized VR training components or reusable interactive templates could simplify content creation while maintaining adaptability \cite{ipsita2024authoring}.

The development of VR applications typically requires creating resources from scratch. To reduce time and effort during the development process, open-source content repositories should be encouraged to promote easy sharing and collaboration among developers. Similar to platforms like GrabCAD that allow the sharing of open-source CAD models, online repositories could support the exchange of animations and physics-based models for manufacturing tools and interactions that can be seamlessly integrated into the development of manufacturing use cases \cite{wang2022constructing}. Furthermore, complex manufacturing processes are quite detailed and physics-based in nature; thus, replicating realistic processes requires substantial time and effort. In terms of technology to support content creation, AI-based methods can play an important role \cite{kamal2024generative}. Automated content creation using reinforcement learning to generate virtual assets, interactions, and environments can reduce the time and effort required to create content \cite{ ashour2021connected}. Large language models and visual language models can assist in understanding users to interactively develop applications with human-in-the-loop \cite{alkhayat2024leveraging}.

As indicated by prior literature, most VR-based learning content is designed with cognitive and behavioristic pedagogical aims \cite{marougkas2023virtual, pavlov1927conditioned, anderson1995cognitive, bushra2018comparative}. In general, most novice educational designers typically lean toward these approaches for framing teaching and learning in this space. This is primarily because these theories are the simplest to design for and likely resonate with the designer's own educational experiences. Simplistic examples include multiple-choice tests \cite{ipsita2022towards} and flashcards, but more complicated solutions in VR might involve haptic feedback when a user strays past a boundary, for instance. This could also include models where the machine is trying to mimic human intelligence through workflows or paths in thinking. By contrast, less work has been done to incorporate what we know about the science of learning and the creation of socially and culturally impactful learning experiences with VR. Future work should explore more sociocultural \cite{vygotsky1978mind, rogoff1990apprenticeship}, constructionist \cite{papert2020mindstorms, piaget1973understand}, and embodied ways of learning \cite{gallagher2006body}, with a specific focus on how these approaches could be effectively implemented in VR.

Many times, VR applications face interoperability issues with VR hardware, software, and real machines, which means the content designed for one system can not be used in another without major modifications. In such cases, ontology architecture can support standardization efforts across the development, collaboration, and support of VR content for various manufacturing use cases. Standardized libraries and SDKs can also facilitate easier updates of applications and promote the effective reuse of VR assets. Finally, Industry 4.0 technologies can enhance the applicability of content across various machines and workplace environments \cite{yun2022immersive}. As a result, this can support interoperability between VR-based systems, humans, and manufacturing processes.

In terms of supporting content accessibility to learners at various levels and age groups, MOOC-based platforms could be supported through collaboration between academia and industry leaders \cite{wang2022constructing}. This can spread awareness about emerging skills in manufacturing and provide remote and accessible training at low cost.


\textbf{Hardware:} In terms of concerns related to hardware, several generic concerns affect VR adoption in manufacturing education such as cumbersome setup, bulky hardware, and interoperability issues. Complex systems for manufacturing training can be cumbersome to set up and require regular maintenance. CAVE systems, in particular, are more difficult to set up compared to HMD systems \cite{ismail2024immersive}. Frequent hardware upgrades can also affect the working of the application and thus impact teaching and learning. Interoperability issues may render the same application unusable across different devices. Although the weight of HMDs has reduced over time, they are still bulky and can be difficult for the user to wear for longer time periods. To address such barriers, organizations can invest in cloud platforms supported by edge computing and 5G to allow for seamless real-time VR experiences \cite{hazarika2023towards}. Efforts can be made to adopt cross-compatible VR hardware to future-proof adoption. Hardware companies that manufacture devices should try to allow integration of such efforts. Rather than making device-specific training applications, device-agnostic content should be developed. This could be achieved through partnerships between hardware companies and software development efforts to produce device-agnostic SDKs and APIs to support such integration.

In manufacturing education e.g., in welding and riveting, real-time sensory feedback is key to learning as indicated by prior literature. In such cases, a one-size-fits-all controller impacts skill learning, as such controllers may not provide the haptic feedback essential for building muscle memory \cite{ipsita2022towards}. To be well aligned with manufacturing needs, VR training requires strong haptic interfaces \cite{zhang2023active, wang2020design}. These can enable students and learners to benefit from augmented feedback and/or virtual training to develop the right touch for the work. In such cases, rather than relying solely on default controllers and trackers, process-specific hardware for manufacturing should be prioritized. This could include creating wearables and hand tools for haptic feedback that provide sensations similar to real tools. VR studios with trackers can be designed in the form of labs to provide realistic sensations similar to those found in real manufacturing workplace environments. The creation of haptic interfaces and environments would require collective efforts in bringing together teams with expertise in kinesiology along with industry experts in modeling human touch in order to reduce costs of training (especially as a large percentage of the workforce seeks to retire and take with them critical knowledge of the industry).

\textbf{Collaboration:} Firstly, the collaboration required for creating effective learning content is hard to achieve due to limited guidelines in this area. The collaboration requires communication between an interdisciplinary team ranging across the fields of manufacturing, learning sciences, organizational psychology, and computing. Finding experts and investing in resources to support such collaboration is costly and takes time. Possible ways to facilitate such collaboration can include relying on frameworks for collaborative design \cite{wang2022constructing} and adopting established principles from backward design \cite{wiggins2005understanding, ipsita2022towards}, multimedia learning, gamified learning theories, and adult education theories to target goal-oriented and learner-centered development of the learning content. Once such a team gets formed, it can be difficult to identify and communicate requirements due to limited prototyping tools for VR applications. Not only is the co-design affected, but limited understanding also poses a risk to the relevance of content developed for learning. Therefore, there should be more research efforts to develop tools for rapid prototyping and ideation \cite{nebeling2019360proto, krauss2022elements}. This would help in gaining better clarity on customer requirements as well as making communication clear at each stage of prototyping.

Second, limited collaboration between education and industry affects VR adoption due to: (1) limiting exposure of academia to real-world problems, and (2) delays in reaching industries about the potential learning outcomes using VR. In such cases, efforts should facilitate VR consortiums and industry-academia partnerships \cite{ismail2024immersive}. Such collaboration can lead increased accessibility to industry resources such as digital twins that could be integrated with VR developments in academia for industry-standard simulations, providing authentic contexts to students to apply learned skills.

Furthermore, the efforts to promote VR adoption can also benefit from other collaborative efforts. For example, industries focusing on similar manufacturing skills or connected in the manufacturing processes can co-develop unified training curricula for VR in manufacturing. Hardware companies can organize frequent workshops to update stakeholders on new developments so that new advancements in the technology can be made applicable in real-world use cases.

In the context of learning, VR in manufacturing education should also prioritize collaborative knowledge construction \cite{dillenbourg2006sharing}. For example, VR experiences can be designed that allow multiple learners to interact simultaneously, mirroring real manufacturing environments where knowledge is socially constructed. Such experiences can enable learners to demonstrate and explain processes to each other within the virtual space. This can also facilitate creation of scenarios requiring group problem-solving and shared decision-making \cite{kim2024behavioral}. Being able to model the social aspects in the virtual environment can also help connect workers across longer distances and aid in institutional memory. A few possibilities in this space could be putting virtual post-it notes on things so that learning and meta-cognition could be shared between employees. In this way, learning can be social and fun \cite{rumsey2020manufacturing}. Small observations, such as something working well or not, can be shared across shifts or days to improve social enjoyment and motivation but also allows for learning over time. This could also allow for distributed cognition across a company to inform tough problems like cybersecurity in manufacturing environments that can have solutions distributed over time and over larger spaces \cite{ismail2024immersive}.

\begin{sidewaystable*}
\centering
\includegraphics[origin=c,width=\textwidth]{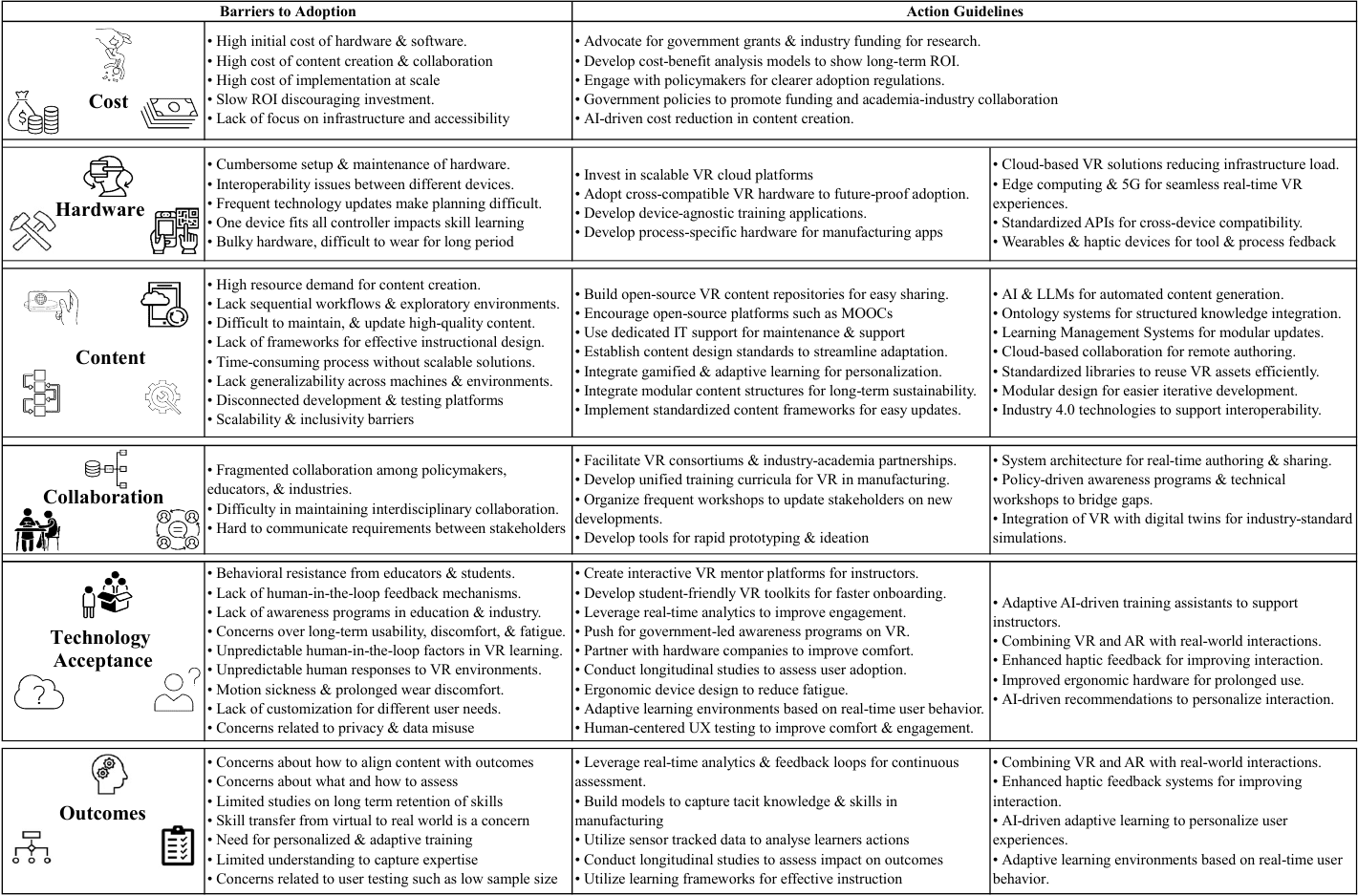}
\caption{Barriers to adoption of VR in manufacturing education and actionable insights to address them}
\label{tab:barriers_to_VR}
\end{sidewaystable*}

\textbf{Technology acceptance:} This barrier in technology acceptance likely stems from two major factors: (1) Behavioral resistance, (2) User experience and engagement. 

In case of behavioral resistance, the major issue lies in the skepticism around the use of VR for skill training outcomes in manufacturing education. This can be possibly due to the limited reach and awareness among subject matter experts. To increase awareness about the technology, government-led awareness programs could be organized on VR. In other cases, even after the technology is known and its potential is understood, behavioral resistance to accepting these technologies in education can exist. In higher education, faculty have a way of teaching particular subjects over time and it may be easier to keep the courses and training as it has existed. Learning management systems and workflows to adopt VR seamlessly in curricula can be useful. In some extreme cases, people consider these new technologies as disruptive factors that could potentially lead to job losses or replace in-person instructors. As a solution, efforts should highlight the benefits of integrated training where VR can complement rather than replace in-person training as indicated by prior literature \cite{lassiter2023welding, stone2013full}.

From the user experience point of view, concerns arise due to unpredictability in training systems. Unpredictability outcomes can leave users clueless and affect motivation towards teaching and learning using the content if something abnormal happens during their interaction. This can happen with learners and instructors who are novices in using VR, as well as during interaction with new content. To enhance accessibility of such training for educators and learners, efforts can be made to create interactive VR mentor platforms for instructors, as well as develop student-friendly VR toolkits for faster onboarding. AI-driven and Large Language Model (LLM)-based training assistants can be useful to support instructors and students during teaching, learning, assessment, and proctoring. These systems could utilize sensor-collected data to leverage real-time analytics to understand user preferences and improve engagement. 

The rigid training content in the training applications usually lacks diverse pathways to accommodate user needs and thus prevents exploratory learning. In manufacturing, people often tend to learn by making mistakes and through trial-and-error methods. However, in VR applications, the training systems can not handle and may become unpredictable due to unexpected responses from users. In such cases, Open-ended learning frameworks \cite{studer2024open} or adaptive learning environments \cite{hoover2021designing} can provide diverse learning pathways based on real-time behavior, and user expertise and experience with the content. Such recommendations can also personalize interaction and make learning more exploratory and fun.

Furthermore, learners may have varied preferences regarding the system elements and consequently the experience may differ. Human-centered UX testing and comparative studies should be conducted to assess usability and preferences for using different gamified elements in the application. Such preferences can provide developers with knowledge about elements that seem more suitable for learning purposes \cite{bushra2018comparative}. Furthermore, preferences may also exist in terms of usage of different learning platforms. Some students may prefer non-immersive learning, whereas others may prefer immersive systems. Based on user preferences, a combination of desktop-based, VR and AR with real-world interactions could be made available to students \cite{chang2024efficient}. Training applications may also collect user-related information, which could be a particularly sensitive topic and impact VR adoption. Policies at the organizational level laying out proper guidelines for data collection and use as well as using advanced technology such as blockchain in enhancing data security can be useful in this context \cite{uddin2024exploring}.

Finally, from a behavioral perspective, gamification strategies have been continually utilized to enhance motivation and engagement in VR-based manufacturing training. These strategies tend to incorporate small nudges to keep learners motivated and enable progress tracking over time \cite{hoover2021designing}. Such motivation-driven design informed by well-established principles in behavioral psychology provides valuable insights into habit formation, incentivizing desired behaviors, and offering formative feedback. While such strategies can effectively encourage participation and build social connections \cite{chase2001want}, it alone may not be sufficient for long-term engagement. Deeper and more sustained motivation in VR training will likely stem from collaborative and collective learning experiences, as well as opportunities for creative problem-solving and innovation \cite{dillenbourg2006sharing, yamada2017educational}. The ability to engage in social interactions, collaborate on tasks, and contribute to the design and development of new solutions will be key drivers of motivation and satisfaction in these learning environments. Therefore, the development of VR in manufacturing education should tap into this field to enhance technology acceptance.


\textbf{Outcomes:} Above all, some areas still require validation to confirm the impact of VR on learning outcomes. We discuss this from both design-based (e.g., aligning content to goals) and assessment-based (e.g., defining measurable learning outcomes and skill transfer) perspectives. In particular, the following approaches are recommended: (1) mapping VR activities directly to specific manufacturing competencies in the curriculum, (2) embedded assessment mechanisms within VR to measure progress toward learning outcomes, and (3) building in reflection opportunities to discuss when things go wrong and why. 

An effective way to design outcome-oriented content is through backward design \cite{wiggins2005understanding, ipsita2022towards}, though it is often underutilized in manufacturing. At its core, the design involves first defining a small set of learning objectives (i.e., 3-5) and making the goals clear and observable. Looking at the goals, appropriate assessment methods are determined that can count as evidence of learning. From there, the learning experiences are designed to lead to the outcomes. The learning experience are then implemented with the learners, assessments are utilized, and the outcomes are measured. If the learning outcomes are not met, the issue often lies with misalignment; the assessment either does not accurately measure the learning activities (i.e., it is measuring something else) or it is not well aligned to the learning objectives and needs refinement in the next iteration.

VR opens the possibility for embedded and invisible assessments that can be visualized on a dashboard over time, providing just in time feedback. This real-time feedback is shown to be highly effective for learning, rather than no feedback or, worse, delayed feedback where students risk learning the wrong way of doing something, making relearning difficult. The principles of game design and behavioral science align well with this approach. However, embedded assessments should also be designed to be more social in nature. For example, they could remind learners to take breaks when needed, support visualization of collaborative contributions to the problem to support recognition of hard work and reflection over time, etc. While there is lot of potential in using assessment to improve the social and cultural aspects, there is limited prior research in this area.


Determining what to assess (and what counts as learning) is dependent on the guiding theory that drives learning towards set competencies. For example in welding training, from a cognitive perspective, assessment could focus on performing the correct type of weld. From a social perspective, the assessment might focus on correctly diagnosing the problem by having a conversation and deciding which weld to use based on the information provided. Other examples could assess language used in the workforce and finding ways to create a more supportive environment. Currently, most assessments focus on low-hanging fruit. For example, ensuring students can weld without any breaks or burns. More complex problem solving skills that might lead to higher quality and bigger picture solutions over time often go unmeasured. Designing for small problems can also result in needing to design for smaller problems which drives up costs, rather than bigger consistent challenges that might be more cross-cutting across manufacturing contexts and workforce needs. In manufacturing, VR can be used to design new manufacturing solutions so that the employee is brought into the loop of the design \cite{rumsey2020manufacturing}, rather than learning how to use preexisting machinery efficiently. This would be a solution that small and medium-sized enterprises would resonate with, as they are in greater need for innovation \cite{yamada2017educational}. An analogy can be drawn from the use of VR for Broadway set design so the actors can perform in the set prior to its being built, as well as the set can be built reciprocally around the needs of the actors and the script \cite{shadowcast2020}. A challenge is that VR technology has generally been designed for 1:1 rather than 1:many experiences. Therefore, to achieve these pedagogical goals, technological advancements are needed to support multi-user engagement and collaboration in VR-based manufacturing environments \cite{parsley2024enhancing}.



\label{sec:8}
\section{Conclusion}
VR has been explored in a number of manufacturing areas and has been shown to have the potential to improve the skills and abilities of learners. Its ability to provide interactive and immersive experiences for complex manufacturing processes, motivate students to learn, and facilitate collaboration at co-located and distributed locations at low cost, training time, and equipment usage demonstrates its potential for learning. Despite its potential, the technology is still in its early stage of adoption. This review aims to identify the barriers surrounding the use of VR in manufacturing education by performing a holistic review of the topic. By developing a taxonomy and conducting a systematic review and analysis of 108 articles, the trends, gaps, and opportunities of VR in the field of manufacturing education are identified. Although the review confirms the benefits of the technology in education, several concerns arise that prevent its widespread adoption. Action guidelines to address such barriers are discussed. Such efforts can prepare education and training to utilize VR to rapidly equip learners with the skills required to meet the growing demands of a skilled workforce in the U.S. manufacturing industry.

\section{Acknowledgements}
We thank the reviewers for their invaluable feedback. Any opinions, findings, and conclusions or recommendations expressed in this material are those of the authors and do not necessarily reflect the views of the funding agency. We acknowledge Asim Unmesh for his support on the review process. We would like to thank Rahul Jain, Jingyu Shi, Dizhi Ma, Xiyun Hu, Chenfei Zhu and Shao-Kang Hsia for their thoughtful suggestions on the review process.






 \bibliographystyle{elsarticle-num} 
 \bibliography{references}

\begin{thebibliography}{100}
\expandafter\ifx\csname url\endcsname\relax
  \def\url#1{\texttt{#1}}\fi
\expandafter\ifx\csname urlprefix\endcsname\relax\def\urlprefix{URL }\fi
\expandafter\ifx\csname href\endcsname\relax
  \def\href#1#2{#2} \def\path#1{#1}\fi

\bibitem{todd2001manufacturing}
R.~H. Todd, W.~E. Red, S.~P. Magleby, S.~Coe, Manufacturing: A strategic opportunity for engineering education, Journal of Engineering Education 90~(3) (2001) 397--405.

\bibitem{zhao2019developing}
R.~Zhao, F.~Aqlan, L.~J. Elliott, H.~C. Lum, Developing a virtual reality game for manufacturing education, in: Proceedings of the 14th International Conference on the Foundations of Digital Games, 2019, pp. 1--4.

\bibitem{nyt_ai_aging_shift_2024}
T.~N.~Y. Times, \href{https://www.nytimes.com/2024/10/17/opinion/economy-us-aging-work-force-ai.html}{America is sleepwalking into an economic storm} (2024).
\newline\urlprefix\url{https://www.nytimes.com/2024/10/17/opinion/economy-us-aging-work-force-ai.html}

\bibitem{economist_manufacturing_delusion_2023}
T.~Economist, \href{https://www.economist.com/finance-and-economics/2023/07/13/the-world-is-in-the-grip-of-a-manufacturing-delusion}{The world is in the grip of a manufacturing delusion} (2023).
\newline\urlprefix\url{https://www.economist.com/finance-and-economics/2023/07/13/the-world-is-in-the-grip-of-a-manufacturing-delusion}

\bibitem{nyt_trump_harris_2024}
T.~N.~Y. Times, \href{https://www.nytimes.com/2024/10/03/business/economy/manufacturing-jobs-trump-harris.html}{To revive manufacturing, how much can a president do?} (2024).
\newline\urlprefix\url{https://www.nytimes.com/2024/10/03/business/economy/manufacturing-jobs-trump-harris.html}

\bibitem{nyt_manufacturing_good_old_days_2024}
T.~N.~Y. Times, \href{https://www.nytimes.com/2024/10/10/opinion/manufacturing-harris-trump-immigration-college.html}{The good old days of manufacturing are long gone} (2024).
\newline\urlprefix\url{https://www.nytimes.com/2024/10/10/opinion/manufacturing-harris-trump-immigration-college.html}

\bibitem{economist_manufacturing_inefficiency_2023}
T.~Economist, \href{https://www.economist.com/finance-and-economics/2023/11/09/why-american-manufacturing-is-becoming-less-efficient}{Why american manufacturing is becoming less efficient} (2023).
\newline\urlprefix\url{https://www.economist.com/finance-and-economics/2023/11/09/why-american-manufacturing-is-becoming-less-efficient}

\bibitem{lowell2024applying}
V.~L. Lowell, W.~Yan, Applying systems thinking for designing immersive virtual reality learning experiences in education, TechTrends 68~(1) (2024) 149--160.

\bibitem{Lowell2024}
V.~L. Lowell, \href{https://doi.org/10.59668/2033.19042}{Extended reality (xr) for authentic learning: New frontiers in educational technology}, The Journal of Applied Instructional Design 13~(4) (December 2024).
\newblock \href {https://doi.org/10.59668/2033.19042} {\path{doi:10.59668/2033.19042}}.
\newline\urlprefix\url{https://doi.org/10.59668/2033.19042}

\bibitem{naranjo2020scoping}
J.~E. Naranjo, D.~G. Sanchez, A.~Robalino-Lopez, P.~Robalino-Lopez, A.~Alarcon-Ortiz, M.~V. Garcia, A scoping review on virtual reality-based industrial training, Applied Sciences 10~(22) (2020) 8224.

\bibitem{guo2020applications}
Z.~Guo, D.~Zhou, Q.~Zhou, X.~Zhang, J.~Geng, S.~Zeng, C.~Lv, A.~Hao, Applications of virtual reality in maintenance during the industrial product lifecycle: A systematic review, Journal of Manufacturing Systems 56 (2020) 525--538.

\bibitem{leu2013cad}
M.~C. Leu, H.~A. ElMaraghy, A.~Y. Nee, S.~K. Ong, M.~Lanzetta, M.~Putz, W.~Zhu, A.~Bernard, Cad model based virtual assembly simulation, planning and training, CIRP Annals 62~(2) (2013) 799--822.

\bibitem{yang2023use}
Y.~Yang, S.~Deb, M.~He, M.~H. Kobir, The use of virtual reality in manufacturing education: State-of-the-art and future directions, Manufacturing Letters 35 (2023) 1214--1221.

\bibitem{berg2017industry}
L.~P. Berg, J.~M. Vance, Industry use of virtual reality in product design and manufacturing: a survey, Virtual reality 21 (2017) 1--17.

\bibitem{radianti2020systematic}
J.~Radianti, T.~A. Majchrzak, J.~Fromm, I.~Wohlgenannt, A systematic review of immersive virtual reality applications for higher education: Design elements, lessons learned, and research agenda, Computers \& education 147 (2020) 103778.

\bibitem{rojas2023systematic}
M.~A. Rojas-S{\'a}nchez, P.~R. Palos-S{\'a}nchez, J.~A. Folgado-Fern{\'a}ndez, Systematic literature review and bibliometric analysis on virtual reality and education, Education and Information Technologies 28~(1) (2023) 155--192.

\bibitem{price2019using}
A.~Price, M.~Kuttolamadom, S.~Obeidat, Using virtual reality welding to improve manufacturing process education, in: 2019 CIEC, 2019.

\bibitem{aqlan2019integrating}
F.~Aqlan, R.~Zhao, H.~Lum, L.~J. Elliott, Integrating simulation games and virtual reality to teach manufacturing systems concepts, in: ASEE Annual Conference proceedings, 2019.

\bibitem{chiou2024virtual}
R.~Chiou, I.~Singh, A.~K.~S. Kavitha, T.-l.~B. Tseng, M.~F. Rahman, N.~Vasudevan, Virtual reality wind turbine for learning green energy manufacturing, in: 2024 ASEE Annual Conference \& Exposition, 2024.

\bibitem{ostrander2020evaluating}
J.~K. Ostrander, C.~S. Tucker, T.~W. Simpson, N.~A. Meisel, Evaluating the use of virtual reality to teach introductory concepts of additive manufacturing, Journal of Mechanical Design 142~(5) (2020) 051702.

\bibitem{abele2017learning}
E.~Abele, G.~Chryssolouris, W.~Sihn, J.~Metternich, H.~ElMaraghy, G.~Seliger, G.~Sivard, W.~ElMaraghy, V.~Hummel, M.~Tisch, et~al., Learning factories for future oriented research and education in manufacturing, CIRP annals 66~(2) (2017) 803--826.

\bibitem{carruth2017virtual}
D.~W. Carruth, Virtual reality for education and workforce training, in: 2017 15th International Conference on Emerging eLearning Technologies and Applications (ICETA), IEEE, 2017, pp. 1--6.

\bibitem{gonzalez2017immersive}
M.~Gonzalez-Franco, R.~Pizarro, J.~Cermeron, K.~Li, J.~Thorn, W.~Hutabarat, A.~Tiwari, P.~Bermell-Garcia, Immersive mixed reality for manufacturing training, Frontiers in Robotics and AI 4 (2017) 3.

\bibitem{poyade2021transferable}
M.~Poyade, C.~Eaglesham, J.~Trench, M.~Reid, A transferable psychological evaluation of virtual reality applied to safety training in chemical manufacturing, ACS Chemical Health \& Safety 28~(1) (2021) 55--65.

\bibitem{azzam2024virtual}
I.~Azzam, K.~El~Breidi, F.~Breidi, C.~Mousas, Virtual reality in fluid power education: Impact on students’ perceived learning experience and engagement, Education Sciences 14~(7) (2024) 764.

\bibitem{el2016assessment}
H.~El-Mounayri, C.~Rogers, E.~Fernandez, J.~C. Satterwhite, Assessment of stem e-learning in an immersive virtual reality (vr) environment, American Society for Engineering Education, 2016.

\bibitem{badamasi2022drivers}
A.~A. Badamasi, K.~R. Aryal, U.~U. Makarfi, M.~Dodo, Drivers and barriers of virtual reality adoption in uk aec industry, Engineering, Construction and Architectural Management 29~(3) (2022) 1307--1318.

\bibitem{scott2020investigation}
H.~Scott, D.~Baglee, R.~O'Brien, R.~Potts, An investigation of acceptance and e-readiness for the application of virtual reality and augmented reality technologies to maintenance training in the manufacturing industry, International Journal of Mechatronics and Manufacturing Systems 13~(1) (2020) 39--58.

\bibitem{jalo2021state}
H.~Jalo, H.~Pirkkalainen, O.~Torro, The state of augmented reality, mixed reality and virtual reality adoption and use in european small and medium-sized manufacturing companies in 2020: Vam realities survey report (2021).

\bibitem{fernandez2017augmented}
M.~Fernandez, Augmented virtual reality: How to improve education systems., Higher Learning Research Communications 7~(1) (2017) 1--15.

\bibitem{liagkou2019realizing}
V.~Liagkou, D.~Salmas, C.~Stylios, Realizing virtual reality learning environment for industry 4.0, Procedia Cirp 79 (2019) 712--717.

\bibitem{doolani2020review}
S.~Doolani, C.~Wessels, V.~Kanal, C.~Sevastopoulos, A.~Jaiswal, H.~Nambiappan, F.~Makedon, A review of extended reality (xr) technologies for manufacturing training, Technologies 8~(4) (2020) 77.

\bibitem{dreyfus2022virtual}
P.-A. Dreyfus, F.~Psarommatis, G.~May, D.~Kiritsis, Virtual metrology as an approach for product quality estimation in industry 4.0: a systematic review and integrative conceptual framework, International Journal of Production Research 60~(2) (2022) 742--765.

\bibitem{cali2022opportunities}
U.~Cali, M.~Kuzlu, E.~Karaarslan, V.~Jovanovic, Opportunities and challenges in metaverse for industry 4.0 and beyond applications, in: 2022 IEEE 1st Global Emerging Technology Blockchain Forum: Blockchain \& Beyond (iGETblockchain), IEEE, 2022, pp. 1--6.

\bibitem{basu20226g}
D.~Basu, U.~Ghosh, R.~Datta, 6g for industry 5.0 and smart cps: a journey from challenging hindrance to opportunistic future, in: 2022 IEEE Silchar Subsection Conference (SILCON), IEEE, 2022, pp. 1--6.

\bibitem{awal2018ontology}
A.~Awal, A.~Mishra, G.~M. Usman, A.~AbdulG, Ontology development for the domain of software requirement elicitation technique, International Journal of Engineering Research \& Technology (IJERT) 7 (2018) 334--338.

\bibitem{elshennawy2015manufacturing}
A.~K. Elshennawy, G.~S. Weheba, Manufacturing processes \& materials, Society of Manufacturing Engineers (SME), 2015.

\bibitem{ManufacturingEducation}
S.~D.~I. Software, Manufacturing education, \url{https://www.plm.automation.siemens.com/global/en/our-story/glossary/manufacturing-education/28572}.

\bibitem{raman2014manufacturing}
S.~Raman, A.~Wadke, A.~Badiru, Manufacturing technology, Handbook of Industrial and Systems Engineering, (2014) 337--349.

\bibitem{groover2020fundamentals}
M.~P. Groover, Fundamentals of modern manufacturing: materials, processes, and systems, John Wiley \& Sons, 2020.

\bibitem{mourtzis2018cyber}
D.~Mourtzis, E.~Vlachou, G.~Dimitrakopoulos, V.~Zogopoulos, Cyber-physical systems and education 4.0--the teaching factory 4.0 concept, Procedia manufacturing 23 (2018) 129--134.

\bibitem{spak2013us}
G.~T. Spak, Us advanced manufacturing skills gap: Innovative education solutions, Procedia-Social and Behavioral Sciences 106 (2013) 3235--3245.

\bibitem{chryssolouris2013manufacturing}
G.~Chryssolouris, D.~Mavrikios, D.~Mourtzis, Manufacturing systems: skills \& competencies for the future, Procedia CIRp 7 (2013) 17--24.

\bibitem{gaspar2020research}
H.~Gaspar, L.~Morgado, H.~Mamede, T.~Oliveira, B.~Manj{\'o}n, C.~G{\"u}tl, Research priorities in immersive learning technology: the perspectives of the ilrn community, Virtual Reality 24 (2020) 319--341.
\newblock \href {https://doi.org/10.1007/s10055-019-00393-x} {\path{doi:10.1007/s10055-019-00393-x}}.

\bibitem{cassola2021novel}
F.~Cassola, M.~Pinto, D.~Mendes, L.~Morgado, A.~Coelho, H.~Paredes, A novel tool for immersive authoring of experiential learning in virtual reality, in: 2021 IEEE Conference on Virtual Reality and 3D User Interfaces Abstracts and Workshops (VRW), IEEE, 2021, pp. 44--49.
\newblock \href {https://doi.org/10.1109/VRW52623.2021.00014} {\path{doi:10.1109/VRW52623.2021.00014}}.

\bibitem{page2021prisma}
M.~J. Page, J.~E. McKenzie, P.~M. Bossuyt, I.~Boutron, T.~C. Hoffmann, C.~D. Mulrow, L.~Shamseer, J.~M. Tetzlaff, E.~A. Akl, S.~E. Brennan, et~al., The prisma 2020 statement: an updated guideline for reporting systematic reviews, bmj 372 (2021).

\bibitem{CIP2025}
{National Center for Education Statistics}, \href{https://nces.ed.gov/ipeds/cipcode/Default.aspx?y=56}{Classification of instructional programs (cip)}, accessed: 2025-02-07 (2025).
\newline\urlprefix\url{https://nces.ed.gov/ipeds/cipcode/Default.aspx?y=56}

\bibitem{ISCED2025}
{UNESCO Institute for Statistics}, \href{https://uis.unesco.org/en/topic/international-standard-classification-education-isced}{International standard classification of education (isced)}, accessed: 2025-02-07 (2025).
\newline\urlprefix\url{https://uis.unesco.org/en/topic/international-standard-classification-education-isced}

\bibitem{NAICS2025}
{U.S. Census Bureau}, \href{https://www.census.gov/naics/}{North american industry classification system (naics)}, accessed: 2025-02-07 (2025).
\newline\urlprefix\url{https://www.census.gov/naics/}

\bibitem{elmaraghy2021evolution}
H.~ElMaraghy, L.~Monostori, G.~Schuh, W.~ElMaraghy, Evolution and future of manufacturing systems, CIRP Annals 70~(2) (2021) 635--658.

\bibitem{mourtzis2018development}
D.~Mourtzis, Development of skills and competences in manufacturing towards education 4.0: A teaching factory approach, in: Proceedings of 3rd International Conference on the Industry 4.0 Model for Advanced Manufacturing: AMP 2018 3, Springer, 2018, pp. 194--210.

\bibitem{oztemel2020literature}
E.~Oztemel, S.~Gursev, Literature review of industry 4.0 and related technologies, Journal of intelligent manufacturing 31~(1) (2020) 127--182.

\bibitem{hozdic2023evolution}
E.~Hozdi{\'c}, I.~Makovec, Evolution of the human role in manufacturing systems: On the route from digitalization and cybernation to cognitization, Applied System Innovation 6~(2) (2023) 49.

\bibitem{adel2024convergence}
A.~Adel, The convergence of intelligent tutoring, robotics, and iot in smart education for the transition from industry 4.0 to 5.0, Smart Cities 7~(1) (2024) 325--369.

\bibitem{li2024reskilling}
L.~Li, Reskilling and upskilling the future-ready workforce for industry 4.0 and beyond, Information Systems Frontiers 26~(5) (2024) 1697--1712.

\bibitem{mavrikios2013industrial}
D.~Mavrikios, N.~Papakostas, D.~Mourtzis, G.~Chryssolouris, On industrial learning and training for the factories of the future: a conceptual, cognitive and technology framework, Journal of Intelligent Manufacturing 24 (2013) 473--485.

\bibitem{studer2024open}
K.~Studer, H.~Lie, Z.~Zhao, B.~Thomson, D.~G. Turakhia, J.~Liu, An open-ended system in virtual reality for training machining skills, in: Extended Abstracts of the CHI Conference on Human Factors in Computing Systems, 2024, pp. 1--5.

\bibitem{lie2023training}
H.~Lie, K.~Studer, Z.~Zhao, B.~Thomson, D.~G. Turakhia, J.~Liu, Training for open-ended drilling through a virtual reality simulation, in: 2023 IEEE International Symposium on Mixed and Augmented Reality (ISMAR), IEEE, 2023, pp. 366--375.

\bibitem{park2023work}
J.~Park, R.~Islam, C.~A. King, L.~Jiang, X.~Peng, B.~Yalvac, Work in progress: Virtual reality for manufacturing equipment training for future workforce development, in: 2023 ASEE Annual Conference \& Exposition, 2023.

\bibitem{moreland2020integrating}
J.~Moreland, J.~Estrada, E.~Mosquera, K.~Toth, A.~K. Silaen, C.~Q. Zhou, Integrating fluid simulation with virtual die casting machine for industry 4.0 and operator training, in: Light Metals 2020, Springer, 2020, pp. 1026--1031.

\bibitem{moreland2022development}
J.~Moreland, K.~Toth, J.~Estrada, J.~Chen, N.~Zhu, C.~Zhou, Development of virtual die casting simulator for workforce development, in: REWAS 2022: Developing Tomorrow’s Technical Cycles (Volume I), Springer, 2022, pp. 515--521.

\bibitem{white2011low}
S.~A. White, M.~Prachyabrued, T.~L. Chambers, C.~W. Borst, D.~Reiners, Low-cost simulated mig welding for advancement in technical training, Virtual reality 15 (2011) 69--81.

\bibitem{ipsita2022towards}
A.~Ipsita, L.~Erickson, Y.~Dong, J.~Huang, A.~K. Bushinski, S.~Saradhi, A.~M. Villanueva, K.~A. Peppler, T.~S. Redick, K.~Ramani, Towards modeling of virtual reality welding simulators to promote accessible and scalable training, in: Proceedings of the 2022 CHI conference on human factors in computing systems, 2022, pp. 1--21.

\bibitem{mclaurin2012comparison}
E.~J. McLaurin, R.~T. Stone, Comparison of virtual reality training vs. integrated training in the development of physical skills, in: Proceedings of the Human Factors and Ergonomics Society Annual Meeting, Vol.~56, SAGE Publications Sage CA: Los Angeles, CA, 2012, pp. 2532--2536.

\bibitem{stone2013full}
R.~Stone, E.~McLaurin, P.~Zhong, K.~Watts, Full virtual reality vs. integrated virtual reality training in welding (2013).

\bibitem{stone2011physical}
R.~T. Stone, K.~P. Watts, P.~Zhong, C.-S. Wei, Physical and cognitive effects of virtual reality integrated training, Human factors 53~(5) (2011) 558--572.

\bibitem{stone2011virtual}
R.~Stone, K.~Watts, P.~Zhong, Virtual reality integrated welder training (2011).

\bibitem{byrd2015use}
A.~Byrd, R.~Stone, R.~Anderson, The use of virtual welding simulators to evaluate experienced welders (2015).

\bibitem{lassiter2023welding}
T.~Lassiter, C.~Collier, K.~R. Fleischmann, S.~R. Greenberg, Welding instructors' perspectives on using ai technology in welding training, Proceedings of the Association for Information Science and Technology 60~(1) (2023) 233--243.

\bibitem{ipsita2024design}
A.~Ipsita, R.~Duan, H.~Li, S.~Chidambaram, Y.~Cao, M.~Liu, A.~Quinn, K.~Ramani, The design of a virtual prototyping system for authoring interactive virtual reality environments from real-world scans, Journal of Computing and Information Science in Engineering 24~(3) (2024).

\bibitem{ipsita2021vrfromx}
A.~Ipsita, H.~Li, R.~Duan, Y.~Cao, S.~Chidambaram, M.~Liu, K.~Ramani, Vrfromx: from scanned reality to interactive virtual experience with human-in-the-loop, in: Extended Abstracts of the 2021 CHI Conference on Human Factors in Computing Systems, 2021, pp. 1--7.

\bibitem{ye2023robot}
Y.~Ye, T.~Zhou, J.~Du, Robot-assisted immersive kinematic experience transfer for welding training, Journal of Computing in Civil Engineering 37~(2) (2023) 04023002.

\bibitem{ye2024user}
Y.~Ye, T.~Uthai, P.~Xia, T.~Zhou, J.~Du, User experience and workload evaluation in robot-assisted virtual reality welding training, in: Construction Research Congress 2024, pp. 99--108.

\bibitem{belga2022carousel}
J.~Belga, T.~D. Do, R.~Ghamandi, R.~P. McMahan, J.~J. LaViola, Carousel: Improving the accuracy of virtual reality assessments for inspection training tasks, in: Proceedings of the 28th ACM Symposium on Virtual Reality Software and Technology, 2022, pp. 1--10.

\bibitem{srinivasa2021virtual}
A.~R. Srinivasa, R.~Jha, T.~Ozkan, Z.~Wang, Virtual reality and its role in improving student knowledge, self-efficacy, and attitude in the materials testing laboratory, International Journal of Mechanical Engineering Education 49~(4) (2021) 382--409.

\bibitem{bowling2010evaluating}
S.~R. Bowling, Evaluating the effectiveness of a priori information on process measures in a virtual reality inspection task, Journal of Industrial Engineering and Management 3~(1) (2010) 221--248.

\bibitem{dodoo2018evaluating}
E.~R. Dodoo, B.~Hill, A.~Garcia, A.~Kohl, A.~MacAllister, J.~Schlueter, E.~Winer, Evaluating commodity hardware and software for virtual reality assembly training, Electronic Imaging 30 (2018) 1--6.

\bibitem{adas2013virtual}
H.~A. Adas, S.~Shetty, S.~K. Hargrove, Virtual and augmented reality based assembly design system for personalized learning, in: 2013 Science and Information Conference, IEEE, 2013, pp. 696--702.

\bibitem{cecil2013virtual}
J.~Cecil, P.~Ramanathan, M.~Mwavita, Virtual learning environments in engineering and stem education, in: 2013 IEEE Frontiers in Education Conference (FIE), IEEE, 2013, pp. 502--507.

\bibitem{fitton2024watch}
I.~S. Fitton, E.~Dark, M.~M. Oliveira~da Silva, J.~Dalton, M.~J. Proulx, C.~Clarke, C.~Lutteroth, Watch this! observational learning in vr promotes better far transfer than active learning for a fine psychomotor task, in: Proceedings of the CHI Conference on Human Factors in Computing Systems, 2024, pp. 1--19.

\bibitem{aqlan2020multiplayer}
F.~Aqlan, Multiplayer physical and virtual reality games for team-based manufacturing simulation, in: ASEE Annual Conference proceedings, 2020.

\bibitem{zhu2021eye}
R.~Zhu, F.~Aqlan, R.~Zhao, H.~Yang, Eye-track modeling of problem-solving in virtual manufacturing environments, in: ASEE annual conference proceedings, 2021.

\bibitem{zhu2022sensor}
R.~Zhu, F.~Aqlan, R.~Zhao, H.~Yang, Sensor-based modeling of problem-solving in virtual reality manufacturing systems, Expert Systems with Applications 201 (2022) 117220.

\bibitem{hartleb2023exploring}
T.~Hartleb, H.~Kim, R.~Zhao, F.~Aqlan, H.~Yang, Exploring magic interactions for collaboration in virtual reality learning factory, in: ASEE Annual Conference and Exposition, Conference Proceedings, 2023.

\bibitem{kim2024behavioral}
H.~Kim, T.~Hartleb, K.~Bello, F.~Aqlan, R.~Zhao, H.~Yang, Behavioral modeling of collaborative problem solving in multiplayer virtual reality manufacturing simulation games, Journal of Computing and Information Science in Engineering 24~(3) (2024).

\bibitem{dwivedi2018manual}
P.~Dwivedi, D.~Cline, C.~Joe, R.~Etemadpour, Manual assembly training in virtual environments, in: 2018 IEEE 18th International Conference on Advanced Learning Technologies (ICALT), IEEE, 2018, pp. 395--399.

\bibitem{etemadpour2019visualization}
R.~Etemadpour, O.~Kozlenko, P.~Dwivedi, A visualization tool for analyzing an assembly training in vr, in: 2019 IEEE 19th International Conference on Advanced Learning Technologies (ICALT), Vol. 2161, IEEE, 2019, pp. 150--152.

\bibitem{de2019effects}
D.~Y. de~Moura, A.~Sadagic, The effects of stereopsis and immersion on bimanual assembly tasks in a virtual reality system, in: 2019 IEEE conference on virtual reality and 3D user interfaces (VR), IEEE, 2019, pp. 286--294.

\bibitem{ma2020approach}
W.~Ma, D.~Kaber, M.~Zahabi, An approach to human motor skill training for uniform group performance, International Journal of Industrial Ergonomics 75 (2020) 102894.

\bibitem{chang2024efficient}
E.~Chang, Y.~Lee, M.~Billinghurst, B.~Yoo, Efficient vr-ar communication method using virtual replicas in xr remote collaboration, International Journal of Human-Computer Studies (2024) 103304.

\bibitem{al2016development}
A.~M. Al-Ahmari, M.~H. Abidi, A.~Ahmad, S.~Darmoul, Development of a virtual manufacturing assembly simulation system, Advances in Mechanical Engineering 8~(3) (2016) 1687814016639824.

\bibitem{sharma2019collaborative}
S.~Sharma, S.-T. Bodempudi, M.~Arrolla, A.~Upadhyay, Collaborative virtual assembly environment for product design, in: 2019 International Conference on Computational Science and Computational Intelligence (CSCI), IEEE, 2019, pp. 606--611.

\bibitem{kim2010itrain}
O.~Kim, U.~Jayaram, S.~Jayaram, L.~Zhu, Itrain: Ontology-based integration of information and methods in computer-based training (cbt) and immersive training (imt) for assembly simulations, in: International Design Engineering Technical Conferences and Computers and Information in Engineering Conference, Vol. 44113, 2010, pp. 1299--1308.

\bibitem{moreland2021development}
J.~Moreland, N.~Zhu, M.~Changoluisa, L.~Raygadas-Lara, G.~Page, Y.~Krotov, C.~Zhou, Development of armss: Augmented reality maintenance and safety system, in: Proceedings of AISTech 2021 Iron \& Steel Technology Conference, 2021.

\bibitem{hannah2024results}
R.~Hannah, Results of integrating short vr exercises into traditional cbts, Research in Learning Technology 32 (2024).

\bibitem{bushra2018comparative}
N.~Bushra, D.~Carruth, S.~Deb, A comparative study of virtual ui for risk assessment and evaluation, in: Advances in Visual Computing: 13th International Symposium, ISVC 2018, Las Vegas, NV, USA, November 19--21, 2018, Proceedings 13, Springer, 2018, pp. 226--236.

\bibitem{ISLAM2024103648}
M.~S. Islam, S.~J.~N. Zahabi, S.~Kim, N.~Lau, M.~A. Nussbaum, S.~Lim, \href{https://www.sciencedirect.com/science/article/pii/S0169814124001045}{Changes in forklift driving performance and postures among novices resulting from training using a high-fidelity virtual reality simulator: An exploratory study}, International Journal of Industrial Ergonomics 104 (2024) 103648.
\newblock \href {https://doi.org/https://doi.org/10.1016/j.ergon.2024.103648} {\path{doi:https://doi.org/10.1016/j.ergon.2024.103648}}.
\newline\urlprefix\url{https://www.sciencedirect.com/science/article/pii/S0169814124001045}

\bibitem{tang2024evaluation}
M.~Tang, M.~Nikolaenko, E.~Boerwinkle, S.~Obafisoye, A.~Kumar, M.~Rezayat, S.~Lehmann, T.~Lorenz, Evaluation of the effectiveness of traditional training vs. immersive training: A case study of building safety and emergency training, in: International Conference on Extended Reality, Springer, 2024, pp. 99--109.

\bibitem{gupta2019viis}
S.~Gupta, L.~Owens, K.~Tsiakas, F.~Makedon, viis: a vocational interactive immersive storytelling framework for skill training and performance assessment, in: Proceedings of the 12th ACM International Conference on PErvasive Technologies Related to Assistive Environments, 2019, pp. 411--415.

\bibitem{doolani2020vis}
S.~Doolani, L.~Owens, C.~Wessels, F.~Makedon, vis: An immersive virtual storytelling system for vocational training, Applied sciences 10~(22) (2020) 8143.

\bibitem{azzam2023virtual}
I.~Azzam, F.~Breidi, P.~Soudah, Virtual reality: A learning tool for promoting learners’ engagement in engineering technology, ASEE Conferences, 2023.

\bibitem{mccusker2018virtual}
J.~R. McCusker, M.~A. Almaghrabi, B.~Kucharski, Is a virtual reality-based laboratory experience a viable alternative to the real thing?, in: 2018 ASEE Annual Conference \& Exposition, 2018.

\bibitem{zhou2016comprehensive}
C.~Zhou, G.~Tang, J.~Wang, D.~Fu, T.~Okosun, A.~Silaen, B.~Wu, Comprehensive numerical modeling of the blast furnace ironmaking process, JOM 68 (2016) 1353--1362.

\bibitem{chen2010virtual}
G.~Chen, J.~Moreland, D.~Ratko, L.~Jin, H.~Shen, B.~Wu, C.~Q. Zhou, Virtual reality development for engineering applications, in: ASME World Conference on Innovative Virtual Reality, Vol. 49088, 2010, pp. 127--135.

\bibitem{jalilvand2024vr}
I.~Jalilvand, J.~Jang, B.~Gopaluni, A.~S. Milani, Vr/mr systems integrated with heat transfer simulation for training of thermoforming: A multicriteria decision-making user study, Journal of Manufacturing Systems 72 (2024) 338--359.

\bibitem{mathur2022identifying}
J.~Mathur, S.~R. Miller, T.~W. Simpson, N.~A. Meisel, Identifying the effects of immersion on design for additive manufacturing evaluation of designs of varying manufacturability, in: International Design Engineering Technical Conferences and Computers and Information in Engineering Conference, Vol. 86250, American Society of Mechanical Engineers, 2022, p. V005T05A003.

\bibitem{mathur2023designing}
J.~Mathur, S.~R. Miller, T.~W. Simpson, N.~A. Meisel, Designing immersive experiences in virtual reality for design for additive manufacturing training, Additive Manufacturing 78 (2023) 103875.

\bibitem{mathur2024effects}
J.~Mathur, S.~R. Miller, T.~W. Simpson, N.~A. Meisel, Effects of immersion on knowledge gain and cognitive load in additive manufacturing process education, 3D Printing and Additive Manufacturing 11~(2) (2024) e787--e800.

\bibitem{mathur2024using}
J.~Mathur, S.~R. Miller, T.~W. Simpson, N.~A. Meisel, Using virtual reality to orient parts for additive manufacturing and its effects on manufacturability and experiential outcomes, Additive Manufacturing 94 (2024) 104421.

\bibitem{aryal2024imvr}
M.~R. Aryal, S.~Deshpande, J.~Aurisano, S.~Anand, Imvr: Enabling immersive design exploration and process integration for additive manufacturing of complex organic geometries, in: International Manufacturing Science and Engineering Conference, Vol. 88100, American Society of Mechanical Engineers, 2024, p. V001T01A001.

\bibitem{mogessie2020work}
M.~Mogessie, S.~D. Wolf, M.~Barbosa, N.~Jones, B.~M. McLaren, Work-in-progress—a generalizable virtual reality training and intelligent tutor for additive manufacturing, in: 2020 6th international conference of the immersive learning research network (iLRN), IEEE, 2020, pp. 355--358.

\bibitem{rafa2024enhancing}
R.~R. Rafa, T.~Rahman, M.~H. Kobir, Y.~Yang, S.~Deb, Enhancing experiential learning through virtual reality: System design and a case study in additive manufacturing, Human Factors and Ergonomics in Manufacturing \& Service Industries 34~(6) (2024) 649--666.

\bibitem{novoa2022new}
M.~Novoa, B.~Howell, J.~W. Hoftijzer, J.~M. Rodriguez, W.~Zhang, N.~Kramer, et~al., New collaborative workflows-immersive co-design from sketching to 3d cad and production, in: DS 117: Proceedings of the 24th International Conference on Engineering and Product Design Education (E\&PDE 2022), London South Bank University in London, UK. 8th-9th September 2022, 2022.

\bibitem{conesa2023influence}
J.~Conesa, F.~J. Mula, K.~A. Bartlett, F.~Naya, M.~Contero, The influence of immersive and collaborative virtual environments in improving spatial skills, Applied Sciences 13~(14) (2023) 8426.

\bibitem{kobir2023human}
M.~H. Kobir, T.~Rahman, Y.~Yang, S.~Deb, A human factors approach to improve layout design for a virtual reality-based training platform, in: Proceedings of the Human Factors and Ergonomics Society Annual Meeting, Vol.~67, SAGE Publications Sage CA: Los Angeles, CA, 2023, pp. 1439--1444.

\bibitem{ismail2024immersive}
Y.~Ismail, O.~P. Ojajuni, B.~Warren, F.~Dawan, A.~Lawson, Immersive engineering learning and workforce development: Pushing the boundaries of knowledge acquisition in a cave, Immersive Engineering Learning and Workforce Development: Pushing the Boundaries of Knowledge Acquisition in a CAVE (2024).

\bibitem{ertekin2023board}
Y.~Ertekin, R.~Chiou, Board 126: Work in progress: Incorporating virtual programming concepts in an advanced robotics course for machining processing and quality inspection of cnc machines and industrial robots, in: 2023 ASEE Annual Conference \& Exposition, 2023.

\bibitem{tram2023intuitive}
M.~Q. Tram, J.~M. Cloud, W.~J. Beksi, Intuitive robot integration via virtual reality workspaces, in: 2023 IEEE International Conference on Robotics and Automation (ICRA), IEEE, 2023, pp. 11654--11660.

\bibitem{kuts2022digital}
V.~Kuts, J.~A. Marvel, M.~Aksu, S.~L. Pizzagalli, M.~Sarkans, Y.~Bondarenko, T.~Otto, Digital twin as industrial robots manipulation validation tool, Robotics 11~(5) (2022) 113.

\bibitem{shi2020affordable}
Z.~Shi, C.~L. McGhan, Affordable virtual reality setup for educational aerospace robotics simulation and testing, Journal of Aerospace Information Systems 17~(1) (2020) 66--69.

\bibitem{robotlearning2018}
Y.-h. Chang, K.~Devine, A tale of the robot: Will virtual reality enhance student learning of industrial robotics?, 2018.
\newblock \href {https://doi.org/10.18260/1-2--30118} {\path{doi:10.18260/1-2--30118}}.

\bibitem{chang2020exploring}
Y.-h.~I. Chang, K.~L. Devine, G.~K. Klitzing, Exploring the vr-based pbd programming approach to teach industrial robotics in manufacturing education, in: 2020 ASEE Virtual Annual Conference Content Access, 2020.

\bibitem{chang2021using}
Y.-h. Chang, K.~Devine, G.~Klitzing, Using uirtual reality for industrial robot programming: A preliminary study., Journal of Engineering Technology 38~(1) (2021).

\bibitem{theofanidis2017varm}
M.~Theofanidis, S.~I. Sayed, A.~Lioulemes, F.~Makedon, Varm: Using virtual reality to program robotic manipulators, in: Proceedings of the 10th International Conference on PErvasive Technologies Related to Assistive Environments, 2017, pp. 215--221.

\bibitem{darmoul2015virtual}
S.~Darmoul, M.~H. Abidi, A.~Ahmad, A.~M. Al-Ahmari, S.~M. Darwish, H.~M. Hussein, Virtual reality for manufacturing: A robotic cell case study, in: 2015 International Conference on Industrial Engineering and Operations Management (IEOM), IEEE, 2015, pp. 1--7.

\bibitem{mitchell2020safety}
D.~Mitchell, H.~Choi, J.~M. Haney, Safety perception and behaviors during human-robot interaction in virtual environments, in: Proceedings of the Human Factors and Ergonomics Society Annual Meeting, Vol.~64, SAGE Publications Sage CA: Los Angeles, CA, 2020, pp. 2087--2091.

\bibitem{srinivasan2021biomechanical}
M.~Srinivasan, S.~T. Mubarrat, Q.~Humphrey, T.~Chen, K.~Binkley, S.~K. Chowdhury, The biomechanical evaluation of a human-robot collaborative task in a physically interactive virtual reality simulation testbed, in: Proceedings of the Human Factors and Ergonomics Society Annual Meeting, Vol.~65, SAGE Publications Sage CA: Los Angeles, CA, 2021, pp. 403--407.

\bibitem{lor2024enabling}
M.~A. Lor, S.-C. Chen, M.-L. Shyu, Y.~Tao, S.~Vassigh, Enabling intelligent immersive learning using deep learning-based learner confidence estimation, in: 2024 IEEE International Conference on Information Reuse and Integration for Data Science (IRI), IEEE, 2024, pp. 55--60.

\bibitem{ryan2022immersive}
M.~Ryan, Y.~Wang, Q.~Xiao, R.~Liu, Y.~Zhang, Immersive virtual reality training with error management for cnc milling set-up, in: International Manufacturing Science and Engineering Conference, Vol. 85819, American Society of Mechanical Engineers, 2022, p. V002T06A027.

\bibitem{rogers2018assessment}
C.~B. Rogers, H.~El-Mounayri, T.~Wasfy, J.~Satterwhite, Assessment of stem e-learning in an immersive virtual reality (vr) environment, ASEE, 2018.

\bibitem{chiou2019virtual}
R.~Y. Chiou, M.~G. Mauk, I.~C.~Husanu, T.-L. Tseng, S.~Sowmithran, Virtual reality laboratory: Green robotic ultrasonic welding, in: ASME International Mechanical Engineering Congress and Exposition, Vol. 59421, American Society of Mechanical Engineers, 2019, p. V005T07A033.

\bibitem{wang2019virtual}
Q.~Wang, W.~Jiao, R.~Yu, M.~T. Johnson, Y.~Zhang, Virtual reality robot-assisted welding based on human intention recognition, IEEE Transactions on Automation Science and Engineering 17~(2) (2019) 799--808.

\bibitem{wang2020digital}
Q.~Wang, W.~Jiao, P.~Wang, Y.~Zhang, Digital twin for human-robot interactive welding and welder behavior analysis, IEEE/CAA Journal of Automatica Sinica 8~(2) (2020) 334--343.

\bibitem{abujelala2018collaborative}
M.~Abujelala, S.~Gupta, F.~Makedon, A collaborative assembly task to assess worker skills in robot manufacturing environments, in: Proceedings of the 11th PErvasive Technologies Related to Assistive Environments Conference, 2018, pp. 118--119.

\bibitem{cecil2024study}
J.~Cecil, V.~Gannina, S.~K. Tentu, Study of perception and cognition in immersive digital twins for robotic assembly processes, in: International Conference on Human-Computer Interaction, Springer, 2024, pp. 147--158.

\bibitem{zhu2023learniotvr}
Z.~Zhu, Z.~Liu, Y.~Zhang, L.~Zhu, J.~Huang, A.~M. Villanueva, X.~Qian, K.~Peppler, K.~Ramani, Learniotvr: An end-to-end virtual reality environment providing authentic learning experiences for internet of things, in: Proceedings of the 2023 CHI Conference on Human Factors in Computing Systems, 2023, pp. 1--17.

\bibitem{jun2021human}
M.~B. Jun, H.~Yun, E.~Kim, Human expertise inspired smart sensing and manufacturing, in: 2021 International conference on electronics, communications and information technology (ICECIT), IEEE, 2021, pp. 1--7.

\bibitem{yun2022immersive}
H.~Yun, M.~B. Jun, Immersive and interactive cyber-physical system (i2cps) and virtual reality interface for human involved robotic manufacturing, Journal of Manufacturing Systems 62 (2022) 234--248.

\bibitem{osti_10293626}
F.~Aqlan, M.~Alabsi, E.~Baxter, R.~Sreekanth, \href{https://par.nsf.gov/biblio/10293626}{A small-scale implementation of industry 4.0}, Proceedings of the 2020 IISE Annual Conference.
\newline\urlprefix\url{https://par.nsf.gov/biblio/10293626}

\bibitem{rahman2023workforce}
M.~Rahman, T.~Rahman, R.~R. Rafa, S.~Deb, M.~R. Raihan, Workforce development for composite manufacturing based on immersive technology, CAMX 2023 (2023).

\bibitem{kamali2020virtual}
R.~Kamali-Sarvestani, P.~Weber, M.~Clayton, M.~Meyers, S.~Slade, Virtual reality to improve nanotechnology education: development methods and example applications, IEEE Nanotechnology Magazine 14~(4) (2020) 29--38.

\bibitem{wang2020towards}
F.~Wang, X.~Xu, W.~Feng, J.~A. Bueno-Vesga, Z.~Liang, S.~Murrell, Towards an immersive guided virtual reality microfabrication laboratory training system, in: 2020 IEEE Conference on Virtual Reality and 3D User Interfaces Abstracts and Workshops (VRW), IEEE, 2020, pp. 796--797.

\bibitem{frank2021green}
K.~Frank, A.~E. Gardner, I.~N. Ciobanescu~Husanu, R.~Y. Chiou, R.~Ruane, Green stem: Virtual reality renewable energy laboratory for remote learning, in: ASME International Mechanical Engineering Congress and Exposition, Vol. 85659, American Society of Mechanical Engineers, 2021, p. V009T09A018.

\bibitem{chiou2021developing}
R.~Chiou, H.~V. Nguyen, I.~N.~C. Husanu, T.-L.~B. Tseng, Developing vr-based solar cell lab module in green manufacturing education, in: Proceedings of the 2021 ASEE Virtual Annual Conference Content Access, Virtual/Washington, DC, USA, 2021, pp. 26--29.

\bibitem{chiou2020project}
R.~Chiou, T.~Fegade, Y.-C.~J. Wu, T.-L.~B. Tseng, M.~G. Mauk, I.~N.~C. Husanu, Project-based learning with implementation of virtual reality for green energy manufacturing education, in: 2020 ASEE Virtual Annual Conference Content Access, 2020.

\bibitem{borst2016virtual}
C.~W. Borst, K.~A. Ritter, T.~L. Chambers, Virtual energy center for teaching alternative energy technologies, in: 2016 IEEE Virtual Reality (VR), IEEE, 2016, pp. 157--158.

\bibitem{ritter2016work}
K.~A. Ritter, T.~L. Chambers, C.~W. Borst, Work in progress: Networked virtual reality environment for teaching concentrating solar power technology, in: 2016 ASEE Annual Conference \& Exposition, 2016.

\bibitem{ritter2018virtual}
K.~A. Ritter~III, C.~W. Borst, T.~L. Chambers, Virtual solar energy center case studies, Comput Educ J 9~(3) (2018) 1--7.

\bibitem{borst2018teacher}
C.~W. Borst, N.~G. Lipari, J.~W. Woodworth, Teacher-guided educational vr: Assessment of live and prerecorded teachers guiding virtual field trips, in: 2018 IEEE conference on virtual reality and 3D user interfaces (VR), IEEE, 2018, pp. 467--474.

\bibitem{woodworth2023study}
J.~W. Woodworth, C.~W. Borst, Y.~Rahman, A.~Kulshreshth, Study of visual guidance cues in vr field trips at high schools, in: Proceedings of the 29th ACM Symposium on Virtual Reality Software and Technology, 2023, pp. 1--2.

\bibitem{kula2024development}
B.~Kula, A.~L. Roxas, K.~S. Cetin, A.~Anctil, G.~Berghorn, Development and evaluation of an energy assessment process using virtual reality technology, in: Construction Research Congress 2024, pp. 487--496.

\bibitem{earle2021wicked}
A.~G. Earle, D.~I. Leyva-de~la Hiz, The wicked problem of teaching about wicked problems: Design thinking and emerging technologies in sustainability education, Management Learning 52~(5) (2021) 581--603.

\bibitem{han2023virtual}
B.~Han, D.~J. Weeks, F.~Leite, Virtual reality-facilitated engineering education: A case study on sustainable systems knowledge, Computer Applications in Engineering Education 31~(5) (2023) 1174--1189.

\bibitem{shetty2018strategies}
D.~Shetty, J.~Xu, Strategies to address “design thinking” in engineering curriculum, in: ASME International Mechanical Engineering Congress and Exposition, Vol. 52064, American Society of Mechanical Engineers, 2018, p. V005T07A046.

\bibitem{mcgrath2014cloud}
J.~McGrath, Cloud based mobile tool to enable collaboration and mentoring opportunities in the mime capstone design class at oregon state university, in: IIE Annual Conference. Proceedings, Institute of Industrial and Systems Engineers (IISE), 2014, p. 3629.

\bibitem{evans2021prototyping}
P.~Evans, C.~S{\"o}derlund, et~al., Prototyping remotely together with 2d, 3d and immersive virtual reality design tools, in: DS 110: Proceedings of the 23rd International Conference on Engineering and Product Design Education (E\&PDE 2021), VIA Design, VIA University in Herning, Denmark. 9th-10th September 2021, 2021.

\bibitem{ashour2021connected}
O.~Ashour, J.~Cunningham, C.~Tucker, Connected learning and integrated course knowledge (click) approach, in: ASEE annual conference, 2021.

\bibitem{renner2015virtual}
A.~Renner, J.~Holub, S.~Sridhar, G.~Evans, E.~Winer, A virtual reality application for additive manufacturing process training, in: International Design Engineering Technical Conferences and Computers and Information in Engineering Conference, Vol. 57045, American Society of Mechanical Engineers, 2015, p. V01AT02A033.

\bibitem{parsley2024enhancing}
D.~Parsley, Enhancing engineering education through hands-on virtual reality training experiences: Developing skills in the continuous improvement of manufacturing systems, in: 2024 ASEE Annual Conference \& Exposition, 2024.

\bibitem{ma2019efficacy}
J.~Ma, R.~Jaradat, O.~Ashour, M.~Hamilton, P.~Jones, V.~L. Dayarathna, Efficacy investigation of virtual reality teaching module in manufacturing system design course, Journal of Mechanical Design 141~(1) (2019) 012002.

\bibitem{lopez2020click}
C.~Lopez, O.~Ashour, J.~Cunningham, C.~Tucker, P.~Lynch, The click approach and its impact on learning introductory probability concepts in an industrial engineering course, in: ASEE Annual Conference proceedings, 2020.

\bibitem{yamada2017educational}
K.~Yamada, A.~Tsumaya, T.~Taura, K.~Shimada, T.~Kaihara, Y.~Yokokohji, R.~Sato, et~al., An educational method for enhancing the ability to design innovative products, in: DS 87-9 Proceedings of the 21st International Conference on Engineering Design (ICED 17) Vol 9: Design Education, Vancouver, Canada, 21-25.08. 2017, 2017, pp. 049--058.

\bibitem{hoover2021designing}
M.~Hoover, E.~Winer, Designing adaptive extended reality training systems based on expert instructor behaviors, IEEE Access 9 (2021) 138160--138173.

\bibitem{rumsey2020manufacturing}
A.~Rumsey, C.~A. Le~Dantec, Manufacturing change: the impact of virtual environments on real organizations, in: Proceedings of the 2020 CHI Conference on Human Factors in Computing Systems, 2020, pp. 1--12.

\bibitem{guerrero2023integration}
V.~Guerrero-Hernandez, G.~Reyes-Morales, P.~Jacome-Onofre, J.~A.~O. Moody, F.-A. Matacapan-Toto, M.~A.~M. Herrera, Integration of an industrial control to a digital twin at the industrial level, in: 2023 3rd International Conference on Electrical, Computer, Communications and Mechatronics Engineering (ICECCME), IEEE, 2023, pp. 1--8.

\bibitem{lowell2023virtual}
V.~L. Lowell, A.~C. Ilobinso, Virtual reality in online higher education for learner engagement, interaction, and experiential learning, The SAGE handbook of online higher education (2023) 458.

\bibitem{ipsita2024authoring}
A.~Ipsita, M.~Patel, A.~Unmesh, K.~Ramani, Authoring instructional flow in ivr learning units to promote outcome-oriented learning, Computers \& Education: X Reality 5 (2024) 100074.

\bibitem{roussou2000immersive}
M.~Roussou, Immersive interactive virtual reality and informal education, in: Proceedings of user interfaces for all: interactive learning environments for children, Citeseer, 2000, pp. 1--9.

\bibitem{ghafar2024practical}
Z.~N. Ghafar, Practical applications of virtual reality in continuing education: A review of the case of secondary occupational education, Int. J. Glob. Sustain. Res 2 (2024) 119--132.

\bibitem{serna2022leveraging}
S.~Serna, D.~Weninger, L.~Ranno, K.~Cicek, P.~Brown, S.~Bechtold, J.~J.~A. Uribe, J.~A.~L. Preciado, S.~Saini, S.~Rayyan, et~al., Leveraging moocs in a hybrid learning bootcamp model for training technicians and engineers in stem manufacturing, in: 2022 IEEE Learning with MOOCS (LWMOOCS), IEEE, 2022, pp. 223--226.

\bibitem{chandramouli2019mooc}
M.~Chandramouli, G.~Jin, D.~Cubillos, Mooc videos in project maneuver, in: 2019 IEEE Learning With MOOCS (LWMOOCS), IEEE, 2019, pp. 84--89.

\bibitem{wang2022constructing}
M.~Wang, H.~Yu, Z.~Bell, X.~Chu, Constructing an edu-metaverse ecosystem: A new and innovative framework, IEEE Transactions on Learning Technologies 15~(6) (2022) 685--696.

\bibitem{kamal2024generative}
M.~Kamal, B.~Prabhakaran, Generative ai for 3-d point clouds, IEEE MultiMedia 31~(2) (2024) 5--6.

\bibitem{alkhayat2024leveraging}
A.~Alkhayat, B.~Ciranni, R.~S. Tumuluri, R.~S. Tulasi, Leveraging large language models for enhanced vr development: Insights and challenges, in: 2024 IEEE Gaming, Entertainment, and Media Conference (GEM), IEEE, 2024, pp. 1--6.

\bibitem{marougkas2023virtual}
A.~Marougkas, C.~Troussas, A.~Krouska, C.~Sgouropoulou, Virtual reality in education: a review of learning theories, approaches and methodologies for the last decade, Electronics 12~(13) (2023) 2832.

\bibitem{pavlov1927conditioned}
I.~P. Pavlov, Conditioned reflexes: Oxford university press, London, UK [Google Scholar] (1927).

\bibitem{anderson1995cognitive}
J.~R. Anderson, J.~Crawford, Cognitive psychology and its implications (1995).

\bibitem{vygotsky1978mind}
L.~S. Vygotsky, M.~Cole, Mind in society: Development of higher psychological processes, Harvard university press, 1978.

\bibitem{rogoff1990apprenticeship}
B.~Rogoff, Apprenticeship in thinking: Cognitive development in social context, Oxford university press, 1990.

\bibitem{papert2020mindstorms}
S.~A. Papert, Mindstorms: Children, computers, and powerful ideas, Basic books, 2020.

\bibitem{piaget1973understand}
J.~Piaget, To understand is to invent: The future of education (1973).

\bibitem{gallagher2006body}
S.~Gallagher, How the body shapes the mind, Clarendon press, 2006.

\bibitem{hazarika2023towards}
A.~Hazarika, M.~Rahmati, Towards an evolved immersive experience: Exploring 5g-and beyond-enabled ultra-low-latency communications for augmented and virtual reality, Sensors 23~(7) (2023) 3682.

\bibitem{zhang2023active}
Z.~Zhang, Z.~Xu, L.~Emu, P.~Wei, S.~Chen, Z.~Zhai, L.~Kong, Y.~Wang, H.~Jiang, Active mechanical haptics with high-fidelity perceptions for immersive virtual reality, Nature Machine Intelligence 5~(6) (2023) 643--655.

\bibitem{wang2020design}
F.~Wang, Z.~Qian, Y.~Lin, W.~Zhang, Design and rapid construction of a cost-effective virtual haptic device, IEEE/ASME Transactions on Mechatronics 26~(1) (2020) 66--77.

\bibitem{wiggins2005understanding}
G.~P. Wiggins, J.~McTighe, Understanding by design, Ascd, 2005.

\bibitem{nebeling2019360proto}
M.~Nebeling, K.~Madier, 360proto: Making interactive virtual reality \& augmented reality prototypes from paper, in: Proceedings of the 2019 CHI Conference on Human Factors in Computing Systems, 2019, pp. 1--13.

\bibitem{krauss2022elements}
V.~Krau{\ss}, M.~Nebeling, F.~Jasche, A.~Boden, Elements of xr prototyping: Characterizing the role and use of prototypes in augmented and virtual reality design, in: Proceedings of the 2022 CHI Conference on Human Factors in Computing Systems, 2022, pp. 1--18.

\bibitem{dillenbourg2006sharing}
P.~Dillenbourg, D.~Traum, Sharing solutions: Persistence and grounding in multimodal collaborative problem solving, The Journal of the Learning Sciences 15~(1) (2006) 121--151.

\bibitem{uddin2024exploring}
M.~Uddin, M.~Obaidat, S.~Manickam, S.~U.~A. Laghari, A.~Dandoush, H.~Ullah, S.~S. Ullah, Exploring the convergence of metaverse, blockchain, and ai: A comprehensive survey of enabling technologies, applications, challenges, and future directions, Wiley Interdisciplinary Reviews: Data Mining and Knowledge Discovery 14~(6) (2024) e1556.

\bibitem{chase2001want}
R.~B. Chase, S.~Dasu, Want to perfect your company's service? use behavioral science., Harvard business review 79~(6) (2001) 78--84.

\bibitem{shadowcast2020}
{Donald Bren School of Information and Computer Sciences, University of California, Irvine}, \href{https://www.informatics.uci.edu/shadowcast-novel-virtual-reality-platform-brings-broadway-dreams-to-life/}{Shadowcast: Novel virtual reality platform brings broadway dreams to life}, accessed: 2025-03-18 (March 2020).
\newline\urlprefix\url{https://www.informatics.uci.edu/shadowcast-novel-virtual-reality-platform-brings-broadway-dreams-to-life/}

\end{thebibliography}






\end{document}